%% file: main.tex
\documentclass[12pt]{article}

\usepackage{arxiv}
\usepackage[utf8]{inputenc} 
\usepackage[T1]{fontenc}    
\usepackage[colorlinks=true, allcolors=blue]{hyperref}       
\usepackage{bookmark}
\usepackage{url}            
\usepackage{booktabs}       
\usepackage{nicefrac}       
\usepackage{microtype}      
\usepackage{lipsum}
\usepackage{graphicx}
\usepackage{tikz}
\usetikzlibrary{fit}
\usepackage{amsmath}
\usepackage{amssymb}
\usepackage{amsfonts,amsthm}
\theoremstyle{plain}
\newtheorem{proposition}{Proposition}

\theoremstyle{remark}
\newtheorem{remark}{Remark}
\usepackage{mathtools}
\usepackage{cancel}
\usepackage{bm}
\usepackage{array}
\usepackage{natbib}
\usepackage{colortbl}
\usepackage{float}
\usepackage{xcolor}
\usepackage{todonotes}
\usepackage{svg}
\usepackage{setspace}
\usepackage{pgfplots}
\usepackage{subcaption}
\usepackage{siunitx}
\usepackage{adjustbox}
\usepackage{caption}
\usepackage{algorithm}
\usepackage{algpseudocode}
\usepackage{orcidlink}
\usepackage{chngcntr}
\usepackage[stable]{footmisc}
\graphicspath{ {./img/} }

\makeatletter
\renewcommand\subsubsection{\@startsection{subsubsection}{3}{\z@}%
  {-3.25ex\@plus -1ex \@minus -.2ex}%
  {1.5ex \@plus .2ex}%
  {\normalfont\normalsize\bfseries}}
\makeatother

\newcommand{\mediqr}[2]{\begin{tabular}[t]{@{}c@{}}#1\\[-1pt]{\scriptsize[#2]}\end{tabular}}

\newcommand{\zenodorepo}{\href{https://doi.org/10.5281/zenodo.18876633}{10.5281/zenodo.18876633}} 

\title{Bayesian Inference for Discrete Markov Random Fields Through Coordinate Rescaling}

\author{
 Giuseppe Arena \\
  Department of Psychology\\
  University of Amsterdam\\
  \texttt{g.arena@uva.nl} \\
  \And
 Maarten Marsman \\
  Department of Psychology\\
  University of Amsterdam\\
  \texttt{m.marsman@uva.nl} \\
}

\begin{document}
\maketitle
\vspace{-3.25em}
\begin{center}
\textbf{This manuscript has not yet been peer-reviewed.} 
\end{center}
\vspace{-0.5em}
\correspondence{Giuseppe Arena\orcidlink{0000-0001-5204-3326}, Department of Psychology, University of Amsterdam -- Nieuwe Achtergracht 129-B, PO Box 15906, 1001 NK Amsterdam, The Netherlands -- E-mail:  \texttt{g.arena@uva.nl}. }

\vspace{-0.75em}
\vspace{-0.5em}
\begin{abstract}
Discrete Markov random fields are undirected graphical models that capture complex conditional dependencies between discrete variables. Conducting exact posterior inference in these models is often computationally challenging because evaluating their normalizing constant requires summation over all possible state configurations, and the size of this state space grows exponentially with the number of variables and their possible states. As a result, exact likelihood-based inference is infeasible in many practical settings, and existing methods, such as Double Metropolis-Hastings or pseudo-likelihood approximations, either scale poorly to large systems or underestimate posterior variability. To address these limitations, we propose a new class of coordinate-rescaling sampling methods that transform pseudo-likelihood-based posteriors to approximate the covariance structure of the full likelihood posterior while preserving computational efficiency. In a simulation study and an empirical application, the coordinate-rescaling samplers correct the underestimated posterior uncertainty of the pseudo-likelihood and reproduce the Bayes factors of a full-likelihood reference, with the accuracy of post hoc calibration of the pseudo-posterior and at the cost of the pseudo-likelihood sampler, providing a scalable and reliable approach to Bayesian inference in discrete MRFs.

\end{abstract}

\clearpage 

\section{Introduction}
\label{sec:introduction}

Markov Random Fields (MRFs) are undirected graphical models in which conditional dependencies are encoded by an undirected graph \citep{Besag1974,Kindermann1980}. In this graph, nodes represent random variables; for instance, in genetics, they may reflect gene or protein expression levels \citep{SchaferStrimmer2004, DobraEtAl2004}, and, in psychology, the severity of mental health symptoms \citep{Borsboom2008, CramerEtAl_2010_BBS}. Edges, in turn, encode conditional dependencies: two variables are conditionally independent given the remaining variables if no edge connects them. We restrict attention to MRFs with at most pairwise interactions, referred to as pairwise MRFs \citep{Besag1974, lauritzen1996}.  Standard examples of pairwise MRFs include the Ising model for binary variables \citep{Ising1925} and the Gaussian graphical model (GGM) for continuous variables \citep{Dempster_1972}. In these models, conditional dependencies are parameterized through pairwise interaction parameters that represent partial associations between variables.

In this paper, we focus on the Bayesian analysis of MRFs for discrete variables. Discrete MRFs often give rise to doubly intractable posteriors \citep{Murray2006}: the likelihood and posterior both involve an intractable normalizing constant. By contrast, continuous MRFs like the GGM have a tractable likelihood, though double intractability may arise through priors with intractable normalizing constants (e.g., the G-Wishart prior; \citealp{Roverato_2002}). We present a computationally efficient method for addressing double intractability in Bayesian inference of discrete MRFs.

The computational difficulty in discrete MRFs results from the intractable normalizing constant in the likelihood. The likelihood for parameter vector $\bm{\eta}$ is given by
\[
    f(\mathbf{X};\bm{\eta}) = \frac{1}{Z(\bm{\eta})}\exp{\left\lbrace -E(\mathbf{X};\bm{\eta})\right\rbrace},
\]
where $E(\mathbf{X};\bm{\eta})$ denotes a real-valued energy function defined on configurations $\mathbf{X}$. The normalizing constant in the denominator
\[ Z(\bm{\eta}) = \sum_{\mathbf{X}^\prime \in \mathcal{X}}{\exp{\left\lbrace-E(\mathbf{X}^\prime;\bm{\eta})\right\rbrace}}, \]
sums the exponential of the energy function over the entire state space $\mathcal{X}$. Computing $Z(\bm{\eta})$ therefore requires enumeration of all possible configurations. Because the size of the state space grows exponentially with the number of variables and their possible states, this cost quickly becomes prohibitive. For instance, $10$ variables measured on two categories yield $2^{10} = 1{,}024$ possible states, whereas $10$ variables measured on four categories yield $4^{10} = 1{,}048{,}576$ states. 

We obtain the posterior distribution by combining the likelihood $f(\mathbf{X};\bm{\eta})$ with a prior distribution $\pi(\bm{\eta})$ through Bayes' rule. The posterior distribution of the MRF parameters is given by
\[
    \pi(\bm{\eta}\mid{}\mathbf{X}) = \frac{f(\mathbf{X};\bm{\eta})\pi(\bm{\eta})}{f(\mathbf{X})},
\]    
where the marginal likelihood 
\[
f(\mathbf{X}) = \int_{\mathbb{R}^{|\bm{\eta}|}}{f(\mathbf{X};\bm{\eta})\pi(\bm{\eta}) \ d\bm{\eta}}
\]
is intractable, integrating a likelihood that itself contains an intractable normalizing constant. As a result, the posterior distribution is doubly intractable.

\begin{figure}[tp]
\centering
\begin{tikzpicture}[
    node style/.style={circle, draw=gray!70, minimum size=1cm},
    label style/.style={font=\small, text=black, inner sep=1pt},
    thick_arrow/.style={->, thick, black},
    edge/.style={draw=gray},
    edgeweak/.style={edge, line width=0.8pt, gray!30},
    edgemed/.style={edge, line width=1.4pt, gray!60},
    edgestrong/.style={edge, line width=2.2pt, gray!90}
  ]
    \foreach [count=\idx] \lab in {i,j,k,l,m,n}{
      \node[node style] (X\idx) at ({360/6 * (\idx-1)}:2cm) {$X_{\lab}$};
    }
    \draw[edgestrong] (X1) --  (X2) node[midway, right=3pt, label style] (thetaij) {$\theta_{ij}$};
    \draw[edgeweak] (X1) -- (X3);
    \draw[edgeweak] (X1) -- (X4);
    \draw[edgemed] (X1) -- (X5);
    \draw[edgeweak] (X1) -- (X6);
    \draw[edgeweak] (X2) -- (X3);
    \draw[edgeweak] (X2) -- (X4);
    \draw[edgemed] (X2) -- (X5);
    \draw[edgeweak] (X2) -- (X6);
    \draw[edgemed] (X3) -- (X4);
    \draw[edgeweak] (X3) -- (X5);
    \draw[edgeweak] (X3) -- (X6);
    \draw[edgemed] (X4) -- (X5);
    \draw[edgemed] (X4) -- (X6);
    \draw[edgeweak] (X5) -- (X6);
  \node[draw=gray!70, thick, inner sep=0pt] (imganchor) at (8.2,0) {
  \def\svgwidth{0.50\textwidth}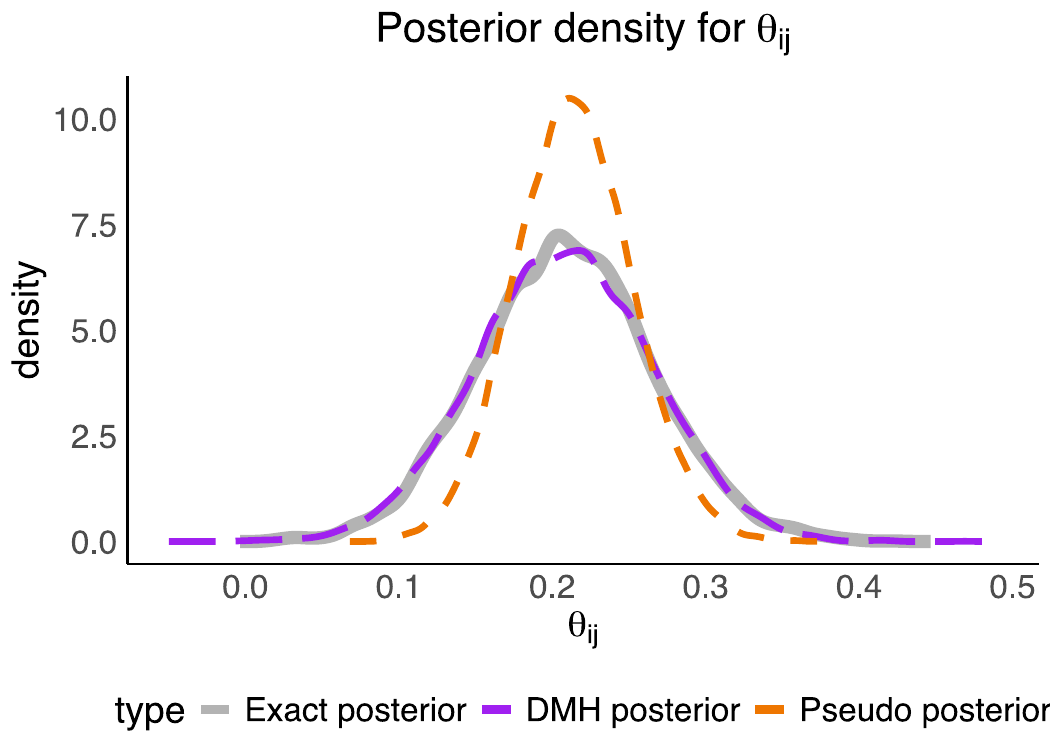};
  \draw[gray!70] (thetaij.east) -- (imganchor.west);
\end{tikzpicture}
\caption{\textit{(Left)} Example of an undirected network of six variables. Edges indicate pairwise conditional associations between variables. Here, the parameter $\theta_{ij}$ denotes the conditional dependence between variables $X_i$ and $X_j$ given the remaining variables in the network. Grayscale intensity and edge thickness illustrate the strength of the associations, with stronger associations corresponding to darker and thicker edges. 
\textit{(Right)} Posterior density of the pairwise association $\theta_{ij}$ between two of the six dysphoria items of the RADS-2 data ($n = 917$; Section \ref{sec:empirical-application}). The gray solid line shows the posterior based on the full likelihood (exact posterior), the orange dashed line the posterior based on the pseudo-likelihood function (pseudo-posterior), and the purple dashed line the posterior obtained using the DMH algorithm (DMH posterior).}
\label{fig:exact_vs_pseudo_vs_dmh_posterior_density_comparison}
\end{figure}

Several strategies have been proposed to address double intractability in discrete MRFs \citep[e.g., see][for a review]{ParkHaran_2018}. A prominent example is the Double Metropolis-Hastings (DMH) algorithm of \citet{Liang_2010}, which \citet{ParkHaran_2018} recommend as a natural starting point due to its ease of implementation and typically good computational efficiency. DMH approximates the ratio of intractable normalizing constants in the Metropolis-Hastings acceptance ratio by replacing the exact samples used in single-variable exchange proposals \citep{Murray2006} with MCMC samples. Although DMH can produce accurate posterior estimates, the additional MCMC sampling required incurs substantial computational cost (runtime $\approx 15$ minutes\footnote{\label{fn:runtime}Time performance for $25{,}000$ iterations ($20{,}000$ posterior draws after burn-in) with the settings of Section~\ref{sec:simulation-study}, based on C\texttt{++}-coded functions, run on a Macbook Air (2024) with Apple M3 chip, 8-CPUs and 16Gb RAM.} for the example in Figure \ref{fig:exact_vs_pseudo_vs_dmh_posterior_density_comparison}), which limits its scalability to large networks. 

Pseudo-likelihood functions \citep{Besag_1975, Lindsay1988} offer an alternative to full-likelihood-based inference by approximating the likelihood with a product of conditional distributions, thereby avoiding the need to compute the normalizing constant. 
For discrete MRFs, estimators based on the pseudo-likelihood are consistent, providing a principled and computationally feasible alternative to full likelihood-based inference \citep[e.g., ][]{ArnoldStrauss_1991, GeysEtAl_2007, Miller2021}.
In this setting, pseudo-likelihood estimation is fast (runtime $\approx 4$ seconds\footref{fn:runtime} for the example in Figure \ref{fig:exact_vs_pseudo_vs_dmh_posterior_density_comparison}) and yields parameter estimates whose bias is comparable to that of maximum likelihood estimation, as shown for the Ising model by \citet{Keetelaar2024}. Although pseudo-likelihood performs poorly in other model classes, such as exponential random graph models (ERGMs; \citealp{vanDuijn2009,Schmid2017}) and GGMs \citep{Huth2025}, this limitation does not arise for discrete MRFs. 
However, pseudo-likelihood-based posterior distributions systematically underestimate posterior variability \citep{Miller2021}, as illustrated in Figure \ref{fig:exact_vs_pseudo_vs_dmh_posterior_density_comparison}.

To address the systematic underestimation of posterior variability associated with pseudo-likelihood inference, \citet{Bouranis2017} proposed a post hoc rescaling of the pseudo-posterior in the context of ERGMs. Such rescaling belongs to a broader class of covariance adjustments for composite- and pseudo-likelihood posteriors, which correct the over-concentration of the composite posterior by rescaling it so that its covariance matches the inverse of the Godambe (sandwich) information \citep{Ribatet2012, Pauli2011}. Their method extends beyond ERGMs and applies to discrete MRFs, where pseudo-posterior distributions likewise underestimate variability, albeit without the location bias observed in ERGMs.

To address this challenge, we make two contributions. First, we introduce a family of MCMC samplers, named \emph{coordinate-rescaling} (CoRe) methods, that embed a covariance adjustment of the pseudo-posterior directly within the sampling scheme. Through a linear transformation of the parameters, the samplers target a transformed pseudo-posterior whose covariance approximates that of the full likelihood posterior, at the computational cost of the pseudo-likelihood sampler. In contrast to post hoc calibration, the adjustment is applied within the sampler rather than to stored draws; in contrast to MCMC preconditioning, it modifies the target distribution rather than only the proposal. Second, we compare the coordinate-rescaling methods with post hoc calibration \citep{Bouranis2017} and DMH sampling \citep{Liang_2010} in a simulation study with networks of up to $22$ variables, using the exact posterior or, for larger networks, DMH as reference, and in an application to a depression scale with $25$ items.

The remainder of the paper is organized as follows. Section 2 introduces the full likelihood and pseudo-likelihood formulations for ordinal MRFs \citep{SuggalaEtAl_2017_OrdinalTale, Marsman2025}, a class of discrete MRFs for variables measured on an ordered scale. We focus on ordinal MRFs because they pose a substantially more severe computational challenge than binary MRFs, while the proposed methodology readily extends to general discrete MRFs. Section 3 introduces the \textit{coordinate-rescaling} methods and details their theoretical foundation, including how rescaling the pseudo-posterior approximates the scale of the full posterior. Section 4 describes the competing methods, the reference posterior against which all methods are evaluated, and the evaluation metrics. Section 5 compares the proposed methodology with the competing methods in a simulation study, highlighting their relative strengths and limitations, and Section 6 applies all methods to the Reynolds Adolescent Depression Scale \citep[RADS-2;][]{Reynolds2002}. Finally, Section 7 discusses the results, summarizes the advantages and limitations of the proposed methods, and outlines directions for future research.

\section{Discrete Markov Random Fields}

A discrete Markov random field models the conditional dependencies between discrete random variables. Here, we restrict attention to the ordinal case. Let $\mathbf{X} = \left(\mathbf{X}_1,\ldots, \mathbf{X}_n\right)$ denote the observed data for $n$ samples on $p$ ordinal variables, where each variable assumes $m+1$ categories. For each sample $\nu=1,\ldots,n$, let $\mathbf{X}_{\nu} = \left(X_{\nu 1},\ldots,X_{\nu p}\right)$ denote the $p$-variate vector of observed values.

The full likelihood for a sample of $n$ observations is given by the product of the likelihood contribution of each single observation:
\begin{equation}
\begin{split}
f(\mathbf{X};\bm{\eta}) &= \prod_{v=1}^{n} {f(\mathbf{X}_{v};\bm{\eta})} = \frac{1}{Z(\bm{\eta})^{n}}\exp{\left\lbrace-\sum_{\nu=1}^{n}{E(\mathbf{X}_{\nu};\bm{\eta})}\right\rbrace}.
\end{split}
    \label{eq:sample_model_ordinal_mrf_model}
\end{equation}
Here $\bm{\eta}\in\mathbb{R}^{|\bm{\eta}|}$ denotes the vector of model parameters, comprising marginal effects of the $p$ variables and their pairwise associations. The observed configuration in the $\nu\text{-th}$ sample is denoted by $\mathbf{X}_{\nu}\in\mathcal{X}$, where $\mathcal{X} = \left\lbrace0,1,\ldots,m\right\rbrace^{p}$. The ordered structure of $\mathcal{X}$ induces an ordinal Markov random field (OMRF; \citealp{Marsman2025, MarsmanEtAl_2025_ttest}). 

The sum of the energy $E(\cdot)$ over all observed states is
\[
-\sum_{\nu=1}^{n}{E(\mathbf{X}_{\nu};\bm{\eta})}= \sum_{i=1}^{p}{\sum_{h=1}^{m}{\mu_{ih}\left(\sum_{v=1}^{n}{\mathcal{I}(X_{\nu i} = h)}\right)}} + \sum_{i<j}{\theta_{ij}\left(\sum_{v=1}^{n}{X_{\nu i} X_{\nu j}}\right)} =
\mathbf{s}\left(\mathbf{X}\right)^{\intercal}\bm{\eta},
\]
the inner product of the sufficient statistics $\mathbf{s}\left(\mathbf{X}\right)$ and the parameter vector $\bm{\eta}$. 

The sufficient statistics consist of two types: $p\times m$ observed category frequencies for each variable (excluding the baseline categories) and $p\times(p-1)/2$ summed cross products between variable pairs. The corresponding parameter vector $\bm{\eta}$ comprises the threshold parameters $\mu_{ih}$, which quantify the tendency of variable $i$ to take on a specific non-baseline category $h$ that is not explained by its interactions with other variables, and the interaction parameters $\theta_{ij}$, which capture the pairwise conditional dependency between variables $i$ and $j$. When the $p$ variables are measured on a binary scale, the OMRF reduces to the Ising model \citep{Ising1925}. 

The full likelihood function of the OMRF in (\ref{eq:sample_model_ordinal_mrf_model}) accounts for all pairwise dependencies between variables in the network. Bayesian inference using this likelihood is computationally expensive because of its intractable normalizing constant. We therefore adopt a composite likelihood approach.

\subsection{The Pseudo-Likelihood Function}
\label{subsec:pseudo-likelihood}
\par The pseudo-likelihood is a specific form of composite likelihood that approximates the full likelihood by expressing it as a product of conditional likelihoods. This construction avoids the intractable normalizing constant of the full likelihood \citep{Lindsay1988, Besag1986}, making Bayesian inference computationally feasible. 

Let $\mathbf{X}_{\nu(-i)}$ denote the variable vector of observation $\nu$ excluding variable $X_i$. The pseudo-likelihood for the single observation $\mathbf{X}_{\nu}$ is defined as the product of the conditional distributions of each variable given the remaining variables,
\begin{equation}
\begin{split} \tilde{f}(\mathbf{X}_{\nu};\bm{\eta}) &= \prod_{i=1}^{p}{f\left(X_{\nu i} \mid \mathbf{X}_{\nu(-i)}, \bm{\eta}\right)}
\end{split}
\label{eq:pseudo_likelihood_mrf}
\end{equation}

The pseudo-likelihood function for a sample of size $n$ is the product of the contributions of the individual observations, 
\begin{equation}
\begin{split}
\tilde{f}(\mathbf{X};\bm{\eta}) &= \prod_{\nu=1}^{n}{\tilde{f}(\mathbf{X}_{\nu};\bm{\eta})}.
\end{split}
    \label{eq:sample_pseudo_likelihood_mrf_model}
\end{equation}
The corresponding normalizing constant factorizes across observations,
\[
\tilde{Z}(\mathbf{X},\bm{\eta}) = \prod_{\nu=1}^{n}\tilde{Z}(\mathbf{X}_{\nu},\bm{\eta}) = \prod_{\nu=1}^{n}{\prod_{i=1}^{p}{\left[1+\sum_{h=1}^{m}{\exp{\left(\mu_{ih}+h\sum_{j\neq i}^{}{\theta_{ij}X_{\nu j}}\right)}}\right]}}.
\]
This factorization is computationally tractable. 

Combined with a prior $\pi(\bm{\eta})$, the pseudo-likelihood defines the \emph{pseudo posterior}
\begin{equation}
  \tilde\pi(\bm{\eta} \mid \mathbf{X}) \propto \tilde f(\mathbf{X};\bm{\eta})\,\pi(\bm{\eta}),
  \label{eq:pseudo-posterior}
\end{equation}
which forms the basis for the inference developed in Section \ref{sec:coordinate-rescaling-methods}.

In the context of OMRFs, the quality of posterior inference depends on balancing computational tractability with accurate posterior uncertainty quantification. The use of the full likelihood is feasible only when the number of variables is sufficiently small to evaluate the normalizing constant within a reasonable time. In most practical settings, the pseudo-posterior $\tilde\pi(\bm{\eta} \mid \mathbf{X})$ provides a computationally scalable approximation to the full posterior, but it typically underestimates posterior variance. In the next section, we address this trade-off by introducing a methodology that retains the scalability of the pseudo-likelihood while improving posterior uncertainty quantification.

\section{Coordinate Rescaling for Posterior Inference}
\label{sec:coordinate-rescaling-methods}
\par Posterior inference for OMRFs relies on sampling-based methods. Sampling from the exact posterior requires evaluating the normalizing constant $Z(\bm{\eta})$ at each iteration. In a standard Metropolis-Hastings algorithm, $ q(\bm{\eta}^\prime\mid \bm{\eta})$ denotes a proposal distribution that generates a candidate move from $\bm{\eta}$ to $\bm{\eta}^\prime$. The move is accepted with probability
\[
\alpha(\bm{\eta},\bm{\eta}^\prime) = \min\left\lbrace 1, \frac{\pi(\bm{\eta}^\prime\mid \mathbf{X})\,q(\bm{\eta}\mid \bm{\eta}^\prime)}{\pi(\bm{\eta}\mid \mathbf{X})\,q(\bm{\eta}^\prime\mid \bm{\eta})} \right\rbrace = \min\left\lbrace 1, \frac{f(\mathbf{X};\bm{\eta}^\prime)\,\pi(\bm{\eta}^\prime)\,q(\bm{\eta}\mid \bm{\eta}^\prime)}{f(\mathbf{X};\bm{\eta})\,\pi(\bm{\eta})\,q(\bm{\eta}^\prime\mid \bm{\eta})}\right\rbrace,
\]
where the likelihood ratio $\frac{f(\mathbf{X};\bm{\eta}^\prime)}{f(\mathbf{X};\bm{\eta})}$ contains the intractable ratio of normalizing constants $Z(\bm{\eta})/Z(\bm{\eta}^\prime)$. 

Sampling procedures such as single-variable exchange \citep[SVE;][]{Murray2006} eliminate the intractable normalizing constants from the acceptance ratio by introducing auxiliary data. Exact exchange algorithms like SVE require generating auxiliary datasets under the model via perfect sampling \citep{Propp1996}, which is computationally demanding. Double Metropolis-Hastings \citep[DMH;][]{Liang_2010} approximates this procedure by replacing exact auxiliary sampling with an inner MCMC run. 

Under exact exchange sampling, auxiliary data $\mathbf{Y}\sim p(\cdot \mid\bm{\eta}^\prime)$ define the acceptance probability
\begin{align*}
    \alpha(\bm{\eta},\bm{\eta}^\prime) = \min\left\lbrace 1, \frac{f(\mathbf{X};\bm{\eta}^\prime)\,f(\mathbf{Y};\bm{\eta})\,\pi(\bm{\eta}^\prime)\,q(\bm{\eta}\mid\bm{\eta}^\prime)}{f(\mathbf{X};\bm{\eta})\,f(\mathbf{Y};\bm{\eta}^\prime)\,\pi(\bm{\eta})\,q(\bm{\eta}^\prime\mid\bm{\eta})}\right\rbrace \\ 
    = \min\left\lbrace 1, \frac{\exp{\left\lbrace\mathbf{s}\left(\mathbf{X}\right)^{\intercal}\bm{\eta}^\prime\right\rbrace}\,\cancel{Z(\bm{\eta})}\,\exp{\left\lbrace\mathbf{s}\left(\mathbf{Y}\right)^{\intercal}\bm{\eta}\right\rbrace}\,\cancel{Z(\bm{\eta}^\prime)}\,\pi(\bm{\eta}^\prime)\,q(\bm{\eta}\mid\bm{\eta}^\prime)}{\exp{\left\lbrace\mathbf{s}\left(\mathbf{X}\right)^{\intercal}\bm{\eta}\right\rbrace}\,\cancel{Z(\bm{\eta}^\prime)}\,\exp{\left\lbrace\mathbf{s}\left(\mathbf{Y}\right)^{\intercal}\bm{\eta}^\prime\right\rbrace}\,\cancel{Z(\bm{\eta})}\,\pi(\bm{\eta})\,q(\bm{\eta}^\prime\mid\bm{\eta})}\right\rbrace,
\end{align*}
where the normalizing constants cancel from the acceptance ratio. In DMH, auxiliary data $\mathbf{Y}$ are obtained as the endpoint of a finite MCMC run, yielding an approximation to the exact exchange algorithm. Despite this cancellation, exchange-based approaches remain computationally demanding: exact sampling is costly in high-dimensional models, and MCMC-based approximations incur an additional layer of computational burden due to nested sampling. In our implementation, we use a Monte Carlo variant of DMH: instead of cancelling the normalizing constants with a single auxiliary dataset, we estimate their ratio, together with the gradient of the full likelihood required by the Langevin proposal, from $L$ auxiliary draws simulated at the proposed parameters by a short Gibbs run (Appendix \ref{appendix:the_fishermala_sampling_algorithm_and_priors}).

We propose a new sampling strategy that combines the computational efficiency of the pseudo-likelihood with improved posterior uncertainty quantification. The method builds on post hoc calibration of pseudo-posterior samples \citep{Bouranis2017} and is tailored to discrete MRFs. We refer to this method as \emph{coordinate rescaling} (CoRe).

CoRe applies a linear transformation to the parameter space during sampling, resulting in a rotation of the posterior contours. The transformation is determined by a rescaling matrix that approximates the variance-covariance structure of the full posterior.

\subsection{Sampling via Coordinate Rescaling}
\label{subsection:the-core-sampler}
\par Let $\bm{\eta}$ denote the parameters on the pseudo-posterior scale, and let $\bm{\beta}$ denote the parameters on the transformed scale. We define the linear transformation 
\[
\bm{\beta} = \mathbf{A} (\bm{\eta} - \bm{\eta}^{\star}) + \bm{\eta}^{\star}.
\] 
Here $\bm{\eta}^{\star}$ is a point estimate, such as a maximum a posteriori (MAP) or a maximum pseudo-likelihood estimate (MPLE) for $\bm{\eta}$. The matrix $\mathbf{A}$ rescales the pseudo-posterior geometry to approximate the covariance structure of the full posterior. 

The linear transformation aligns with that proposed by \citet{Bouranis2017} for post hoc calibration of pseudo-likelihood-based posterior draws. The key difference is that we embed the transformation within the sampling scheme and therefore do not require iterative estimation of $\bm{\eta}^{\star}$ since the MPLE and the MLE have comparable bias \citep[as shown for the Ising model by][]{Keetelaar2024}. In our simulations, the pseudo-posterior mode of the thresholds lies within $0.07$ posterior standard deviations of the full-posterior mode (Table \ref{tab:simulation-threshold-asymmetry}). For a given rescaling matrix and center, the CoRe target is therefore the distribution of the post hoc calibrated pseudo-posterior draws: the two approaches differ in where the transformation is applied, not in the distribution they approximate. Applying it within the sampler makes the calibrated scale available to every move of the chain, which matters when the sampler contains further steps that depend on the posterior scale, such as the edge-selection moves discussed in Section \ref{sec:discussion}.

Let $\tilde\pi(\bm{\eta} \mid \mathbf{X}) \propto \tilde f(\mathbf{X};\bm{\eta})\,\pi(\bm{\eta})$ be the pseudo-posterior of Section~\ref{subsec:pseudo-likelihood} (Eq.~\ref{eq:pseudo-posterior}). CoRe defines a distribution $\pi(\bm{\beta} \mid \mathbf{X})$ on the transformed parameters $\bm{\beta}$ through the change of variables $\bm{\eta} \to \bm{\beta}$, with inverse $\bm{\eta}(\bm{\beta}) = \mathbf{A}^{-1}(\bm{\beta} - \bm{\eta}^\star) + \bm{\eta}^\star$:
\begin{equation}
  \pi(\bm{\beta} \mid \mathbf{X})
  = \tilde\pi\big(\bm{\eta}(\bm{\beta}) \mid \mathbf{X}\big)\,\left|\det\left(\mathbf{A}\right)^{-1}\right|
  \propto\tilde f\big(\mathbf{X};\bm{\eta}(\bm{\beta})\big)\,\pi\big(\bm{\eta}(\bm{\beta})\big),
  \label{eq:core-target}
\end{equation}
where the Jacobian $\left|\det\left(\mathbf{A}\right)^{-1}\right|$ is constant because $\mathbf{A}^{-1}$ is fixed during sampling. The density carried through the change of variables is the \emph{pseudo}-posterior $\tilde\pi$. Hence, $\pi(\bm{\beta} \mid \mathbf{X})$ is a transformed pseudo-posterior, not the full likelihood posterior. It is constructed to approximate the full posterior in location and scale, made precise in Section~\ref{sec:core-interp} (Proposition~\ref{prop:secondorder}).

The sampler targets the transformed pseudo-posterior $\pi(\bm{\beta}\mid\mathbf{X})$ of
Eq. \eqref{eq:core-target}  from a proposal distribution $q(\bm{\beta}^\prime\mid\bm{\beta})$ on the $\bm{\beta}$-scale. The Metropolis-Hastings acceptance probability is
\begin{equation*}
\begin{aligned}
\alpha(\bm{\beta},\bm{\beta}^\prime) = \min\left\lbrace 1, \frac{\pi(\bm{\beta}^\prime\mid\mathbf{X})\,q(\bm{\beta}\mid\bm{\beta}^\prime)}{\pi(\bm{\beta}\mid\mathbf{X})\,q(\bm{\beta}^\prime\mid\bm{\beta})} \right\rbrace \\ = \min\left\lbrace 1, \frac{\tilde{f}\left(\mathbf{X};\bm{\eta}(\bm{\beta}^\prime)\right)\,\pi(\bm{\eta}(\bm{\beta}^\prime))\,\cancel{\left|\det\left(\mathbf{A}\right)^{-1}\right|}\,q(\bm{\beta}\mid\bm{\beta}^\prime)}{\tilde{f}\left(\mathbf{X};\bm{\eta}(\bm{\beta})\right)\,\pi(\bm{\eta}(\bm{\beta}))\,\cancel{\left|\det\left(\mathbf{A}\right)^{-1}\right|}\,q(\bm{\beta}^\prime\mid\bm{\beta})}\right\rbrace.
\end{aligned}
\end{equation*}

Because the constant Jacobian cancels, the acceptance ratio reduces to the standard Metropolis-Hastings ratio for the target $\pi(\bm{\beta} \mid \mathbf{X})$, which gives the following.

\begin{proposition}[Detailed balance]
\label{prop:detailed-balance}
Let $q(\cdot \mid \cdot)$ be any proposal on the $\bm{\beta}$-scale, and let the CoRe
kernel accept a move $\bm{\beta} \to \bm{\beta}^\prime$ with the probability
$\alpha(\bm{\beta},\bm{\beta}^\prime)$ of Algorithm \ref{algorithm:core_sampler}. Then the kernel satisfies detailed
balance with respect to the transformed pseudo-posterior $\pi(\bm{\beta} \mid \mathbf{X})$ in Eq. \eqref{eq:core-target}. Consequently, $\pi(\bm{\beta} \mid \mathbf{X})$ is its invariant distribution.
\end{proposition}

\begin{proof}
Write $\pi(\bm{\beta} \mid \mathbf{X}) = c\,\tilde f(\mathbf{X};\bm{\eta}(\bm{\beta}))\,\pi(\bm{\eta}(\bm{\beta}))$, where $c > 0$ combines the constant Jacobian $\left|\det\left(\mathbf{A}\right)^{-1}\right|$ and the normalizing constant, and is identical for every state. With $\alpha(\bm{\beta},\bm{\beta}^\prime) = \min\{1, r\}$ and $r = \tilde f(\mathbf{X};\bm{\eta}(\bm{\beta}^\prime))\,\pi(\bm{\eta}(\bm{\beta}^\prime))\,q(\bm{\beta} \mid \bm{\beta}^\prime) \,/\, [\tilde f(\mathbf{X};\bm{\eta}(\bm{\beta}))\,\pi(\bm{\eta}(\bm{\beta}))\,q(\bm{\beta}^\prime \mid \bm{\beta})]$, multiply by the positive factor $\pi(\bm{\beta} \mid \mathbf{X})\,q(\bm{\beta}^\prime \mid \bm{\beta})$ and distribute it into the minimum:
\begin{equation}
\pi(\bm{\beta} \mid \mathbf{X})\,q(\bm{\beta}^\prime \mid \bm{\beta})\,\alpha(\bm{\beta},\bm{\beta}^\prime)
= \min\!\big\{\,
    \pi(\bm{\beta} \mid \mathbf{X})\,q(\bm{\beta}^\prime \mid \bm{\beta}),\;
    \pi(\bm{\beta}^\prime \mid \mathbf{X})\,q(\bm{\beta} \mid \bm{\beta}^\prime)
  \,\big\}.
\label{eq:db-sym}
\end{equation}
The second entry follows because $c$ and the Jacobian cancel in $\pi(\bm{\beta} \mid \mathbf{X})\,q(\bm{\beta}^\prime \mid \bm{\beta})\,r = \pi(\bm{\beta}^\prime \mid \mathbf{X})\,q(\bm{\beta} \mid \bm{\beta}^\prime)$. The right-hand side of \eqref{eq:db-sym} is unchanged under the exchange $\bm{\beta} \leftrightarrow \bm{\beta}^\prime$. Applying \eqref{eq:db-sym} with the roles of $\bm{\beta}$ and $\bm{\beta}^\prime$ reversed therefore gives
\begin{equation}
\pi(\bm{\beta} \mid \mathbf{X})\,q(\bm{\beta}^\prime \mid \bm{\beta})\,\alpha(\bm{\beta},\bm{\beta}^\prime)
= \pi(\bm{\beta}^\prime \mid \mathbf{X})\,q(\bm{\beta} \mid \bm{\beta}^\prime)\,\alpha(\bm{\beta}^\prime,\bm{\beta}),
\end{equation}
i.e. the CoRe kernel is reversible with respect to $\pi(\bm{\beta} \mid \mathbf{X})$, which is therefore its invariant distribution \citep{Robert2004}.
\end{proof}

Proposition \ref{prop:detailed-balance} holds for any proposal $q(\cdot \mid \cdot)$ on the $\bm{\beta}$-scale. In particular, the invariant distribution does not depend on the proposal, which affects only how efficiently the chain explores $\pi(\bm{\beta} \mid \mathbf{X})$. Appendix \ref{appendix:random_walk_proposal} illustrates this with a random-walk proposal. Algorithm \ref{algorithm:core_sampler} summarizes the CoRe sampling procedure.
\begin{algorithm}[t]
    \caption{Coordinate Rescaling (CoRe) sampling scheme}
    \label{algorithm:core_sampler}
    \begin{algorithmic}
    \State{\textbf{Input:}} $\bm{\beta}$ and $\bm{\eta}$ at the $s$-th iteration, the observed data $\mathbf{X} = \left(\mathbf{X}_{1},\ldots,\mathbf{X}_n\right)$, the inverse rescaling matrix $\mathbf{A}^{-1}$, and $\bm{\eta}^{\star}$.
    \State Propose $\bm{\beta}^\prime \sim q(\cdot \mid \bm{\beta})$.
    \State Calculate $\bm{\eta}^\prime = \mathbf{A}^{-1}\left(\bm{\beta}^\prime - \bm{\eta}^{\star}\right) + \bm{\eta}^{\star}$.
    \State Accept $\bm{\beta}_{s+1} = \bm{\beta}^\prime$ and $\bm{\eta}_{s+1} = \bm{\eta}^\prime$ with probability $$ \alpha(\bm{\beta},\bm{\beta}^\prime) = \min\left\lbrace 1, \frac{\tilde{f}\left(\mathbf{X};\bm{\eta}^\prime\right)\,\pi(\bm{\eta}^\prime)\,q(\bm{\beta}\mid\bm{\beta}^\prime)}{\tilde{f}\left(\mathbf{X};\bm{\eta}\right)\,\pi(\bm{\eta})\,q(\bm{\beta}^\prime \mid \bm{\beta})}\right\rbrace,$$ otherwise reject and set $\bm{\beta}_{s+1} = \bm{\beta}$ and $\bm{\eta}_{s+1} = \bm{\eta}$.
    \end{algorithmic}
\end{algorithm}

\begin{figure}[t]
\centering
\def\svgwidth{0.9\textwidth}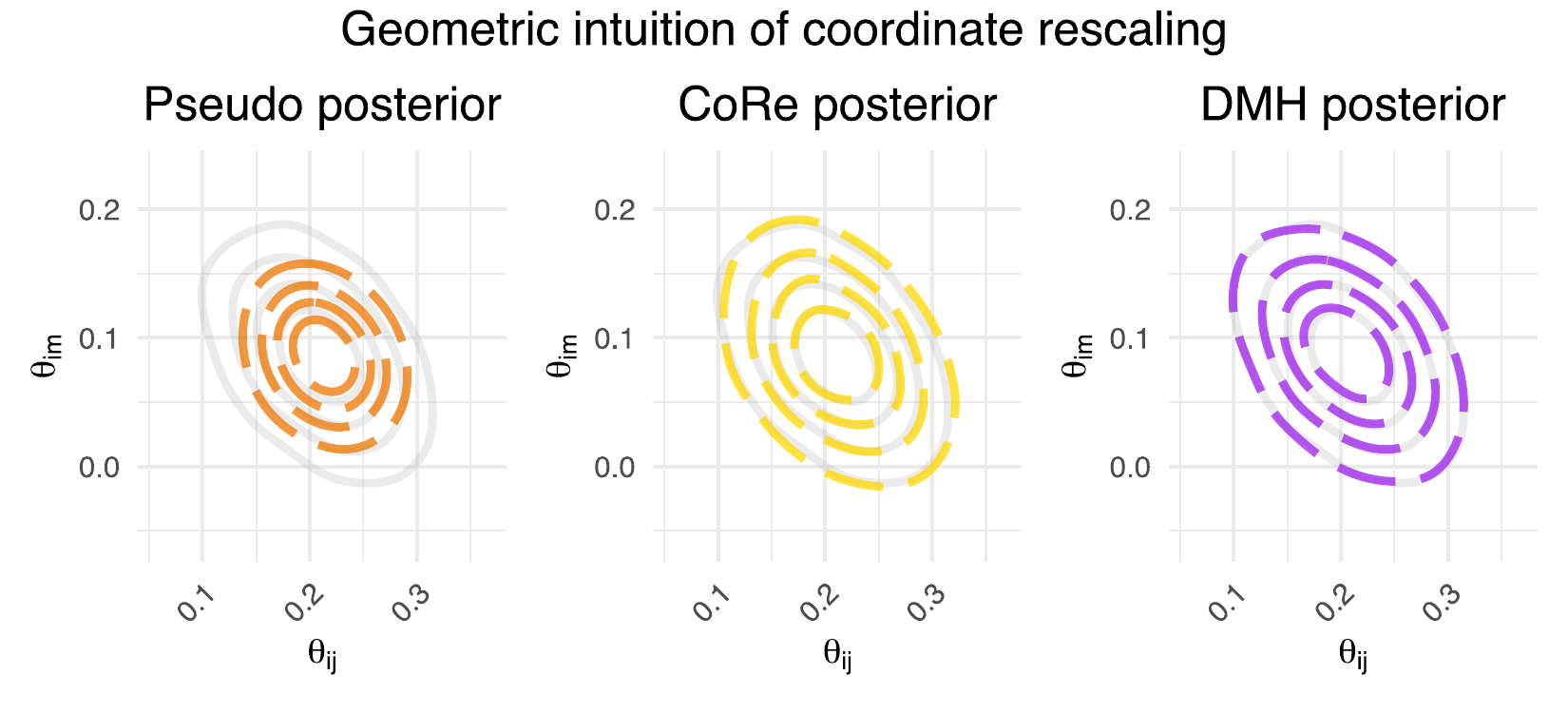
\caption{Contour lines of the joint posterior distribution of two pairwise association parameters, $\theta_{ij}$ of Figure \ref{fig:exact_vs_pseudo_vs_dmh_posterior_density_comparison} and $\theta_{im}$, which shares variable $X_i$ with it. \textit{(Left)} Pseudo-posterior (orange dashed contours) and exact posterior (gray solid contours). \textit{(Center)} CoRe posterior (gold dashed contours) and exact posterior (gray solid contours). \textit{(Right)} DMH posterior (purple dashed contours) and exact posterior (gray solid contours).}
\label{fig:geometric_intuition_target_rescaling}
\end{figure}

Figure \ref{fig:geometric_intuition_target_rescaling} illustrates the joint posterior distribution of two parameters, $\theta_{ij}$ and $\theta_{im}$, introduced in Section \ref{sec:introduction}. The pseudo-posterior (left) exhibits overly concentrated contours relative to the exact posterior (gray), indicating underestimation of posterior variability. The CoRe posterior (center) expands and rotates the contours toward the geometry of the exact posterior, substantially reducing this distortion, although visible discrepancies remain in finite samples. The DMH posterior (right) shows the closest alignment with the exact posterior among the three approximations. 

In addition to correcting posterior scale, both CoRe and DMH recover non-zero posterior covariances between parameters, reflected in the rotation of the posterior contours---an aspect that is largely attenuated under the pseudo-likelihood.

\begin{figure}
    \centering
    \def\svgwidth{0.90\textwidth}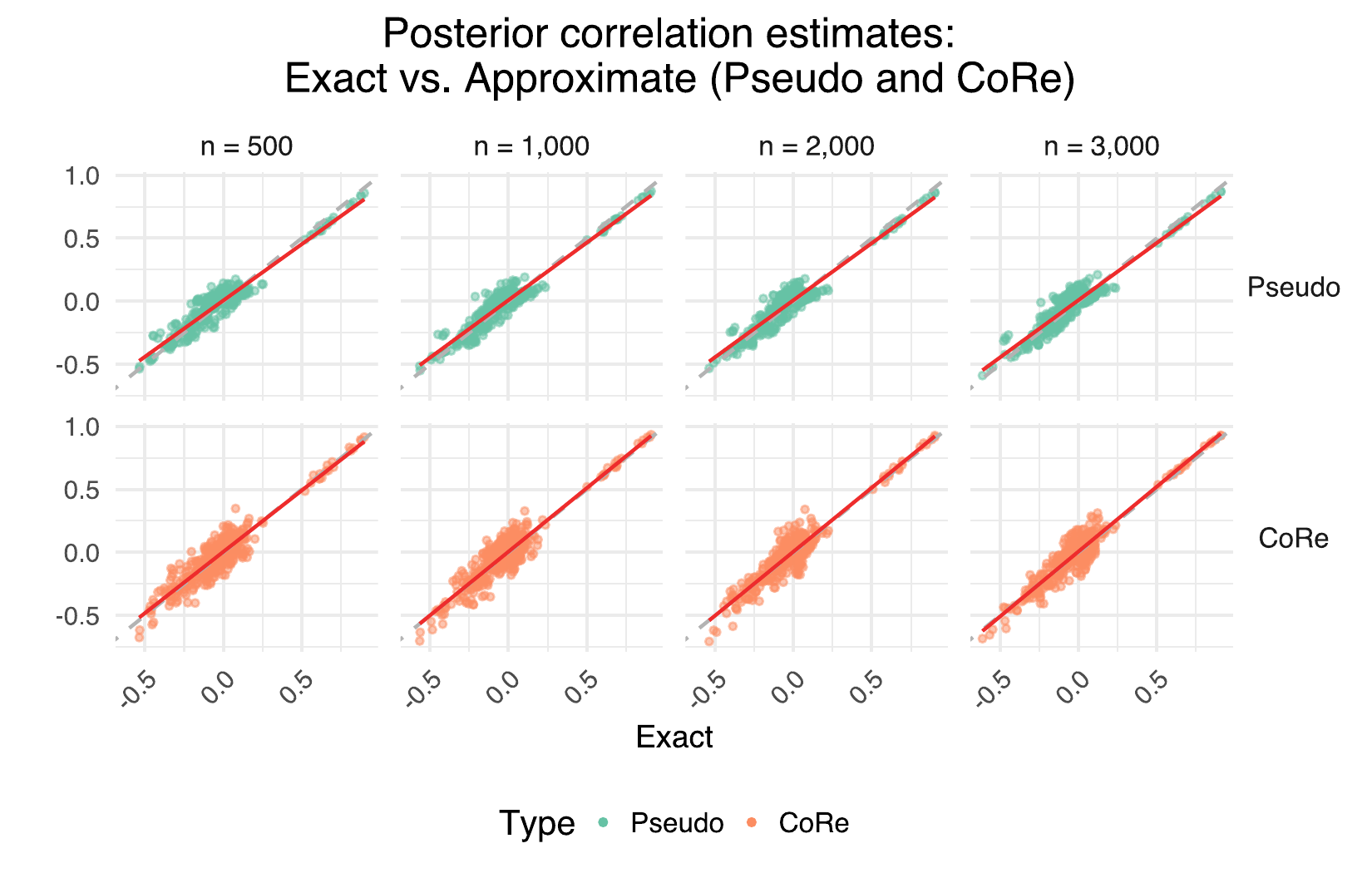
    \captionof{figure}{Comparison of approximate and exact posterior correlations between all pairs of parameters, for samples of size $n$ (by column) of the six dysphoria items of the RADS-2 data (observations sampled with replacement). Red line: least-squares fit; dashed gray line: identity. (\textit{Top row}) Pseudo-posterior correlations are attenuated relative to the exact posterior (fitted slope about $0.9$). (\textit{Bottom row}) CoRe posterior correlations scatter around the identity (fitted slope $0.97$--$1.03$), with a slightly larger spread than the pseudo-posterior correlations.}
    \label{fig:posterior_correlations_comparison}
\end{figure}

Figure \ref{fig:posterior_correlations_comparison} compares posterior correlations obtained from CoRe and the pseudo-posterior across increasing sample sizes ($n$). The CoRe correlations are centered on the exact correlations, whereas the pseudo-posterior correlations are attenuated towards zero at every sample size; the CoRe correlations are slightly more dispersed around the exact values.

\begin{figure}[t]
\centering
\def\svgwidth{0.7\textwidth}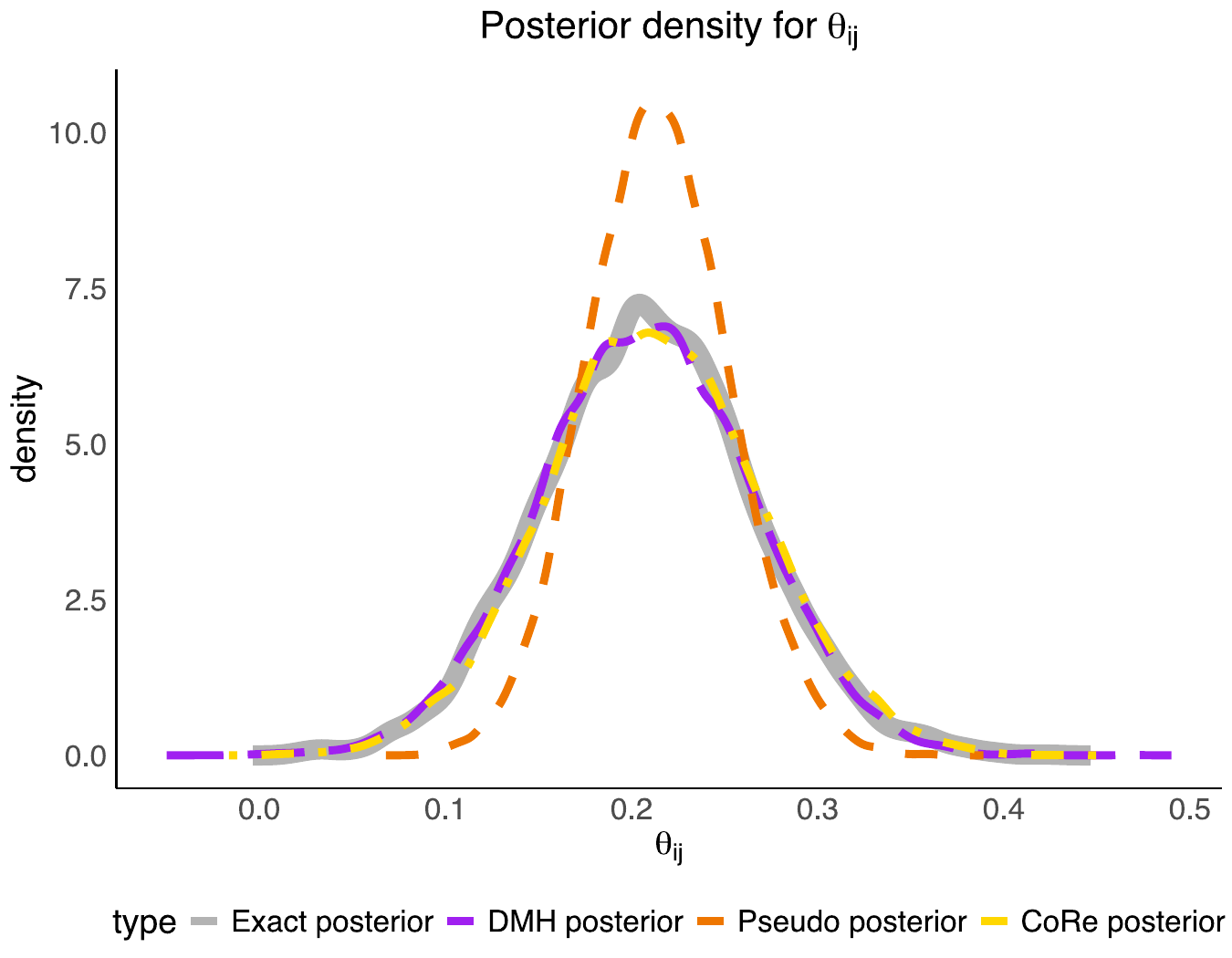
\caption{Posterior density of the pairwise association $\theta_{ij}$ of Figure \ref{fig:exact_vs_pseudo_vs_dmh_posterior_density_comparison}. Gray solid line: exact posterior based on the full likelihood. Orange dashed line: pseudo-posterior based on the pseudo-likelihood. Purple dashed line: posterior obtained using DMH. Gold dot-dashed line: posterior obtained using the CoRe sampler.}
\label{fig:core_posterior_density_comparison}
\end{figure}

Figure \ref{fig:core_posterior_density_comparison} revisits the example from Figure \ref{fig:exact_vs_pseudo_vs_dmh_posterior_density_comparison} and compares the pseudo-posterior, the DMH-posterior, and the CoRe posterior. The proposed method runs in approximately $4$ seconds\footref{fn:runtime} per chain and achieves overlap with the exact posterior comparable to DMH, at substantially lower cost, the same as the pseudo-posterior sampler. 

We now describe the construction of the rescaling matrix $\mathbf{A}$ and its role in approximating the covariance structure of the full posterior. 

\subsection[Construction of the Rescaling Matrix A]{Construction of the Rescaling Matrix $\mathbf{A}$}
\label{subsection:the-rescaling-matrix-A}
\par The rescaling matrix $\mathbf{A}$ plays a central role in the CoRe sampling scheme. Although the sampler operates on its inverse, $\mathbf{A}^{-1}$, we define $\mathbf{A}$ directly to clarify its construction. We define the rescaling matrix as
\[
\mathbf{A} = \mathbf{\Gamma} \mathbf{L}^{\intercal},
\]
where:
\begin{itemize}
    \item $\mathbf{L}^{\intercal}$ is the upper-triangular Cholesky factor of the negative posterior Hessian evaluated at $\bm{\eta}^{\star}$, representing the local curvature of the pseudo-posterior. It is given by
    \[
    \mathbf{L}\mathbf{L}^{\intercal} = -(\mathbf{H} + \mathbf{H}_{\bm{\eta}}),
    \]
    where $\mathbf{H}$ denotes the Hessian of the pseudo-likelihood and $\mathbf{H}_{\bm{\eta}}$ the log-prior curvature, both evaluated at $\bm{\eta}^{\star}$.
    \item $\mathbf{\Gamma}$ is the lower-triangular Cholesky factor of a scale matrix approximating the variance-covariance structure of the full posterior. We define $\mathbf{\Gamma\Gamma}^{\intercal}$ using a robust posterior covariance estimator based on the results of \citet{Godambe1960}, \citet{Huber1967}, and \citet{White1980}, referred to as the Godambe-Huber-White (GHW) posterior scale. Let $\mathbf{\Sigma}_{\text{GHW}}$ denote the GHW covariance estimator evaluated at $\bm{\eta}^{\star}$, 
    \[
    \mathbf{\Sigma}_{\text{GHW}} = \left(-\mathbf{H}^{-1}\right) \times \mathbf{U} \times \left(-\mathbf{H}^{-1}\right)
    \]
    where $\mathbf{U} = \sum_{\nu=1}^{n}\mathbf{u}_{\nu}\mathbf{u}_{\nu}^{\intercal}$ is the variance of the score. 
    The corresponding posterior covariance estimator is then
    \[
    \mathbf{\Gamma}\mathbf{\Gamma}^{\intercal}_{\text{GHW}} = \left(\mathbf{\Sigma}_{\text{GHW}}^{-1} - \mathbf{H}_{\bm{\eta}}\right)^{-1},
    \]
    which accounts for the curvature of the prior through $\mathbf{H}_{\bm{\eta}}$.
\end{itemize} 

In the standard CoRe sampler, implementation requires the inverse of the rescaling matrix, $\mathbf{A}^{-1} = \mathbf{L}^{-\intercal}\mathbf{\Gamma}^{-1}$, which is precomputed before sampling. In the next section, we introduce an adaptive variant, AdaCoRe, in which $\mathbf{A}^{-1}$ is updated iteratively during the burn-in phase.

\subsection{Adaptive Estimation of the Rescaling Matrix}
\par The adaptive version of the CoRe sampler, AdaCoRe, removes the need to specify the rescaling matrix $\mathbf{A}$ a priori. Instead, the components that define the rescaling---namely the local posterior curvature and variability---are updated during sampling based on the running mean $\overline{\bm{\eta}}$. To preserve detailed balance and convergence to the transformed pseudo-posterior, these updates are restricted to the burn-in phase of the Markov chain.

Once the relevant quantities are evaluated at $\overline{\bm{\eta}}$, the inverse scaling matrix $\mathbf{A}^{-1} = \mathbf{L}^{-\intercal} \mathbf{\Gamma}^{-1}$ is updated using linear solves rather than explicit matrix inversion. This improves both numerical stability and computational speed \citep{Sanderson2020}. To avoid reacting to Monte Carlo noise, the rescaling matrix is updated only when the average relative change in local curvature exceeds a threshold of $3/\sqrt{n}$.\footnote{The factor $1/\sqrt{n}$ reflects the typical sampling variability of a Hessian estimated from $n$ observations. The threshold $3/\sqrt{n}$ is a conservative choice that avoids updating the rescaling matrix in response to random fluctuations.} 

Algorithm \ref{algorithm:rescaling_matrix_update} summarizes the update of the inverse rescaling matrix $\mathbf{A}^{-1}$, and Algorithm \ref{algorithm:adacore_sampler} presents the full AdaCoRe sampling procedure. Changes in local curvature are monitored using the Frobenius norm, denoted $\lVert \cdot \rVert_{F}$.

\begin{algorithm}[t]
    \caption{Update of the inverse rescaling matrix $\mathbf{A}^{-1}$}
    \label{algorithm:rescaling_matrix_update}
    \begin{algorithmic}
    \State{\textbf{Input:}} running mean $\overline{\bm{\eta}}_s$ at iteration $s$ and observed data $\mathbf{X} = \left(\mathbf{X}_{1},\ldots,\mathbf{X}_n\right)$.
    \State Calculate the score covariance $\mathbf{U}_{s} = \sum_{\nu=1}^n \mathbf{u}_{\nu}\mathbf{u}_{\nu}^{\intercal}$,
where $\mathbf{u}_{\nu} = \nabla \log \tilde{f}(\mathbf{X}_{\nu};\overline{\bm{\eta}}_s)$.
    \State Calculate model Hessian $\mathbf{H}_s=\sum_{\nu=1}^n{ \nabla^2\log{\tilde{f}}(\mathbf{X}_{\nu};\overline{\bm{\eta}}_s)}$.
    \State Calculate prior Hessian $\mathbf{H}_{\bm{\eta},s}=\text{diag}\left(\nabla^2\log{\pi}(\overline{\bm{\eta}}_s)\right)$.
    \State Solve the linear system $\mathbf{U}_{s}\mathbf{Z}_{s} = -\mathbf{H}_{s}$ for $\mathbf{Z}_{s}$.
    \State Calculate robust posterior covariance estimator $\mathbf{\Gamma}\mathbf{\Gamma}^{\intercal}_{\text{GHW}} =
\left( (-\mathbf{H}_{s})\mathbf{Z}_{s} - \mathbf{H}_{\bm{\eta},s}  \right)^{-1}$.
    \State Find lower triangular $\mathbf{\Gamma}_{\text{GHW}} \gets \text{chol}\left(\mathbf{\Gamma\Gamma}^{\intercal}_{\text{GHW}}\right)$.
    \State Solve $\mathbf{\Gamma}_{\text{GHW}} \tilde{\mathbf{\Gamma}} = \mathbf{I}$ for $\tilde{\mathbf{\Gamma}}$.
    \State Calculate posterior curvature $\mathbf{L}\mathbf{L}^{\intercal} = -\left(\mathbf{H}_{s} + \mathbf{H}_{\bm{\eta},s}\right)$.
    \State Find upper triangular $\mathbf{L}^{\intercal} \gets \text{chol}\left(\mathbf{LL}^{\intercal}\right) $.
    \State Solve $\mathbf{L}^{\intercal} \tilde{\mathbf{L}}= \mathbf{I}$ for $\tilde{\mathbf{L}}$.
    \State Set $\mathbf{A}_{s}^{-1} = \tilde{\mathbf{L}}\tilde{\mathbf{\Gamma}}$ and its transpose $\mathbf{A}_{s}^{-\intercal} = \tilde{\mathbf{\Gamma}}^{\intercal}\tilde{\mathbf{L}}^{\intercal}$ (required by the proposal distribution).
    \end{algorithmic}
\end{algorithm}

\begin{algorithm}[t]
    \caption{AdaCoRe transition step}
    \label{algorithm:adacore_sampler}
    \begin{algorithmic}
\State{\textbf{Input:}} current parameter states $\bm{\beta}$ and $\bm{\eta}$ at iteration $s$, observed data $\mathbf{X} = (\mathbf{X}_1,\ldots,\mathbf{X}_n)$, reference point $\bm{\eta}^{\star}$, and tuning parameters $\xi = 0.05$, $\tau = 3/\sqrt{n}$, and $\varepsilon = 1 \times 10^{-12}$.
\State{\textbf{Stored from iteration $s-1$:}} cumulative sum $\sum_{l=1}^{s-1} \bm{\eta}_l$, square root of the inverse Fisher information $\mathbf{R}_s$ (the FisherMALA preconditioner, Appendix \ref{appendix:the_fishermala_sampling_algorithm_and_priors}; updated at every iteration), its value $\mathbf{R}^\star$ at the last accepted proposal, and exponential moving average $\overline{\delta}_{s-1}$.
    \If{$s < S_\text{burn-in}$} \textsc{(Inverse rescaling matrix update)}
        \State Update cumulative sum $\sum_{l=1}^{s} {\bm{\eta}_l} = \sum_{l=1}^{s-1} {\bm{\eta}_l} + \bm{\eta}_s$
        \If{$s > 1$}
            \State Calculate relative curvature change $\delta_s = \frac{\lVert \mathbf{R}_s - \mathbf{R}^{\star} \rVert_{F}}{\lVert \mathbf{R}^{\star}\rVert_{F}+\varepsilon}$.
            \State Calculate exponential moving average $\overline{\delta}_s = (1-\xi)\overline{\delta}_{s-1} + \xi \delta_s$.
            \If{$\overline{\delta}_s  > \tau$}
                \State Update running mean $\overline{\bm{\eta}}_s = \Sigma_{l=1}^{s} {\bm{\eta}_l}/s$.
                \State Update $\mathbf{A}^{-1}$ and $\mathbf{A}^{-\intercal}$ using \textbf{Algorithm \ref{algorithm:rescaling_matrix_update}}.
            \EndIf
        \EndIf
    \EndIf
    \State Propose $\bm{\beta}^\prime \sim q(\cdot \mid \bm{\beta})$.
    \State Calculate $\bm{\eta}^\prime = \mathbf{A}^{-1}\left(\bm{\beta}^\prime - \bm{\eta}^{\star}\right) + \bm{\eta^{\star}}$.
    \State Update square root of inverse Fisher information $\mathbf{R}^\prime$ (see Appendix \ref{appendix:the_fishermala_sampling_algorithm_and_priors}).
    \State Accept $\bm{\beta}_{s+1} = \bm{\beta}^\prime$, $\bm{\eta}_{s+1} = \bm{\eta}^\prime$ and set $\mathbf{R}^{\star} = \mathbf{R}^\prime$ with probability $$ \alpha(\bm{\beta},\bm{\beta}^\prime) = \min\left\lbrace 1, \frac{\tilde{f}\left(\mathbf{X};\bm{\eta}^\prime\right)\,\pi(\bm{\eta}^\prime)\,q(\bm{\beta}\mid\bm{\beta}^\prime)}{\tilde{f}\left(\mathbf{X};\bm{\eta}\right)\,\pi(\bm{\eta})\,q(\bm{\beta}^\prime\mid\bm{\beta})}\right\rbrace,$$ otherwise reject and set $\bm{\beta}_{s+1} = \bm{\beta}$ and $\bm{\eta}_{s+1} = \bm{\eta}$.
    \end{algorithmic}
\end{algorithm}

The CoRe sampling methods combine the computational scalability of the pseudo-likelihood with a principled correction of the posterior covariance structure. As a result, the sampler explores regions of the parameter space that have low probability under the pseudo-posterior. In the next section, we evaluate the efficiency and performance of CoRe by benchmarking it against established methods using simulated data.

\subsection{Interpretation of the CoRe Target}
\label{sec:core-interp}

The fixed-$\mathbf{A}$ sampler targets the transformed pseudo-posterior $\pi(\bm{\beta} \mid \mathbf{X})$ of Eq. \eqref{eq:core-target}. We now make precise the sense in which this target approximates the full likelihood posterior. CoRe replaces the naive pseudo-posterior scale with the Godambe-Huber-White covariance $\bm{\Gamma}\bm{\Gamma}^\top_{\mathrm{GHW}}$, the correct asymptotic covariance of the maximum-pseudo-likelihood estimator \citep{Godambe1960, Huber1967, White1980}. Its validity as an approximation to the full posterior depends on two asymptotic facts. First, under standard regularity conditions, the full posterior satisfies the Bernstein-von Mises theorem and concentrates on a Gaussian centered at the MLE \citep{Vaart1998, Kleijn2012}. Second, the pseudo-posterior follows an analogous generalized Bernstein-von Mises limit \citep{Miller2021}, concentrating on a Gaussian centered at $\bm{\eta}^\star$ but with an overly optimistic naive scale. Rescaling by $\mathbf{A}$ replaces this scale with the Godambe-Huber-White covariance, which matches the asymptotic covariance of the full posterior up to the efficiency loss of the pseudo-likelihood. As a linear map, $\mathbf{A}$ can match the mean and the covariance of both posteriors, yet it cannot adjust higher-order features of the distribution. This limitation is mild in our setting. Because the pseudo-likelihood consists of a product of multinomial-logistic full conditionals, it is log-concave. Consequently, the pseudo-posterior is unimodal and lacks heavy tails under the regimes studied. The main discrepancy left uncorrected by the linear map is the skewness. A linear transformation preserves the standardized skewness of a distribution, so the CoRe target inherits the skewness of the pseudo-posterior, which is of order $n^{-1/2}$ under the standard Edgeworth expansion of a posterior \citep{Johnson1970}. It therefore vanishes as the sample size grows. We summarize these findings in the following result.

\begin{proposition}[Second-order calibration]
\label{prop:secondorder}
Assume regularity conditions under which (i) the estimator $\bm{\eta}^\star$ is consistent and asymptotically unbiased for the data-generating parameter $\bm{\eta}_0$ \citep[the MPLE and the MLE have comparable bias in the Ising model;][]{Keetelaar2024}, and (ii) both the full posterior and the pseudo-posterior admit Bernstein--von Mises limits \citep{Vaart1998,
Kleijn2012, Miller2021}. Then, as $n \to \infty$, the CoRe target $\pi(\bm{\beta} \mid \mathbf{X})$ of Eq.~\eqref{eq:core-target} satisfies:
\begin{enumerate}
\item[(a)] \emph{(location)} it concentrates at $\bm{\eta}^\star$, which converges to $\bm{\eta}_0$, the same limit as the full-posterior mode;
\item[(b)] \emph{(scale)} its covariance equals $\mathbf{A}\tilde{\bm{\Sigma}}\mathbf{A}^\top = \bm{\Gamma}\bm{\Gamma}^\top_{\mathrm{GHW}}$ to leading order, the Godambe--Huber--White covariance of $\bm{\eta}^\star$;
\item[(c)] \emph{(shape)} it is asymptotically Gaussian up to a skewness term of order $n^{-1/2}$ \citep{Johnson1970};
\end{enumerate}
Consequently $\pi(\bm{\beta} \mid \mathbf{X})$ matches the full posterior in location asymptotically, and in covariance up to the asymptotic relative efficiency of the pseudo-likelihood. When the pseudo-likelihood is efficient, the covariances coincide and the approximation is exact to second order.
\end{proposition}

\begin{proof}
Claim (a) follows from assumption (i) together with the fact that the map $\bm{\beta} = \mathbf{A}(\bm{\eta} - \bm{\eta}^\star) + \bm{\eta}^\star$ fixes $\bm{\beta} = \bm{\eta}^\star$ at $\bm{\eta} = \bm{\eta}^\star$, so the mode is preserved. For (b), under the Gaussian approximation to the pseudo-posterior with covariance $\tilde{\bm{\Sigma}} \approx (\mathbf{L}\mathbf{L}^\top)^{-1}$, the change of variables gives $\mathrm{Cov}(\bm{\beta}) \approx \mathbf{A}\tilde{\bm{\Sigma}}\mathbf{A}^\top$. Substituting $\mathbf{A} = \bm{\Gamma}\mathbf{L}^\top$ and using $\mathbf{L}^\top(\mathbf{L}\mathbf{L}^\top)^{-1}\mathbf{L} = \mathbf{I}$ yields $\bm{\Gamma}\bm{\Gamma}^\top = \bm{\Gamma}\bm{\Gamma}^\top_{\mathrm{GHW}}$, the scale of Section \ref{subsection:the-rescaling-matrix-A} and the correct asymptotic covariance of the maximum-pseudo-likelihood estimator \citep{Godambe1960, White1980}. Claim (c) follows from the generalized Bernstein--von Mises limit for the pseudo-posterior \citep{Miller2021} and the invariance of the skewness order under the linear map $\mathbf{A}$. The final statement is the equality case of the Remark \ref{rem:ghw-efficiency} below: when the pseudo-score is a linear function of the full-likelihood score, $\bm{\Gamma}\bm{\Gamma}^\top_{\mathrm{GHW}} = \mathcal{I}^{-1}$, where $\mathcal{I}$ is the Fisher information.
\end{proof}

\begin{remark}
\label{rem:ghw-efficiency}
Claim (b) refers to the asymptotic Godambe-Huber-White covariance of $\bm{\eta}^\star$. Because the pseudo-score is an unbiased estimating function, this covariance satisfies $\bm{\Gamma}\bm{\Gamma}^\top_{\mathrm{GHW}} \succeq \mathcal{I}^{-1}$ in the Löwner sense \citep{Godambe1960, Lindsay1988, Bhatia1997}, with equality when the pseudo-score is a linear function of the full-likelihood score. In the networks considered, the two are numerically close, so CoRe recovers the correct posterior scale
(cf.\ Figures \ref{fig:posterior_correlations_comparison}--\ref{fig:core_posterior_density_comparison}). The residual gap and the uncorrected skewness of order $n^{-1/2}$ grow for small $n$, densely connected networks, or variables with many categories $m$. The correction is therefore most accurate when $n$ is large relative to the number of parameters.
\end{remark}

\section{Competing Methods and Evaluation Metrics}
\label{sec:performance-comparison-with-competing-approaches}
\par In the previous section, we introduced two sampling strategies aimed at improving posterior inference in discrete MRFs: the CoRe sampler with a fixed rescaling matrix and its adaptive version, the AdaCoRe sampler. In this and the next two sections, we compare the efficiency and performance of these two samplers with existing approaches that are either readily available for discrete MRFs or adaptable to them, first in a simulation study (Section \ref{sec:simulation-study}) and then on empirical data (Section \ref{sec:empirical-application}). This section describes the competing methods, the reference posterior against which all methods are evaluated, and the evaluation metrics. We organize the competing methods into two classes: (i) recalibration methods, which adjust pseudo-posterior covariance structures; and (ii) approximations of the Metropolis-Hastings transition kernel, which estimate gradients and ratios of partition functions via Monte Carlo simulation. All methods considered here are implemented within a common MCMC framework, described in Appendix \ref{appendix:the_fishermala_sampling_algorithm_and_priors}.

\subsection{Competing Methods}
\subsubsection{Recalibration-Based Methods}
\label{subsubsec:recalibration-methods}
We consider two groups of methods that transform the pseudo-posterior distribution to an adjusted covariance structure: post hoc calibration methods, and alternative definitions of CoRe sampling, which are based on the same CoRe sampler in Algorithm \ref{algorithm:core_sampler} but differ in the definition of the rescaling matrix.

\paragraph{Post hoc calibration methods.} \citet{Bouranis2017} proposed a post hoc calibration for exponential random graph models, which transforms pseudo-posterior draws to a new variance-covariance structure. To rescale each draw $\bm{\eta}$ into a new vector $\bm{\eta}_{\text{rescaled}}$, we apply a transformation that standardizes the draws, assigns a different scale to the parameters, and then shifts them back to their original location. The transformation is given by
    \[
    \bm{\eta}_{\text{rescaled}} = \mathbf{\Gamma} \mathbf{L}^{\intercal } (\bm{\eta} - \hat{\bm{\eta}}_{\text{MAP}}) + \bm{\eta}^\star,
    \]
    where $\hat{\bm{\eta}}_{\text{MAP}}$ is the pseudo-posterior mode (MAP), $\mathbf{L}^{\intercal}$ is the upper triangular Cholesky factor of the negative Hessian of the log pseudo-posterior evaluated at $\hat{\bm{\eta}}_{\text{MAP}}$, $\mathbf{\Gamma}$ is the lower triangular Cholesky factor of the calibration covariance matrix $\mathbf{\Gamma\Gamma}^{\intercal}$, and $\bm{\eta}^\star$ is the new location parameter applied after rescaling, which differs between the methods described below. The calibrated draws have covariance $\mathbf{\Gamma\Gamma}^{\intercal}$, which is wider than that of the pseudo-posterior.

    In the original work of \citet{Bouranis2017}, the new location parameter $\bm{\eta}^{\star}$ is the mode of the full posterior, obtained by the Robbins-Monro stochastic approximation method \citep{RobbinsMonro}, which we implement in its Newton-type form with the inverse Monte Carlo Hessian of the full posterior as gain matrix \citep{Ruppert1985, GuZhu2001}, and the calibration covariance is the inverse of the negative Hessian of the full log-posterior at that mode, estimated by Monte Carlo simulation. We refer to this approach as Post Hoc calibration with Robbins-Monro (PH-RM). We also include two post hoc variants that are centered at the pseudo-posterior mode, $\bm{\eta}^\star = \hat{\bm{\eta}}_{\text{MAP}}$, as CoRe is (Section \ref{subsection:the-core-sampler}), and differ in the calibration covariance: (i) the Godambe-Huber-White approach (PH-GHW), which uses the estimator $\mathbf{\Sigma}_{\text{GHW}}$ of Section \ref{subsection:the-rescaling-matrix-A}, and (ii) the Monte Carlo Hessian method (PH-MCH), which estimates the Hessian of the full log-posterior at the pseudo-posterior mode by Monte Carlo simulation. Implementation details are given in Appendix \ref{appendix:post_hoc_calibration_methodological_details_and_implementation}.
    
\paragraph{Alternative definitions of CoRe sampling.} We introduce two variants of the CoRe sampler with a fixed rescaling matrix that differ in the covariance structure used to construct it. CoRe-RM uses the covariance structure of PH-RM, estimated at the Robbins-Monro solution, and CoRe-MCH uses that of PH-MCH, estimated by Monte Carlo simulation at the pseudo-posterior mode. Like CoRe, both variants rescale the pseudo-posterior around its mode. Comparing CoRe-RM and CoRe-MCH with PH-RM and PH-MCH isolates the effect of embedding the rescaling in the sampler rather than applying it after sampling, for the same covariance structure.

\subsubsection{Approximations of the Transition Kernel}
\label{subsubsec:approximations-of-the-transition-kernel}
\par We consider the Double Metropolis-Hastings (DMH) algorithm \citep{Liang_2010}, which replaces the intractable ratio of normalizing constants in the Metropolis-Hastings acceptance probability by an estimate based on auxiliary data simulated from the model at the proposed parameter values. This approach addresses the computational challenges of doubly intractable models by using approximate expectations of the full likelihood. Its sampling scheme is illustrated in Appendix \ref{appendix:the_fishermala_sampling_algorithm_and_priors}: an inner Gibbs sampler generates the auxiliary data, which are used both for the ratio of normalizing constants and for the gradient of the full likelihood required by the Langevin proposal. DMH targets the full posterior, but every iteration requires Monte Carlo simulation, and its accuracy depends on the number of auxiliary draws and inner Gibbs iterations. Table \ref{tab:methods} summarizes all methods, together with the exact sampler used as reference (Section \ref{subsec:reference-posterior}).

\input{tables/methods_summary_table}

\subsection{Reference Posterior}
\label{subsec:reference-posterior}
\par All methods are evaluated against a reference posterior, which closely approximates the full posterior within computational limits. For networks with up to $p = 10$ variables, we sample the exact posterior: the normalizing constant is computed by enumerating all $(m+1)^p$ response patterns (about $10^{6}$ for $p = 10$ variables with four categories), which makes the Metropolis-Hastings acceptance probability and the gradient of the full likelihood available in closed form. For larger networks, enumeration is infeasible and the reference is DMH (Section \ref{subsubsec:approximations-of-the-transition-kernel}). At $p = 10$ both references are available, which allows us to verify that DMH reproduces the exact posterior before relying on it for larger networks (Section \ref{subsubsec:sim-overlap}). The same strategy is used in the empirical application (Section \ref{sec:empirical-application}), where the reference is a more demanding configuration of DMH.

\subsection{Evaluation Metrics}
\label{subsec:evaluation-metrics}
\par We compare each method with the reference posterior on four aspects: (i) the agreement of the marginal and bivariate posterior distributions, measured by the overlap index \citep{Pastore2019}; (ii) the evidence for conditional independence, measured by the Bayes factor based on the Savage-Dickey density ratio \citep{Dickey_1971,WagenmakersEtAl_2010,Bouranis2018,Sekulovski2024}; (iii) posterior uncertainty, measured by the ratio of posterior standard deviations; and (iv) computational efficiency, measured by the effective sample size (ESS) and runtime.

\subsubsection{Posterior Overlap}
\label{subsubsec:metric-overlap}
To measure the agreement between two distributions, we use the distribution-free overlap index introduced by \citet{Pastore2019}. For a model parameter $\theta$, let $\pi^\star(\theta\mid \mathbf{X})$ denote the reference posterior and $\tilde{\pi}(\theta\mid \mathbf{X})$ its approximation under a competing method. We use the normalized version of the index,
\[
\eta\left(\pi^\star,\tilde{\pi}\right) = \frac{\int_{\mathbb{R}}\min{\left[\pi^\star(\theta\mid \mathbf{X}),\tilde{\pi}(\theta\mid \mathbf{X})\right]}\,\text{d}\theta}{\int_{\mathbb{R}}\max{\left[\pi^\star(\theta\mid \mathbf{X}),\tilde{\pi}(\theta\mid \mathbf{X})\right]}\,\text{d}\theta},
\]
which takes values in $[0,1]$, where $0$ indicates no overlap and $1$ identical distributions. Because we rely on posterior draws, both densities are estimated from the posterior draws by kernel density estimation on a common grid of $1{,}024$ points spanning the combined support, and the integrals are approximated by sums over the grid.

\paragraph{Bivariate overlap.} The marginal index does not reflect whether a method reproduces the dependence between parameters, which is what a covariance correction should restore. We therefore also compute the overlap index $\eta_{\text{biv}}$ of the joint posterior of a pair of parameters $(\theta_1, \theta_2)$, defined as above with the integrals taken over $\mathbb{R}^2$ and both joint densities estimated using a bivariate kernel density estimation on a common $128 \times 128$ grid. Because the pseudo-likelihood misrepresents the dependence between parameters that involve the same variable, we consider three types of pairs: two interactions that share a variable, a threshold and an interaction of the same variable, and two thresholds of the same variable. For interaction pairs, only interactions that are present in the data-generating network are selected.

\subsubsection{Savage-Dickey Density Ratio}
\label{subsubsec:metric-sddr}
The Savage-Dickey density ratio computes Bayes factors for nested models \citep{Dickey_1971,WagenmakersEtAl_2010,Bouranis2018,Sekulovski2024}. We test the constrained hypothesis $\mathcal{H}_c : \theta = 0$ against the unconstrained hypothesis $\mathcal{H}_u: \theta \in \mathbb{R}$, for which the Bayes factor equals the posterior-to-prior ratio evaluated at the null value,
\[
\text{BF}_{cu} = \frac{\pi(\theta = 0\mid\mathbf{X})}{\pi(\theta = 0)}.
\]
The numerator is estimated from the posterior draws of each method by kernel density estimation, and the denominator follows from the prior (cf. Appendix \ref{appendix:the_fishermala_sampling_algorithm_and_priors}). The Savage-Dickey representation is valid only if the prior of the nuisance parameters $\bm{\psi}$ under $\mathcal{H}_c$ matches their conditional prior under $\mathcal{H}_u$, $\pi(\bm{\psi} \mid \mathcal{H}_c) = \pi(\bm{\psi} \mid \theta  = 0, \mathcal{H}_u)$. Because the prior factorizes across parameters, both sides reduce to the marginal prior $\pi(\bm{\psi})$, so the condition holds by construction. We use the Savage-Dickey ratio to test conditional independence between pairs of variables; alternative Bayesian approaches are discussed by \citet{Sekulovski2024}. Because $\text{BF}_{cu}$ depends on the posterior only through its height at zero, it directly reflects how well a method reproduces the scale of the posterior: a posterior that is too narrow inflates the evidence for independence when its mode is close to zero and deflates it otherwise.

\subsubsection{Ratio of Posterior Standard Deviations}
\label{subsubsec:metric-sd-ratio}
For each parameter, we compare the posterior standard deviation of a method, $\sigma_{\text{method}}$, with that of the reference posterior, $\sigma_{\text{ref}}$, through the ratio $\sigma_{\text{method}} / \sigma_{\text{ref}}$. Ratios above (below) $1$ indicate overestimation (underestimation) of posterior uncertainty relative to the reference.


\subsubsection{Effective Sample Size and Runtime}
\label{subsubsec:metric-ess}
We compute the ESS of each parameter with the R package \texttt{coda} \citep{Plummer2006} and record the wall-clock time required to generate the posterior draws, including the time spent on the Monte Carlo approximations required by each method (e.g., the Robbins-Monro algorithm for PH-RM and CoRe-RM).


\section{Simulation Study}
\label{sec:simulation-study}

\subsection{Data-Driven Parameterization}
\label{subsec:data-driven-parameterization}
Specifying the parameters of a discrete MRF is not straightforward, as they govern complex dependency structures and have a nonlinear relation with the normalizing constant. We therefore adopt a data-driven approach in which the parameters are first estimated from the observed data and then treated as the true parameters to generate random data.

We use the data analyzed in Section \ref{sec:empirical-application}: the responses of $917$ adolescents to the $25$ items of the Reynolds Adolescent Depression Scale \citep[RADS-2;][]{Reynolds2002, RamosVera2023}, each answered on a four-point Likert scale. For a network of size $p$, we construct a data-generating model in three steps. First, we draw $p$ of the $25$ items at random. Second, we generate a network structure on these items (Section \ref{subsec:simulation-design}). Third, we estimate the thresholds and the interactions of the edges in this structure by maximizing the pseudo-posterior on the observed responses to the selected items, and we set the interactions of the absent edges to zero (see Appendix \ref{appendix:framework_for_dataset_simulation_and_analysis} for details).

\subsection{Simulation Design}
\label{subsec:simulation-design}
\begin{figure}[h]
    \centering
    \begin{subfigure}{0.3\textwidth}
        \centering
        \def\svgwidth{\textwidth}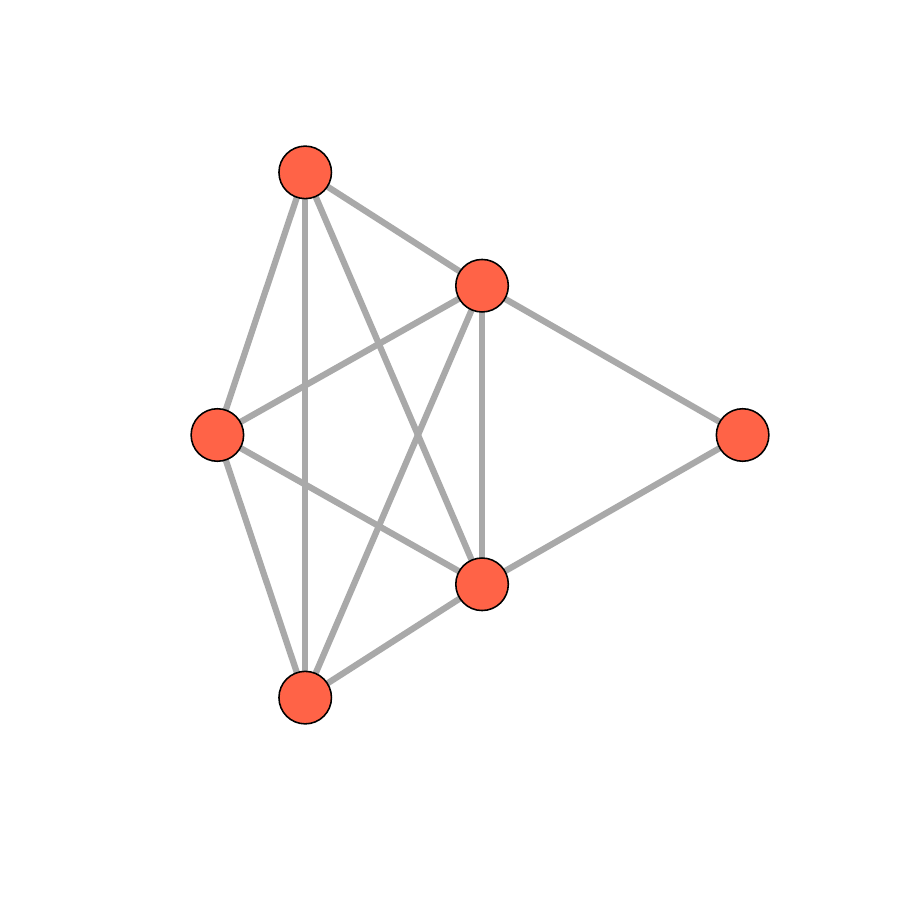
        \caption{Small-world graph}
        \label{fig:smallworld_grah}
    \end{subfigure}
    \hfill
    \begin{subfigure}{0.3\textwidth}
        \centering
        \def\svgwidth{\textwidth}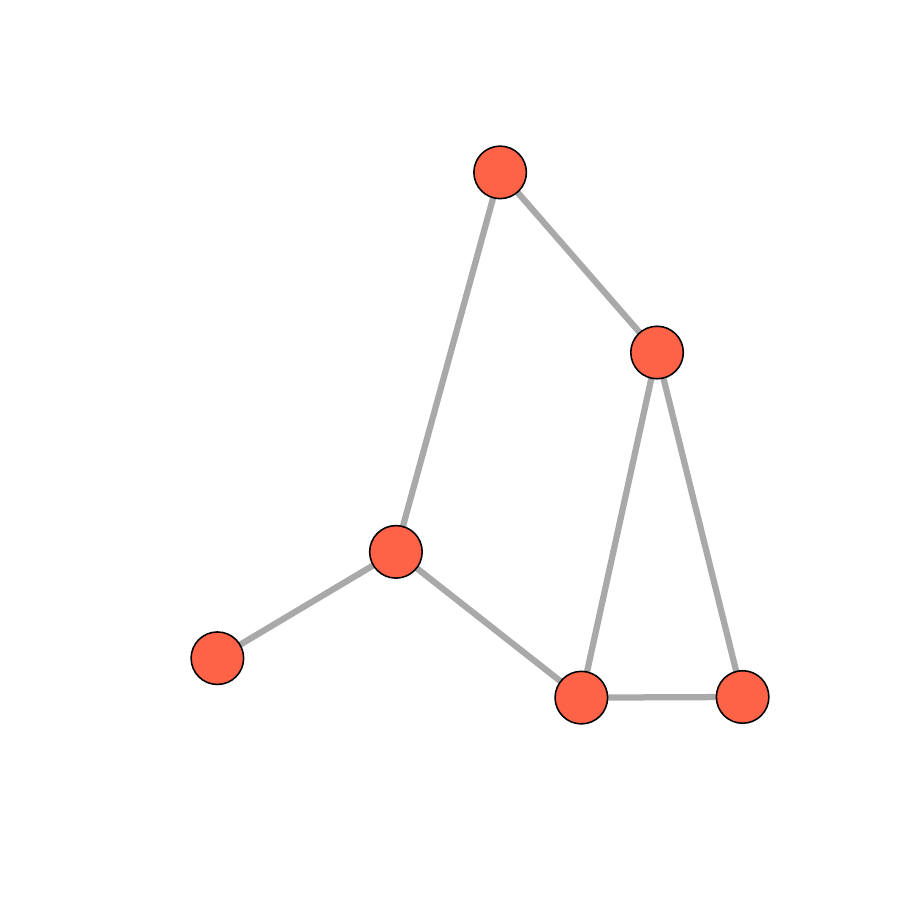
        \caption{Random graph}
        \label{fig:random_graph}
    \end{subfigure}
    \hfill
    \begin{subfigure}{0.3\textwidth}
        \centering
        \def\svgwidth{\textwidth}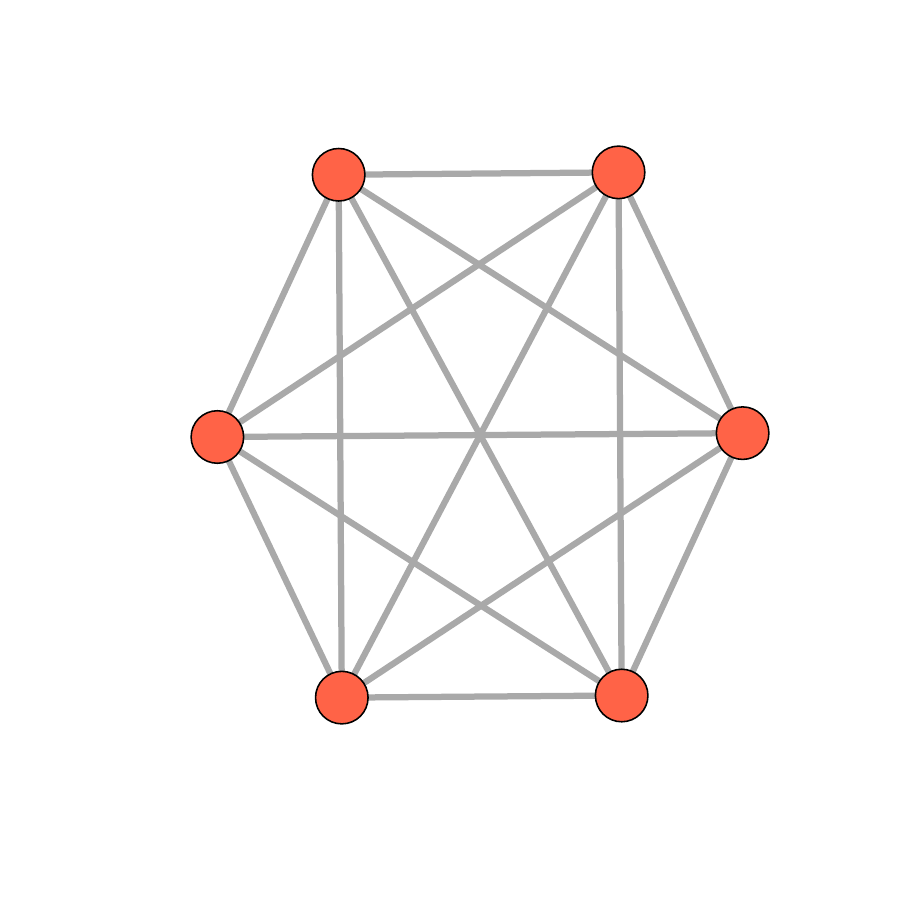
        \caption{Full graph}
        \label{fig:full_graph}
    \end{subfigure}
    \caption{Examples of three graph structures for $p = 6$.}
    \label{fig:example_of_graph_structures}
\end{figure}

We examine how well the methods reproduce the reference posterior, and in particular the evidence for the absence of an interaction (conditional independence) between pairs of variables. We consider three network structures: (1) a small-world graph \citep{Watts1998}, (2) a random graph \citep{Renyi1959} whose edge-inclusion probability is drawn uniformly from $(0,1)$ for each data-generating model, and (3) a fully connected graph (examples with $p = 6$ nodes are shown in Figure \ref{fig:example_of_graph_structures}). We consider four network sizes, $p \in \{6, 10, 15, 22\}$, which range from $33$ to $297$ parameters, and four sample sizes, $n \in \{500, 1000, 2000, 3000\}$.

We design a grid of $48$ conditions defined by the combination of the three network structures, the four network sizes, and the four sample sizes. Under each condition, we generate $10$ data-generating models (random item subsets and structures) and simulate $10$ datasets from each, for a total of $100$ datasets per condition. Datasets are simulated with a Gibbs sampler run for $5{,}000$ iterations from a random starting state (Appendix \ref{appendix:framework_for_dataset_simulation_and_analysis}). For each dataset, we estimate all thresholds and interactions, including those of edges that are absent in the data-generating network. The reference posterior is the exact posterior for $p \leq 10$ and DMH for $p \geq 15$ (Section \ref{subsec:reference-posterior}). DMH is also evaluated as a competing method against the exact posterior for $p = 6$ and $p = 10$.

\subsection{Computational Setup}
\label{subsec:computational-setup}
Every method runs one Markov chain of $25{,}000$ iterations, of which the first $5{,}000$ are discarded as burn-in (including an adaptive stage of $500$ iterations), leaving $20{,}000$ draws. The chains start at the pseudo-posterior mode, and the global step size $\sigma^2$ of the sampler in Appendix \ref{appendix:the_fishermala_sampling_algorithm_and_priors} is initialized at $1.0$ for all methods and adapted throughout the chain, stabilizing well within burn-in (Appendix \ref{appendix:the_global_step_size_parameter}). Appendix \ref{appendix:random_walk_proposal} complements this single-chain setup with a multi-chain analysis for a larger network ($p = 15$), reporting effective sample sizes and $\widehat{R}$ under both the FisherMALA and a random-walk proposal. We assign Beta-Prime priors ($a = b = 0.5$) to the thresholds and Cauchy priors with scale $2.5$ to the interactions. All calibration methods start from the same pseudo-posterior mode, and we apply the post hoc calibrations to the same draws of the pseudo-posterior sampler, so that the methods differ only in how they correct the pseudo-posterior.

DMH uses $L = 25{,}000$ auxiliary draws and $5$ inner Gibbs sweeps per iteration, both as a reference and as a competing method. The methods that approximate the full likelihood outside the sampler (PH-RM, PH-MCH, CoRe-RM and CoRe-MCH) use $100{,}000$ Monte Carlo draws. The Robbins-Monro algorithm starts at the pseudo-posterior mode and runs for $50$ iterations with gain $1/(t+1)$ and the inverse Monte Carlo Hessian of the full posterior as gain matrix (Appendix \ref{appendix:post_hoc_calibration_methodological_details_and_implementation}).

We ran the experiments on the Dutch national supercomputer Snellius\footnote{\url{https://www.surf.nl/en/dutch-national-supercomputer-snellius}}. For each condition and method, we used an exclusive node and analyzed the $100$ datasets in parallel, one per core. The R \citep{R} code for all experiments is available at \zenodorepo{}.

\begin{figure}[t]
\centering
\def\svgwidth{\textwidth}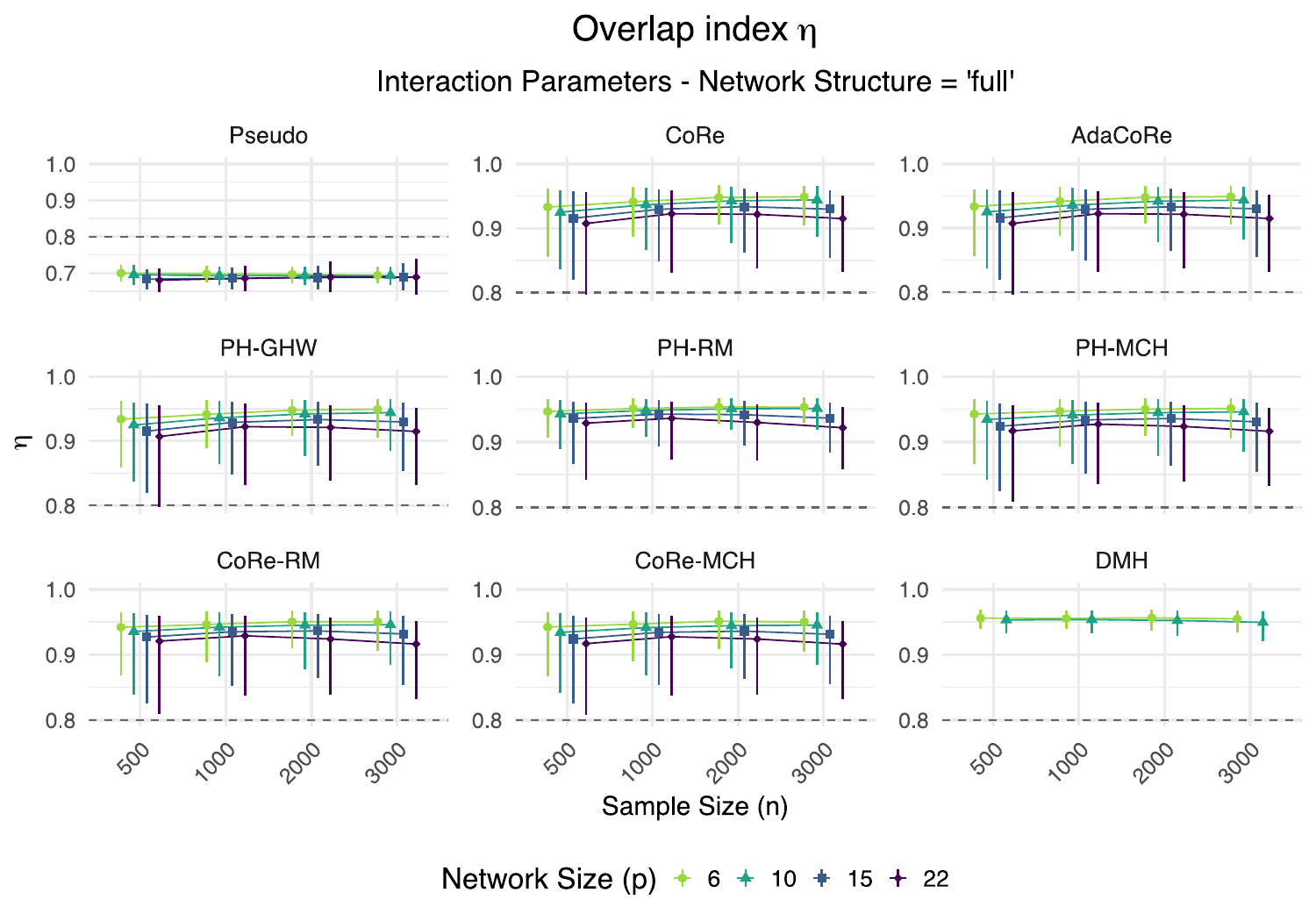
\caption{Overlap index $\eta$ between each method's posterior distribution (panel) and the reference posterior for the interaction parameters in a network with full structure. Points show the median over all interactions and datasets of a condition and bars the $5\%$ and $95\%$ quantiles; colors and symbols correspond to network size ($p$). The reference is the exact posterior for $p \leq 10$ and DMH for $p \geq 15$; DMH is evaluated as a competing method for $p \leq 10$ only. The horizontal dashed gray line indicates a benchmark level of $0.8$.}
\label{fig:simulation_overlap_interactions_full_structure}
\end{figure}

\begin{figure}[t]
\centering
\def\svgwidth{\textwidth}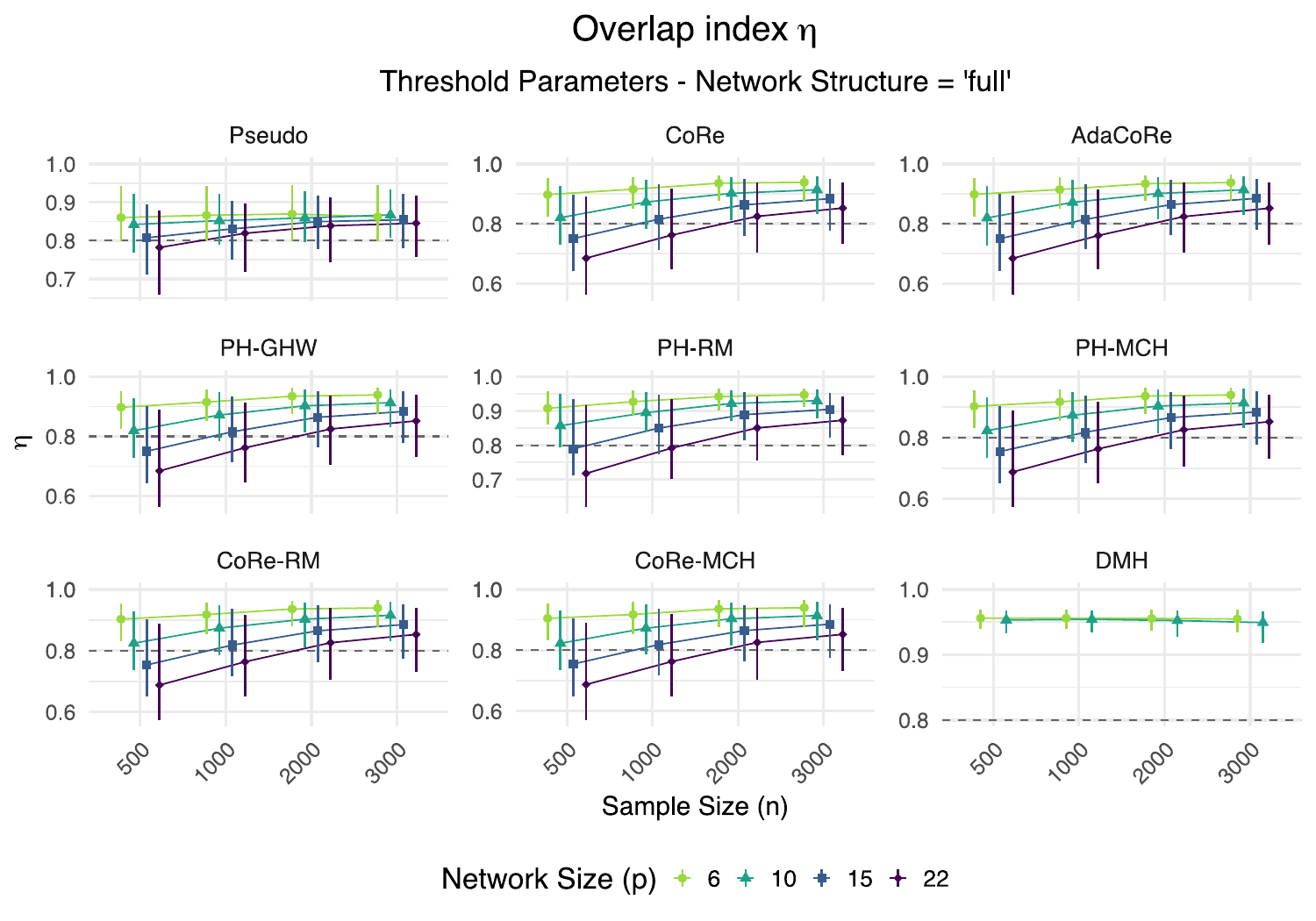
\caption{Overlap index $\eta$ between each method's posterior distribution (panel) and the reference posterior for the threshold parameters in a network with full structure. Points, bars, colors, reference and dashed line as in Figure~\ref{fig:simulation_overlap_interactions_full_structure}.}
\label{fig:simulation_overlap_thresholds_full_structure}
\end{figure}

\input{tables/simulation_ess_table_full}

\input{tables/simulation_runtime_table_full}

\subsection{Simulation Results}
\label{subsection:performance-evaluation-and-simulation-results}
We present results for each metric, focusing on the full network structure for the overlap and on the random structure for the absent edges. The results for the random and small-world structures are almost identical. Figures, ratios of posterior standard deviations, effective sample sizes and runtimes for these structures are reported in the supplementary material.

\subsubsection{Reference Posterior}
\label{subsubsec:sim-overlap}
First, we verify the reference. At $p = 10$, where both references are available, DMH overlaps with the exact posterior with a median $\eta$ between $0.947$ and $0.954$ across the $12$ conditions, for both thresholds and interactions. Replacing the exact posterior by DMH as the reference lowers the median overlap of every method by at most $0.011$ ($0.003$ on average for the interactions) and preserves their ordering (see supplementary material). The overlaps reported for $p \geq 15$ are therefore slightly conservative but comparable across methods.

Because the overlap is estimated from a finite number of autocorrelated draws, it stays below $1$ even for two samples of the same posterior. At $p = 6$, two independent runs of the exact sampler on the same datasets overlap by $0.96$ (marginal) and $0.88$--$0.89$ (bivariate), for every sample size (see supplementary material). These values mark the attainable level of the two indices. DMH reaches it at $p = 6$ ($0.95$--$0.96$) and comes close at $p = 10$. The attainable level decreases with the effective sample size of the chains, which is somewhat lower for the exact reference at $p = 10$ than at $p = 6$ and decreases markedly for the DMH reference at $p \geq 15$ as $n$ grows, from $1{,}440$ to $442$ at $p = 15$ and from $992$ to $220$ at $p = 22$ (Table \ref{tab:ess}). The lower overlaps of all methods at large $p$ and $n$ partly reflect this Monte Carlo error of the reference.

\subsubsection{Posterior Overlap}
\label{subsubsec:sim-marginal-overlap}
Figure \ref{fig:simulation_overlap_interactions_full_structure} shows the overlap index $\eta$ for the interaction parameters. The pseudo-posterior overlaps with the reference by $0.68$--$0.70$ in every condition. All calibration methods improve substantially: their median overlap is $0.93$--$0.95$ for $p \leq 10$, $0.92$--$0.94$ for $p = 15$, and $0.91$--$0.94$ for $p = 22$. The lower values for the larger networks partly reflect the growing Monte Carlo error of the DMH reference at large $n$. For $p \leq 10$, the best corrections are within $0.01$ of the attainable level ($0.96$). PH-RM has the highest overlap and reaches $0.94$--$0.95$ already at $n = 500$ for $p \leq 10$. The other calibration methods are within $0.01$ of each other in every condition. DMH, evaluated for $p \leq 10$, reaches $0.95$--$0.96$.

For the thresholds (Figure \ref{fig:simulation_overlap_thresholds_full_structure}), the corrections improve on the pseudo-posterior for small networks and for large samples, but not for large networks at small sample sizes. At $p = 6$, their median overlap is $0.90$--$0.95$, against $0.86$--$0.87$ for the pseudo-posterior. However, in the full network at $n = 500$, the corrections centered at the pseudo-posterior mode fall below the pseudo-posterior from $p = 10$ onwards, increasingly with $p$: $0.82$ against $0.84$ at $p = 10$, $0.75$ against $0.81$ at $p = 15$, and $0.68$--$0.69$ against $0.78$ at $p = 22$. PH-RM is less affected: it improves on the pseudo-posterior at $p = 10$ ($0.86$) and falls below it at $p = 15$ ($0.79$) and $p = 22$ ($0.72$). The corrections catch up as $n$ increases: they overtake the pseudo-posterior at $n = 1{,}000$ for $p = 10$ and at $n = 2{,}000$ for $p = 15$, and match it at $n = 3{,}000$ for $p = 22$ ($0.85$). In the random and small-world structures, they improve on the pseudo-posterior from $n = 1{,}000$ onwards, and for $p \leq 10$ already from $n = 500$, except in the small-world structure at $p = 22$, where they improve from $n = 2{,}000$.

This pattern does not reflect an error in the scale or in the location of the corrections: their posterior standard deviations for the thresholds match those of the reference, and the pseudo-posterior mode around which they are centered lies close to the mode of the full posterior. It reflects the shape of the full posterior instead. In small samples, the posterior of the thresholds is asymmetric: it has a longer tail towards lower values, so that its mean lies below its mode. The asymmetry is largest for the thresholds of rarely chosen response categories, which are estimated from few observations. In our design, based on the response frequencies of the RADS-2 data, these correspond to the highest categories. The asymmetry grows with the number of parameters and shrinks with $n$ at about the $n^{-1/2}$ rate expected from the Bernstein-von Mises theorem (Table \ref{tab:simulation-threshold-asymmetry}). Because the calibration methods adjust only the location and the covariance of the pseudo-posterior, they cannot reproduce this asymmetry, and their thresholds are displaced in dense networks with many parameters relative to the sample size.

\begin{figure}[htp]
\centering
\def\svgwidth{\textwidth}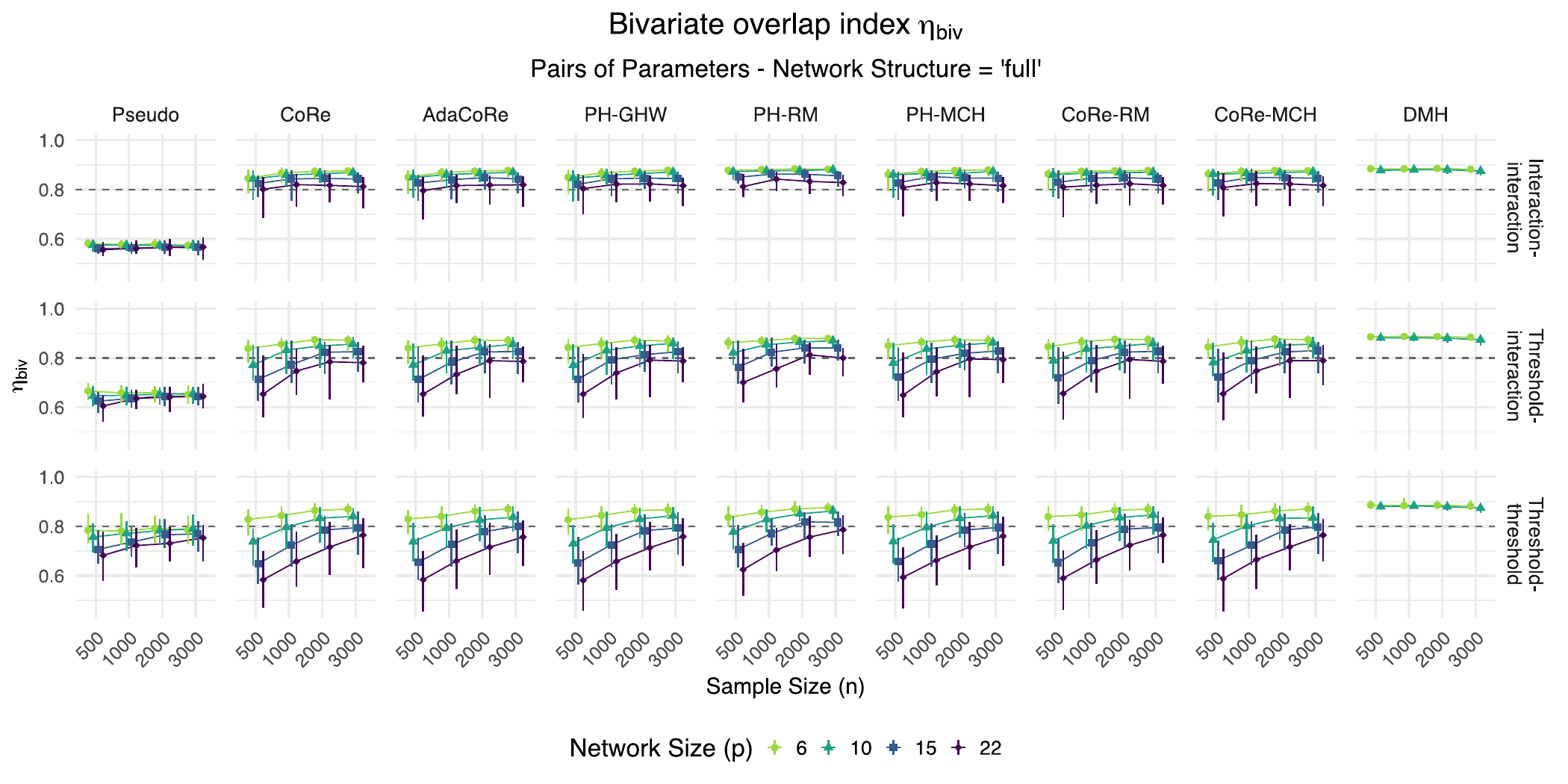
\caption{Bivariate overlap index $\eta_{\text{biv}}$ between each method's joint posterior distribution (column) and the reference posterior for pairs of parameters in a network with full structure: two interactions sharing a variable (top row), a threshold and an interaction of the same variable (middle row), and two thresholds of the same variable (bottom row). One pair of each type per dataset. Points show the median over the datasets of a condition and bars the $5\%$ and $95\%$ quantiles. Colors and symbols correspond to network size ($p$). Reference and dashed line as in Figure \ref{fig:simulation_overlap_interactions_full_structure}.}
\label{fig:simulation_bivariate_overlap_full_structure}
\end{figure}

\input{tables/simulation_threshold_asymmetry_table_full}

\subsubsection{Bivariate Overlap}
\label{subsubsec:sim-bivariate-overlap}
Figure \ref{fig:simulation_bivariate_overlap_full_structure} shows whether the methods reproduce the joint posterior of pairs of parameters involving the same variable. The attainable level of the bivariate overlap is lower than that of the marginal overlap: two runs of the exact sampler overlap by $0.88$--$0.89$, and DMH reaches $0.87$--$0.89$ against the exact posterior. The pseudo-posterior misrepresents the dependence between parameters more than their marginal distributions: its bivariate overlap is $0.56$--$0.58$ for pairs of interactions in every condition, against a marginal overlap of about $0.69$. The calibration methods restore the joint posterior of the interactions to $0.82$--$0.88$ for $p \leq 10$ and $0.80$--$0.87$ for $p \geq 15$, that is, at or close to the attainable level for $p \leq 10$, with PH-RM at $0.87$--$0.88$ from $n = 500$. For pairs of a threshold and an interaction, the corrections improve on the pseudo-posterior ($0.60$--$0.67$) in every condition, less at small $n$ and large $p$ ($0.65$--$0.70$ against $0.61$ at $p = 22$ and $n = 500$). For pairs of thresholds, the asymmetry affects both thresholds of the pair, so the corrections again fall below the pseudo-posterior at small $n$ and large $p$: at $p = 22$ and $n = 500$, they reach $0.58$--$0.62$, against $0.68$ for the pseudo-posterior, and catch up at $n = 3{,}000$ ($0.76$--$0.79$ against $0.75$).

\begin{figure}[t]
\centering
\def\svgwidth{\textwidth}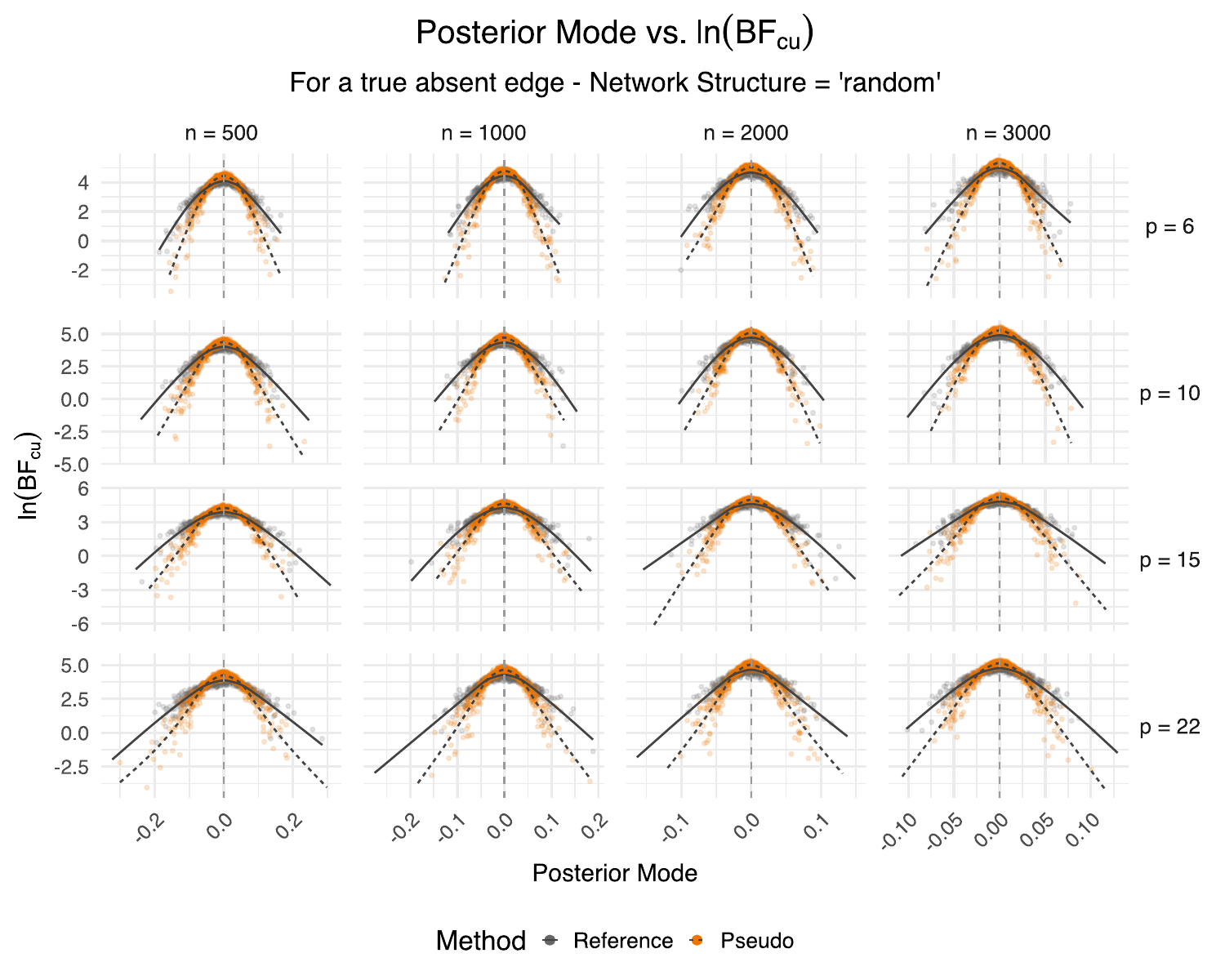
\caption{Posterior mode versus $\ln{\left(\text{BF}_{\text{cu}}\right)}$ (Savage-Dickey log-density ratio; positive values favor conditional independence) for truly absent edges in a random network structure, for the reference posterior (gray points) and the pseudo-posterior (orange points). Rows correspond to network size ($p$) and columns to sample size ($n$). Solid lines show trends for the reference posterior, dashed lines for the pseudo-posterior. The trends use all interactions, and the points are a random subset of at most $600$ points per panel and method. Interactions without posterior mass near zero, whose $\ln\text{BF}_{\text{cu}}$ is at the numerical floor, are omitted ($0.5\%$ of the values of the pseudo-posterior, fewer than $0.1\%$ otherwise).}
\label{fig:simulation_scatterplot_bf_reference_vs_pseudo_random_structure}
\end{figure}

\begin{figure}[t]
\centering
\def\svgwidth{\textwidth}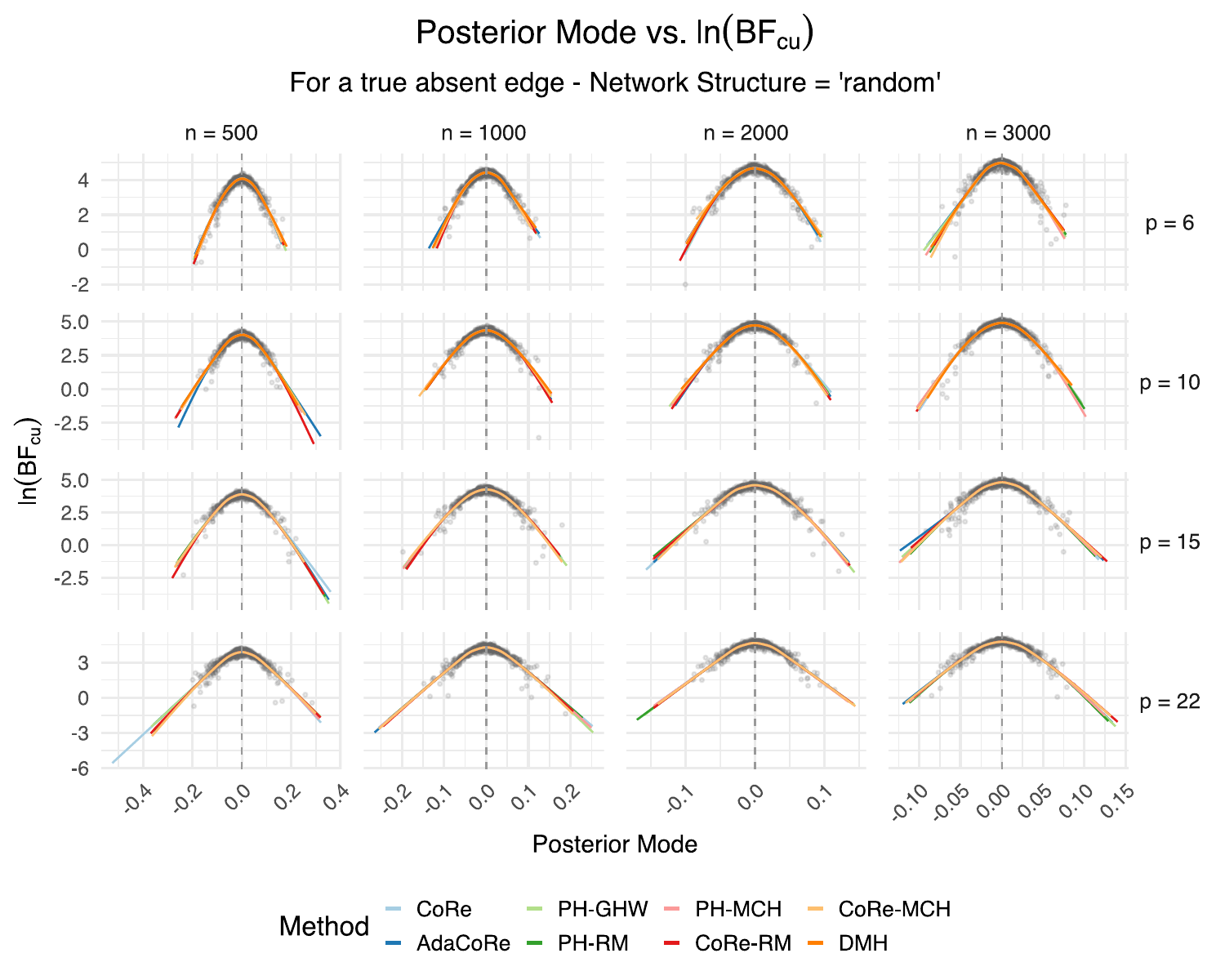
\caption{Trend lines of posterior mode versus $\ln{\left(\text{BF}_{\text{cu}}\right)}$ for the calibration methods and DMH (line colors) for truly absent edges in a random network structure; DMH is evaluated as a competing method for $p \leq 10$ only. Rows correspond to network size ($p$) and columns to sample size ($n$). Gray dots show a random subset of the corresponding values for the reference posterior. Values at the numerical floor are omitted as in Figure~\ref{fig:simulation_scatterplot_bf_reference_vs_pseudo_random_structure}.}
\label{fig:simulation_scatterplot_bf_methods_random_structure}
\end{figure}

\subsubsection{Ratio of Posterior Standard Deviations}
\label{subsubsec:sim-sd-ratio}
For the interactions of truly absent edges, the ratio of pseudo-posterior to reference posterior standard deviations is approximately constant: across all values of $n$ and $p$, the pseudo-posterior standard deviation is $30$--$32\%$ lower than that of the reference (median ratio $0.68$--$0.70$; figures in the supplementary material).
All calibration methods have median ratios between $0.98$ and $1.01$ in every condition. The methods whose covariance is estimated by Monte Carlo from the full likelihood (PH-RM, PH-MCH, CoRe-RM and CoRe-MCH) vary slightly less around the median than those based on $\mathbf{\Sigma}_{\text{GHW}}$ (CoRe, AdaCoRe and PH-GHW) at small sample sizes. This difference vanishes as $n$ increases.

\subsubsection{Savage-Dickey Density Ratio}
\label{subsubsec:sim-sddr}
Figure \ref{fig:simulation_scatterplot_bf_reference_vs_pseudo_random_structure} shows the relationship between the posterior mode of a truly absent edge and its log Bayes factor. The pseudo-posterior exhibits systematic distortion. When the posterior mode is far from zero, the log-density ratio is smaller than that of the reference; when the mode approaches zero, the ratio is inflated. Both effects stem from the underestimated posterior variability: a narrower posterior raises the density at zero when the mode is close to it and lowers it otherwise. Per edge, the log Bayes factor of the pseudo-posterior differs from that of the reference by a median of $0.31$--$0.33$, and by $1.7$ at the $95$th percentile, a factor of about $5.5$ on the Bayes factor. Figure \ref{fig:simulation_scatterplot_bf_methods_random_structure} shows that the calibration methods correct this distortion: their trend lines coincide with each other and with the reference across $n$ and $p$. Per edge, the log Bayes factors of the calibration methods differ from the reference by a median of $0.04$--$0.06$ (PH-RM $0.04$--$0.05$). This is about as close as DMH comes to the exact posterior ($0.03$--$0.04$), so the remaining differences are of the size of the Monte Carlo error of the Savage-Dickey ratio.

\subsubsection{Computational Efficiency}
\label{subsubsec:sim-efficiency}
Tables \ref{tab:ess} and \ref{tab:runtime} report the median and interquartile range of the effective sample size and of the runtime per chain for the full network structure. All methods based on the pseudo-likelihood have the same effective sample size, from about $3{,}300$ at $p = 6$ to $1{,}250$ at $p = 22$ out of $20{,}000$ draws: CoRe and AdaCoRe mix as well as the pseudo-posterior sampler, and the post hoc corrections inherit the pseudo-posterior chains. The effective sample size of DMH decreases with $n$, both as a competing method (from $2{,}949$ to $1{,}776$ at $p = 6$ and from $1{,}979$ to $869$ at $p = 10$) and as the reference.

The runtimes differ by orders of magnitude. CoRe, PH-GHW, PH-MCH and CoRe-MCH cost about as much as the pseudo-posterior itself, between $6$ and $21$ seconds at $p = 6$ and between $53$ and $276$ seconds at $p = 22$: the rescaling adds almost no cost to the sampler. The Robbins-Monro step of PH-RM and CoRe-RM adds between $20$ seconds ($p = 6$) and $90$ seconds ($p = 22$). AdaCoRe, which adapts the rescaling matrix during burn-in, becomes expensive for large networks ($2{,}513$ seconds at $p = 22$ and $n = 3{,}000$, about nine times CoRe). The exact sampler is the fastest method at $p = 6$, where the normalizing constant sums over $4^6$ states, but takes about $40$ minutes at $p = 10$. DMH requires $36$ minutes ($p = 6$) to about an hour ($p = 10$) per chain as a competing method, and $1.6$ ($p = 15$) to $2.5$ hours ($p = 22$) as the reference, with a decreasing effective sample size.

The acceptance rate does not depend on the method or on the size of the network: its median over the datasets of each condition ranges between $0.574$ and $0.576$ for all samplers and all $48$ conditions, because the step size of the sampler is adapted towards an acceptance rate of $0.574$ (Appendix \ref{appendix:the_fishermala_sampling_algorithm_and_priors}). Therefore, rescaling the pseudo-posterior does not lower the acceptance rate, and differences in efficiency appear in the effective sample size instead. AdaCoRe updates its rescaling matrix between about $230$ times ($p = 6$, $n = 500$) and $770$ times ($p = 22$, $n = 3{,}000$) during the $5{,}000$ burn-in iterations (see supplementary material). The number of updates grows with $p$ and with $n$, because the threshold $3/\sqrt{n}$ decreases with the sample size, and each update recomputes the Godambe-Huber-White covariance, which explains the cost of AdaCoRe for large networks. Since the rescaling matrix is fixed after burn-in, the adaptation does not affect the target, and AdaCoRe has the same effective sample size as CoRe.

\subsection{Discussion of Simulation Results}
\label{subsection:summary-of-findings}
The pseudo-posterior underestimates the posterior uncertainty of the interactions by about $30\%$ in every condition, misrepresents the dependence between parameters even more than their marginal distributions, and distorts the evidence for conditional independence, inflating the Savage-Dickey Bayes factor for edges with modes near zero and deflating it otherwise. These distortions do not attenuate with the sample size.

All calibration methods correct them. For the interactions, they closely reproduce the marginal and joint posterior of the reference, their posterior standard deviations match the reference, and their Savage-Dickey Bayes factors agree with the reference up to Monte Carlo error. The differences between the calibration methods are small. The choice of the rescaling matrix matters mainly at small sample sizes, where the covariance estimated by Monte Carlo from the full likelihood is more precise than $\mathbf{\Sigma}_{\text{GHW}}$, and PH-RM, which also moves the location to the mode of the full posterior, is the most accurate calibration method.

The calibration methods correct the location and the covariance of the pseudo-posterior, but not its shape. For the thresholds of categories that are rarely chosen and whose posterior is asymmetric in small samples, this can make the corrections less accurate than the pseudo-posterior in dense networks with many parameters relative to the sample size. This effect disappears as the sample size grows.

The methods differ clearly in cost. CoRe achieves the accuracy of the post hoc corrections at the cost of the pseudo-posterior sampler, with the same effective sample size, and, unlike the post hoc corrections, rescales the posterior within the sampler, which in principle allows it to be combined with procedures that sample the network structure. The Robbins-Monro and Monte Carlo Hessian variants require an additional Monte Carlo approximation of the full likelihood, AdaCoRe becomes costly for large networks, and DMH, although accurate, is about two orders of magnitude slower, with an effective sample size that decreases as the sample size grows.
 
\section{Empirical Application}
\label{sec:empirical-application}
\par We now analyze the observed data used to calibrate the simulation study, fitting the unconstrained model to all $p = 25$ items. The resulting network is larger than any network in the simulation study. We ask three questions: (i) whether the corrections reproduce a full-likelihood reference posterior, (ii) at what computational cost, and (iii) whether the conclusions drawn about the network depend on the correction.

\subsection{Data, Model and Reference Posterior}
\label{subsec:application-data}
\par The data include the responses of $n = 917$ adolescents from Lima (Peru) to the $25$ items retained in the four-factor version of the Reynolds Adolescent Depression Scale \citep[RADS-2;][]{Reynolds2002, RamosVera2023}, each answered on a four-point Likert scale. We keep the item numbering of \citet{RamosVera2023}, whose network analysis grouped the items into four clusters: dysphoria (D3, D6, D7, D8, D16, D21, D26), anhedonia and negative affect (D1, D5, D12), negative self-assessment (D4, D9, D13, D14, D15, D19, D20, D30), and somatic complaints (D11, D17, D18, D22, D24, D27, D28). The OMRF has $75$ thresholds and $300$ interactions, with the prior hyperparameters of the simulation study. Because the normalizing constant sums over $4^{25} \approx 10^{15}$ states, exact inference is infeasible, and we use as reference a demanding DMH configuration that doubles the auxiliary draws ($L = 50{,}000$) and inner Gibbs iterations ($10$) of the simulation setting. All methods run four chains from shared overdispersed starting values, with the same settings of the simulation study. PH-GHW, PH-MCH and the CoRe variants are centered at the pseudo-posterior mode, and PH-RM at the Robbins-Monro estimate of the full-posterior mode. The reference chains mix well (pooled ESS $\geq 5{,}376$, $\hat{R}_{\max} = 1.002$; \citealp{GelmanRubin1992}). Splitting the four reference chains into two pairs (chains 1--2 versus 3--4) gives a median overlap of $\eta = 0.97$ between the pairs ($0.92$ for the bivariate overlap of pairs of interactions). This reflects the Monte Carlo variability of the reference itself and is therefore the highest overlap any method can attain. All analyses were run on a MacBook Air with an Apple M3 chip (eight cores, 16\,GB of memory) under macOS 26 and R 4.5.2 \citep{R}, with the four chains of each method run in parallel.

\subsection{Results}
\label{subsec:application-results}
\input{tables/application_table_main}

\begin{figure}[t]
\centering
\def\svgwidth{\textwidth}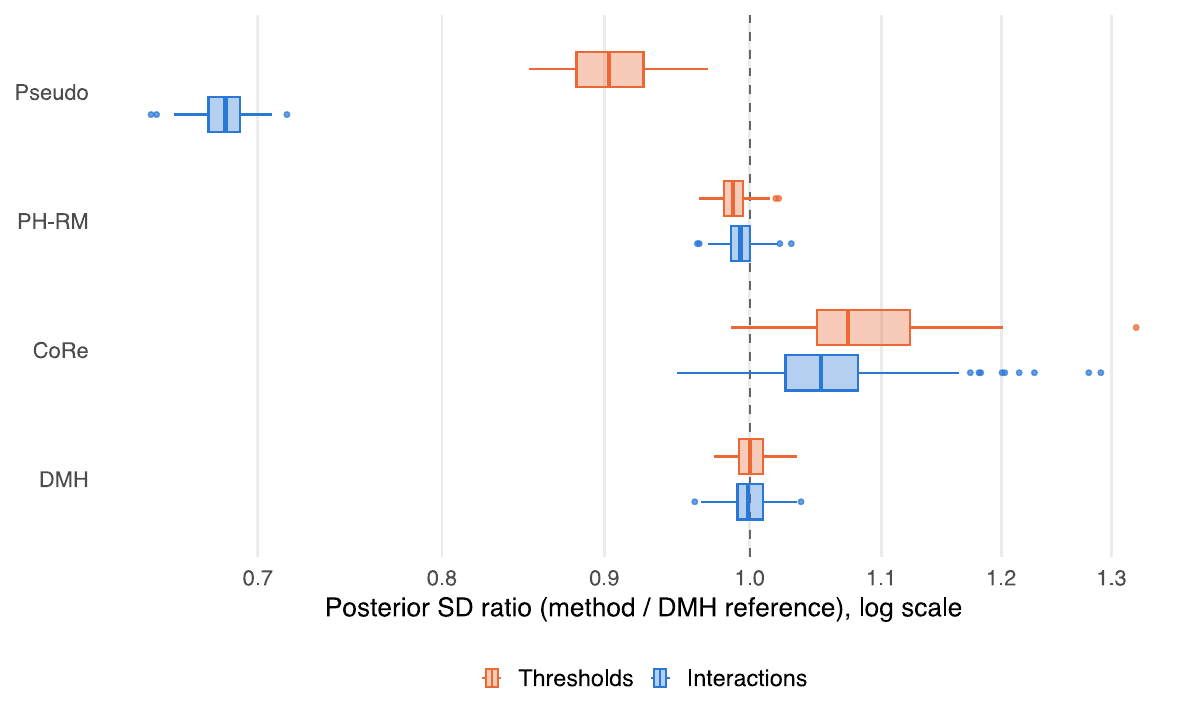
\caption{Empirical application (RADS-2, $25$ items): ratio of posterior standard deviations to the DMH reference for all thresholds and interactions (horizontal axis on a logarithmic scale), shown as standard box plots (the box spans the interquartile range, the line inside it marks the median). DMH uses the setting of the simulation study ($L = 25{,}000$, $5$ inner Gibbs iterations), while the reference DMH doubles both. The dashed line indicates equal posterior standard deviations.}
\label{fig:application_sd_ratio}
\end{figure}

\par Table \ref{tab:application} and Figure \ref{fig:application_sd_ratio} reproduce the pattern of the simulation study (all methods are reported in the supplementary material): the pseudo-posterior underestimates the posterior standard deviation of every interaction by about $32\%$ (median ratio $0.68$, and interquartile range $[0.68, 0.69]$), with median overlap $0.69$. We compare CoRe with PH-RM, the original post hoc calibration of \citet{Bouranis2017}. Both correct the interactions (ratios $1.05$ for CoRe and $0.99$ for PH-RM; overlap $0.91$ and $0.96$) and restore the dependence between them (bivariate overlap from $0.57$ to $0.85$ and $0.91$). However, for the thresholds, PH-RM reaches an overlap of $0.77$, against $0.85$ for CoRe, although its posterior standard deviations match those of the reference (ratio $0.99$). As discussed in Section \ref{sec:simulation-study}, the posterior of the thresholds is asymmetric, and corrections of location and covariance do not reproduce this asymmetry. CoRe is also more efficient: $12.5$ effective draws per second against $7.1$ for PH-RM, whose Robbins-Monro step doubles its runtime, and $0.08$ for DMH. Finally, its chains are better behaved: CoRe, CoRe-RM and CoRe-MCH reach $\hat{R}_{\max} = 1.003$ (AdaCoRe $1.016$), whereas post hoc calibration inherits the chains of the pseudo-posterior ($\hat{R}_{\max} = 1.010$ for PH-RM and $1.019$ for the pseudo-posterior).

\begin{figure}[t]
\centering
\def\svgwidth{\textwidth}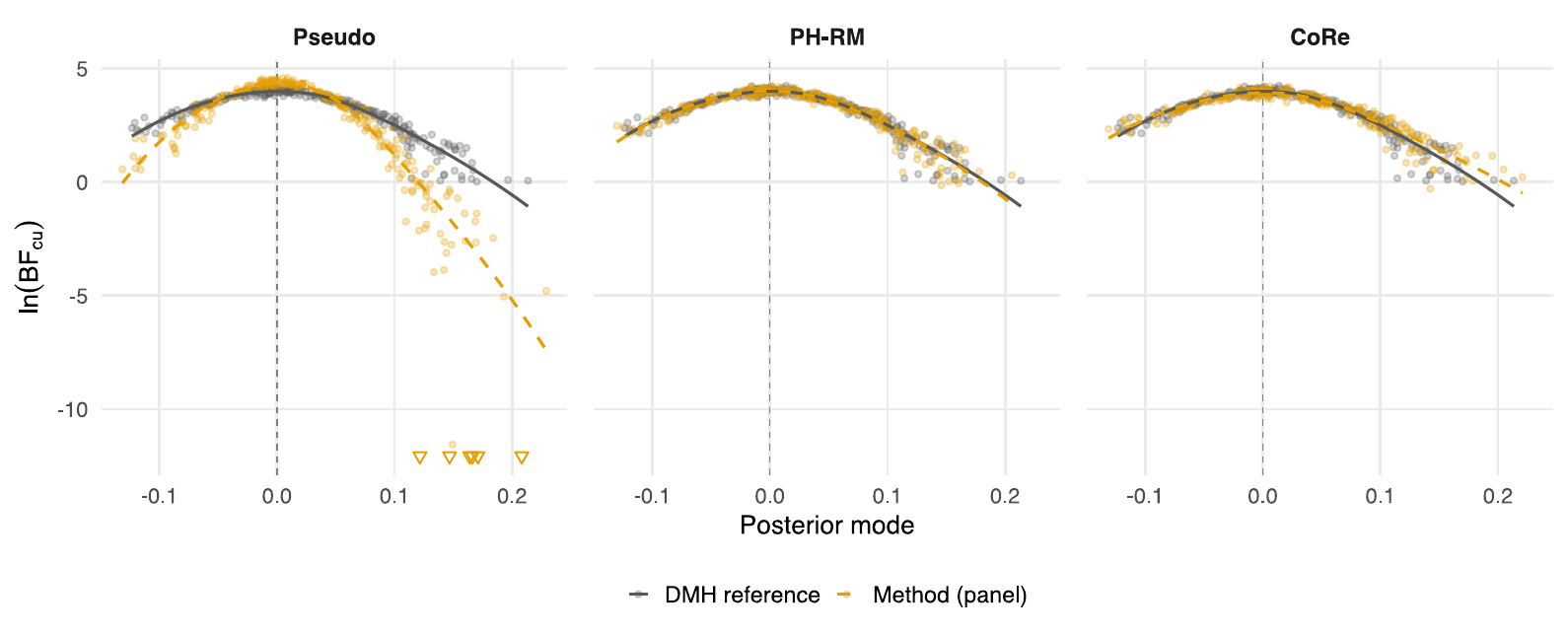
\caption{Empirical application: posterior mode versus $\ln\text{BF}_{\text{cu}}$ for the $257$ interactions for which the reference favors conditional independence ($\ln\text{BF}_{\text{cu}} > 0$). Gray points and solid line: reference; orange points and dashed line: method of the panel (lines represent trends). Triangles mark interactions for which no pseudo-posterior draw lies near zero, so that the estimated density at zero, and hence $\text{BF}_{\text{cu}}$, is effectively zero (they are excluded from the trend estimation).}
\label{fig:application_sddr}
\end{figure}

\par
\begin{figure}[t]
\centering
\def\svgwidth{\textwidth}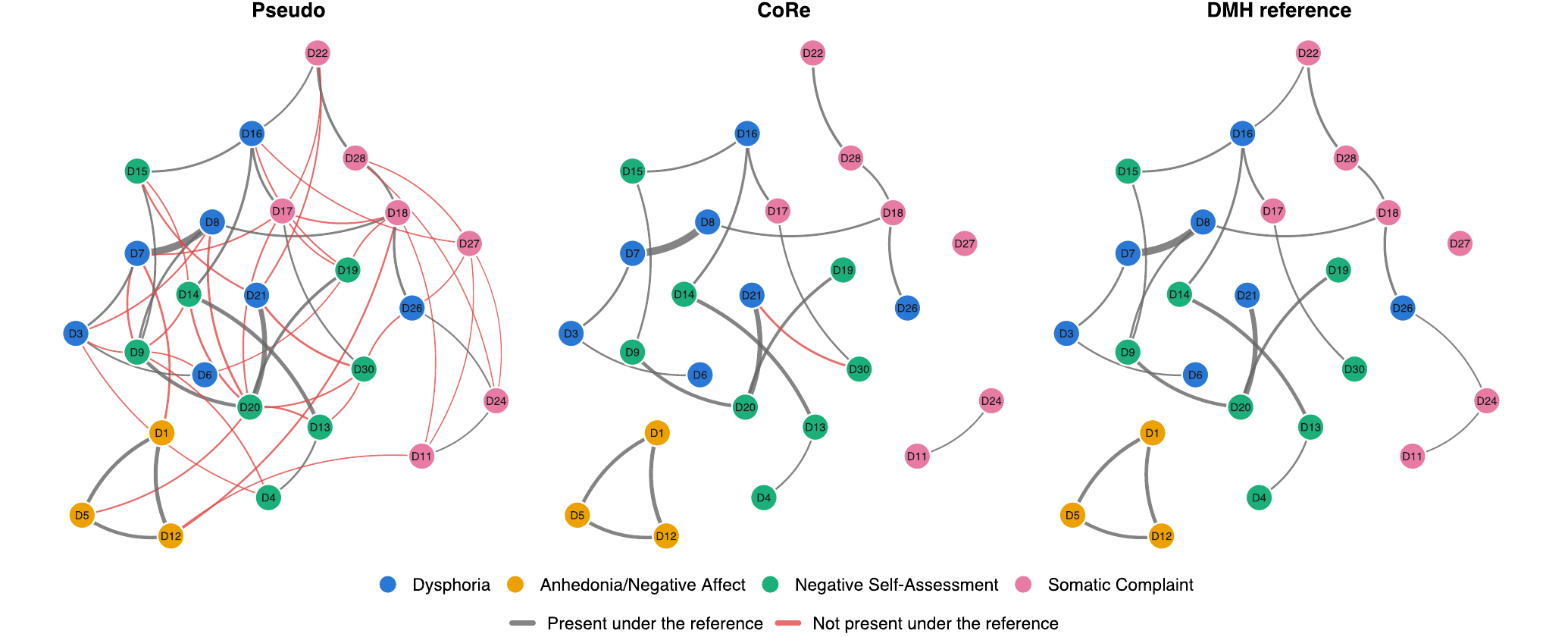
\caption{Empirical application: interactions with strong evidence for presence ($\text{BF}_{\text{cu}} < 1/10$) under the pseudo-posterior, CoRe and the DMH reference. Node positions are computed once from the DMH reference posterior and are the same in all panels. Node colors correspond to the item clusters of \citet{RamosVera2023}. Edge width is proportional to the absolute posterior mode. Gray edges are also present under the DMH reference. Red edges are not present under the DMH reference. Interactions present under the reference but not under a method are not drawn in that method's panel.}
\label{fig:application_network}
\end{figure}

 The corrections matter for inference on conditional independence (Figure \ref{fig:application_sddr}). As in the simulation study, the pseudo-posterior inflates the Savage-Dickey evidence for interactions with modes near zero and deflates it away from zero. For six interactions, it even indicates decisive evidence for an interaction where the reference favors conditional independence. PH-RM and CoRe follow the reference throughout. Classifying interactions as present ($\text{BF}_{\text{cu}} < 1/10$), absent ($\text{BF}_{\text{cu}} > 10$) or inconclusive, the reference finds $24$ present and $206$ absent interactions. The pseudo-posterior agrees with this classification for $75\%$ of the interactions and identifies $60$ as present, $36$ of which are inconclusive under the reference. The corrections agree for $94.7\%$--$98.3\%$ (CoRe $95.7\%$, PH-RM $98.3\%$), and PH-RM matches the agreement between the two DMH configurations ($98.3\%$). 
Figure \ref{fig:application_network} shows that the spurious interactions of the pseudo-posterior span all four item clusters. CoRe identifies $21$ of the $24$ interactions present under the reference and one additional interaction (D21--D30), for which the reference evidence lies just below the threshold ($1/\text{BF}_{\text{cu}} = 9.6$). The three interactions it misses (D8--D9, D16--D22 and D24--D26) are inconclusive under CoRe, with evidence in favor of an interaction that does not reach the threshold ($1/\text{BF}_{\text{cu}}$ between $2.5$ and $6.5$).

\par Corrections that rescale the pseudo-posterior around its mode do not remove all differences from the full posterior. The location of the pseudo-posterior mode agrees with that of the full posterior only asymptotically (Proposition~\ref{prop:secondorder}), so small finite-sample differences in location remain. These are negligible for individual parameters but can accumulate in quantities that aggregate many parameters, such as node centrality. For the expected influence $\sum_{j}\theta_{ij}$ \citep{Robinaugh2016}, the DMH reference identifies D7 as the most central item (posterior probability of being the most central item is $0.74$), followed by D20 ($0.21$). Under the pseudo-likelihood, the interactions of D20 lie slightly above their reference values. When summed over the node, these small shifts raise its expected influence from $2.25$ to about $2.4$ (supplementary material). As a result, the pseudo-posterior and all corrections centered at its mode, including CoRe, rank D7 and D20 as almost tied (posterior probabilities of about $0.4$ and $0.5$). PH-RM, whose draws are centered at the Robbins-Monro estimate of the full-posterior mode, reproduces the reference ranking ($0.75$ and $0.20$; see supplementary material).

\par In summary, the corrections recover the reference posterior of the interactions at a fraction of the cost of DMH and restore the conclusions on conditional independence that the pseudo-posterior distorts.

\section{Discussion}
\label{sec:discussion}

The main objective of this work was to develop sampling methods that retain the computational scalability of the pseudo-likelihood for discrete Markov random fields (MRFs), while providing robust Bayesian uncertainty quantification for network parameters. To this end, we introduced the Coordinate-Rescaling (CoRe) methodology, which embeds a linear transformation of the parameter space within the sampling procedure to correct the posterior covariance structure. In addition to a fixed rescaling variant, we proposed an adaptive extension that updates the rescaling matrix during burn-in, thereby removing the need for external curvature approximations.

Our simulation results and the empirical application indicate that recalibration-based approaches achieve a favorable balance between statistical accuracy and computational efficiency. CoRe attains the accuracy of post hoc calibration at the cost of the pseudo-likelihood sampler, whereas its adaptive extension becomes costly for large networks. Unlike transition-kernel approximations such as DMH, CoRe methods do not rely on nested Monte Carlo approximations of ratios of intractable normalizing constants within the sampling procedure. Instead, they adjust the local geometry of the pseudo-posterior to approximate the covariance structure of the full posterior directly. By separating scale correction from likelihood evaluation, CoRe methods avoid the additional Monte Carlo variability introduced by exchange-based algorithms and maintain the effective sample size of the pseudo-likelihood sampler as the sample size and network complexity increase.

The effectiveness of this approach reflects a structural feature of discrete MRFs: although pseudo-likelihood-based inference yields consistent location estimates, it systematically underestimates posterior variability. In this context, the primary deficiency lies in scale rather than bias. By correcting the geometry of the posterior distribution---rather than approximating ratios of intractable normalizing constants---CoRe methods exploit this feature directly. The resulting samplers retain computational scalability while recovering posterior uncertainty and covariance structures that closely resemble those of the reference posterior, including the dependence between parameters that the pseudo-likelihood misrepresents.

Accurate inference on conditional independence depends on the posterior scale, particularly when using Bayes factors or density ratios. Post hoc calibration corrects stored draws after sampling. CoRe applies the rescaling within the sampler, so it can in principle be combined with samplers that also move over the network structure, such as spike-and-slab or reversible-jump edge-selection schemes, provided a rescaling matrix is available for each visited structure (e.g., as a submatrix of the full-model one). Developing and evaluating such a joint edge-selection procedure is left for future work.

Several limitations of the present study should be acknowledged. First, the corrections considered here adjust the location and the covariance of the pseudo-posterior, but not its shape. For the thresholds of rarely chosen response categories, whose posterior is asymmetric in small samples, this makes the corrections less accurate than the pseudo-posterior in dense networks with many parameters relative to the sample size, although the effect disappears as the sample size grows. Moreover, the exact posterior is available only for small networks; for larger networks, our comparisons rely on DMH as reference, whose Monte Carlo error grows with the sample size.

Second, proposals were constructed on the full parameter vector rather than componentwise. While componentwise Metropolis updates can offer greater robustness in higher-dimensional settings---particularly when global proposals suffer from low acceptance rates---their use in exchange-type algorithms such as DMH would substantially increase computational cost. Each coordinate update would require a separate Monte Carlo approximation of the ratio of normalizing constants, thereby multiplying the cost of the already expensive inner sampling step. Developing efficient block or partially factorized proposal mechanisms for doubly intractable models therefore remains an important direction for future research. 

Third, the performance of recalibration methods depends on accurate estimation of local curvature. In small samples, the Godambe-Huber-White estimator yields slightly more variable posterior standard deviations than the Monte Carlo estimates based on the full likelihood, although the difference vanishes as the sample size grows. This variability may increase when response categories are rarely observed or when the data carry little information about some parameters, which can affect the quality of the rescaling matrix. Addressing this issue will be essential for extending the methodology to larger and more complex networks.

The present work suggests several directions for future research. The geometric perspective underlying coordinate-rescaling can be extended beyond pseudo-likelihoods to other forms of composite or surrogate likelihoods that retain more information from the joint distribution. For instance, composite likelihoods based on maximal cliques or alternative surrogate constructions may yield improved approximations of the full posterior geometry. Moreover, the rescaling principle can be generalized to other model classes in which full likelihood evaluation is infeasible but composite likelihood are available, including exponential random graph models (ERGMs), Potts models, and spatial autoregressive models (SARs).

Finally, implementation of the CoRe methodology in the R package \texttt{bgms} \citep{bgms-package} is planned, with the goal of providing applied researchers with an accessible tool for computationally efficient Bayesian inference in discrete MRFs.

\section*{Acknowledgement}
-- INFORMATION REMOVED FOR BLINDED REVIEW --

\section*{Competing interests}
The authors declare none.

\section*{Supplementary Material}
Supplementary Material associated with this article can be found at \zenodorepo{}.

\appendix 
\counterwithin{figure}{section}
\renewcommand{\thefigure}{\Alph{section}\arabic{figure}}
\include{appendix}


\clearpage
\bibliography{references}

\end{document}

%% file: img/exact_vs_pseudo_vs_dmh_posterior_density_comparison-tex.pdf_tex
\begingroup%
  \makeatletter%
  \providecommand\color[2][]{%
    \errmessage{(Inkscape) Color is used for the text in Inkscape, but the package 'color.sty' is not loaded}%
    \renewcommand\color[2][]{}%
  }%
  \providecommand\transparent[1]{%
    \errmessage{(Inkscape) Transparency is used (non-zero) for the text in Inkscape, but the package 'transparent.sty' is not loaded}%
    \renewcommand\transparent[1]{}%
  }%
  \providecommand\rotatebox[2]{#2}%
  \newcommand*\fsize{\dimexpr\f@size pt\relax}%
  \newcommand*\lineheight[1]{\fontsize{\fsize}{#1\fsize}\selectfont}%
  \ifx\svgwidth\undefined%
    \setlength{\unitlength}{504bp}%
    \ifx\svgscale\undefined%
      \relax%
    \else%
      \setlength{\unitlength}{\unitlength * \real{\svgscale}}%
    \fi%
  \else%
    \setlength{\unitlength}{\svgwidth}%
  \fi%
  \global\let\svgwidth\undefined%
  \global\let\svgscale\undefined%
  \makeatother%
  \begin{picture}(1,0.71428571)%
    \lineheight{1}%
    \setlength\tabcolsep{0pt}%
    \put(0,0){\includegraphics[width=\unitlength,page=1]{exact_vs_pseudo_vs_dmh_posterior_density_comparison-tex.pdf}}%
  \end{picture}%
\endgroup%

%% file: img/geometric_intuition-tex.pdf_tex
\begingroup%
  \makeatletter%
  \providecommand\color[2][]{%
    \errmessage{(Inkscape) Color is used for the text in Inkscape, but the package 'color.sty' is not loaded}%
    \renewcommand\color[2][]{}%
  }%
  \providecommand\transparent[1]{%
    \errmessage{(Inkscape) Transparency is used (non-zero) for the text in Inkscape, but the package 'transparent.sty' is not loaded}%
    \renewcommand\transparent[1]{}%
  }%
  \providecommand\rotatebox[2]{#2}%
  \newcommand*\fsize{\dimexpr\f@size pt\relax}%
  \newcommand*\lineheight[1]{\fontsize{\fsize}{#1\fsize}\selectfont}%
  \ifx\svgwidth\undefined%
    \setlength{\unitlength}{792bp}%
    \ifx\svgscale\undefined%
      \relax%
    \else%
      \setlength{\unitlength}{\unitlength * \real{\svgscale}}%
    \fi%
  \else%
    \setlength{\unitlength}{\svgwidth}%
  \fi%
  \global\let\svgwidth\undefined%
  \global\let\svgscale\undefined%
  \makeatother%
  \begin{picture}(1,0.45454545)%
    \lineheight{1}%
    \setlength\tabcolsep{0pt}%
    \put(0,0){\includegraphics[width=\unitlength,page=1]{geometric_intuition-tex.pdf}}%
  \end{picture}%
\endgroup%

%% file: img/posterior_correlations-tex.pdf_tex
\begingroup%
  \makeatletter%
  \providecommand\color[2][]{%
    \errmessage{(Inkscape) Color is used for the text in Inkscape, but the package 'color.sty' is not loaded}%
    \renewcommand\color[2][]{}%
  }%
  \providecommand\transparent[1]{%
    \errmessage{(Inkscape) Transparency is used (non-zero) for the text in Inkscape, but the package 'transparent.sty' is not loaded}%
    \renewcommand\transparent[1]{}%
  }%
  \providecommand\rotatebox[2]{#2}%
  \newcommand*\fsize{\dimexpr\f@size pt\relax}%
  \newcommand*\lineheight[1]{\fontsize{\fsize}{#1\fsize}\selectfont}%
  \ifx\svgwidth\undefined%
    \setlength{\unitlength}{792bp}%
    \ifx\svgscale\undefined%
      \relax%
    \else%
      \setlength{\unitlength}{\unitlength * \real{\svgscale}}%
    \fi%
  \else%
    \setlength{\unitlength}{\svgwidth}%
  \fi%
  \global\let\svgwidth\undefined%
  \global\let\svgscale\undefined%
  \makeatother%
  \begin{picture}(1,0.63636364)%
    \lineheight{1}%
    \setlength\tabcolsep{0pt}%
    \put(0,0){\includegraphics[width=\unitlength,page=1]{posterior_correlations-tex.pdf}}%
  \end{picture}%
\endgroup%

%% file: img/core_posterior_density_comparison-tex.pdf_tex
\begingroup%
  \makeatletter%
  \providecommand\color[2][]{%
    \errmessage{(Inkscape) Color is used for the text in Inkscape, but the package 'color.sty' is not loaded}%
    \renewcommand\color[2][]{}%
  }%
  \providecommand\transparent[1]{%
    \errmessage{(Inkscape) Transparency is used (non-zero) for the text in Inkscape, but the package 'transparent.sty' is not loaded}%
    \renewcommand\transparent[1]{}%
  }%
  \providecommand\rotatebox[2]{#2}%
  \newcommand*\fsize{\dimexpr\f@size pt\relax}%
  \newcommand*\lineheight[1]{\fontsize{\fsize}{#1\fsize}\selectfont}%
  \ifx\svgwidth\undefined%
    \setlength{\unitlength}{648bp}%
    \ifx\svgscale\undefined%
      \relax%
    \else%
      \setlength{\unitlength}{\unitlength * \real{\svgscale}}%
    \fi%
  \else%
    \setlength{\unitlength}{\svgwidth}%
  \fi%
  \global\let\svgwidth\undefined%
  \global\let\svgscale\undefined%
  \makeatother%
  \begin{picture}(1,0.77777778)%
    \lineheight{1}%
    \setlength\tabcolsep{0pt}%
    \put(0,0){\includegraphics[width=\unitlength,page=1]{core_posterior_density_comparison-tex.pdf}}%
  \end{picture}%
\endgroup%

%% file: tables/methods_summary_table.tex
\begin{table}[t]
\centering
\caption{Summary of the methods. Center: point around which the pseudo-posterior is rescaled ($\hat{\bm{\eta}}_{\text{MAP}}$: pseudo-posterior mode; $\hat{\bm{\eta}}_{\text{RM}}$: Robbins-Monro estimate of the full-posterior mode). Calibration covariance: $\mathbf{\Sigma}_{\text{GHW}}$ is the Godambe-Huber-White covariance, MC Hessian is the inverse of the negative Monte Carlo Hessian of the full log-posterior. Relative runtime: median runtime per chain divided by that of the pseudo-posterior sampler, range over the full-network conditions of Table \ref{tab:runtime} ($>1$ means slower than the pseudo-posterior sampler, $<1$ faster).}
\label{tab:methods}
\footnotesize
\begin{tabular}{@{}l >{\raggedright\arraybackslash}p{2.9cm} l >{\raggedright\arraybackslash}p{3.1cm} >{\raggedright\arraybackslash}p{2.9cm} r@{}}
\toprule
Method & Sampled distribution & Center & Calibration covariance & Full-likelihood Monte Carlo & \begin{tabular}[b]{@{}r@{}}Relative\\runtime\end{tabular} \\
\midrule
Pseudo   & pseudo-posterior & -- & -- & -- & $1$ \\
\addlinespace
CoRe     & transformed pseudo-posterior & $\hat{\bm{\eta}}_{\text{MAP}}$ & $\mathbf{\Sigma}_{\text{GHW}}$ at $\hat{\bm{\eta}}_{\text{MAP}}$ & -- & $0.9$--$1.1$ \\
AdaCoRe  & transformed pseudo-posterior & $\hat{\bm{\eta}}_{\text{MAP}}$ & $\mathbf{\Sigma}_{\text{GHW}}$ at the running mean, updated during burn-in & -- & $1.4$--$9.6$ \\
CoRe-MCH & transformed pseudo-posterior & $\hat{\bm{\eta}}_{\text{MAP}}$ & MC Hessian at $\hat{\bm{\eta}}_{\text{MAP}}$ & once, before sampling & $0.9$--$1.2$ \\
CoRe-RM  & transformed pseudo-posterior & $\hat{\bm{\eta}}_{\text{MAP}}$ & MC Hessian at $\hat{\bm{\eta}}_{\text{RM}}$ & once, before sampling (RM and Hessian) & $1.2$--$4.2$ \\
\addlinespace
PH-GHW   & pseudo-posterior (draws rescaled after sampling) & $\hat{\bm{\eta}}_{\text{MAP}}$ & $\mathbf{\Sigma}_{\text{GHW}}$ at $\hat{\bm{\eta}}_{\text{MAP}}$ & -- & $1.0$ \\
PH-MCH   & pseudo-posterior (draws rescaled after sampling) & $\hat{\bm{\eta}}_{\text{MAP}}$ & MC Hessian at $\hat{\bm{\eta}}_{\text{MAP}}$ & once, after sampling & $1.0$--$1.2$ \\
PH-RM    & pseudo-posterior (draws rescaled after sampling) & $\hat{\bm{\eta}}_{\text{RM}}$ & MC Hessian at $\hat{\bm{\eta}}_{\text{RM}}$ & once, after sampling (RM and Hessian) & $1.3$--$4.2$ \\
\addlinespace
DMH      & full posterior (approximate transition kernel) & -- & -- & every iteration (inner Gibbs sampler) & $34$--$380$ \\
Exact ($p \le 10$) & full posterior & -- & -- & -- (enumeration of the normalizing constant) & $0.1$--$150$ \\
\bottomrule
\end{tabular}
\end{table}

%% file: img/smallworld_network-tex.pdf_tex
\begingroup%
  \makeatletter%
  \providecommand\color[2][]{%
    \errmessage{(Inkscape) Color is used for the text in Inkscape, but the package 'color.sty' is not loaded}%
    \renewcommand\color[2][]{}%
  }%
  \providecommand\transparent[1]{%
    \errmessage{(Inkscape) Transparency is used (non-zero) for the text in Inkscape, but the package 'transparent.sty' is not loaded}%
    \renewcommand\transparent[1]{}%
  }%
  \providecommand\rotatebox[2]{#2}%
  \newcommand*\fsize{\dimexpr\f@size pt\relax}%
  \newcommand*\lineheight[1]{\fontsize{\fsize}{#1\fsize}\selectfont}%
  \ifx\svgwidth\undefined%
    \setlength{\unitlength}{432bp}%
    \ifx\svgscale\undefined%
      \relax%
    \else%
      \setlength{\unitlength}{\unitlength * \real{\svgscale}}%
    \fi%
  \else%
    \setlength{\unitlength}{\svgwidth}%
  \fi%
  \global\let\svgwidth\undefined%
  \global\let\svgscale\undefined%
  \makeatother%
  \begin{picture}(1,1)%
    \lineheight{1}%
    \setlength\tabcolsep{0pt}%
    \put(0,0){\includegraphics[width=\unitlength,page=1]{smallworld_network-tex.pdf}}%
  \end{picture}%
\endgroup%

%% file: img/random_network-tex.pdf_tex
\begingroup%
  \makeatletter%
  \providecommand\color[2][]{%
    \errmessage{(Inkscape) Color is used for the text in Inkscape, but the package 'color.sty' is not loaded}%
    \renewcommand\color[2][]{}%
  }%
  \providecommand\transparent[1]{%
    \errmessage{(Inkscape) Transparency is used (non-zero) for the text in Inkscape, but the package 'transparent.sty' is not loaded}%
    \renewcommand\transparent[1]{}%
  }%
  \providecommand\rotatebox[2]{#2}%
  \newcommand*\fsize{\dimexpr\f@size pt\relax}%
  \newcommand*\lineheight[1]{\fontsize{\fsize}{#1\fsize}\selectfont}%
  \ifx\svgwidth\undefined%
    \setlength{\unitlength}{432bp}%
    \ifx\svgscale\undefined%
      \relax%
    \else%
      \setlength{\unitlength}{\unitlength * \real{\svgscale}}%
    \fi%
  \else%
    \setlength{\unitlength}{\svgwidth}%
  \fi%
  \global\let\svgwidth\undefined%
  \global\let\svgscale\undefined%
  \makeatother%
  \begin{picture}(1,1)%
    \lineheight{1}%
    \setlength\tabcolsep{0pt}%
    \put(0,0){\includegraphics[width=\unitlength,page=1]{random_network-tex.pdf}}%
  \end{picture}%
\endgroup%

%% file: img/full_network-tex.pdf_tex
\begingroup%
  \makeatletter%
  \providecommand\color[2][]{%
    \errmessage{(Inkscape) Color is used for the text in Inkscape, but the package 'color.sty' is not loaded}%
    \renewcommand\color[2][]{}%
  }%
  \providecommand\transparent[1]{%
    \errmessage{(Inkscape) Transparency is used (non-zero) for the text in Inkscape, but the package 'transparent.sty' is not loaded}%
    \renewcommand\transparent[1]{}%
  }%
  \providecommand\rotatebox[2]{#2}%
  \newcommand*\fsize{\dimexpr\f@size pt\relax}%
  \newcommand*\lineheight[1]{\fontsize{\fsize}{#1\fsize}\selectfont}%
  \ifx\svgwidth\undefined%
    \setlength{\unitlength}{432bp}%
    \ifx\svgscale\undefined%
      \relax%
    \else%
      \setlength{\unitlength}{\unitlength * \real{\svgscale}}%
    \fi%
  \else%
    \setlength{\unitlength}{\svgwidth}%
  \fi%
  \global\let\svgwidth\undefined%
  \global\let\svgscale\undefined%
  \makeatother%
  \begin{picture}(1,1)%
    \lineheight{1}%
    \setlength\tabcolsep{0pt}%
    \put(0,0){\includegraphics[width=\unitlength,page=1]{full_network-tex.pdf}}%
  \end{picture}%
\endgroup%

%% file: img/simulation_overlap_interactions_full_structure-tex.pdf_tex
\begingroup%
  \makeatletter%
  \providecommand\color[2][]{%
    \errmessage{(Inkscape) Color is used for the text in Inkscape, but the package 'color.sty' is not loaded}%
    \renewcommand\color[2][]{}%
  }%
  \providecommand\transparent[1]{%
    \errmessage{(Inkscape) Transparency is used (non-zero) for the text in Inkscape, but the package 'transparent.sty' is not loaded}%
    \renewcommand\transparent[1]{}%
  }%
  \providecommand\rotatebox[2]{#2}%
  \newcommand*\fsize{\dimexpr\f@size pt\relax}%
  \newcommand*\lineheight[1]{\fontsize{\fsize}{#1\fsize}\selectfont}%
  \ifx\svgwidth\undefined%
    \setlength{\unitlength}{720bp}%
    \ifx\svgscale\undefined%
      \relax%
    \else%
      \setlength{\unitlength}{\unitlength * \real{\svgscale}}%
    \fi%
  \else%
    \setlength{\unitlength}{\svgwidth}%
  \fi%
  \global\let\svgwidth\undefined%
  \global\let\svgscale\undefined%
  \makeatother%
  \begin{picture}(1,0.7)%
    \lineheight{1}%
    \setlength\tabcolsep{0pt}%
    \put(0,0){\includegraphics[width=\unitlength,page=1]{simulation_overlap_interactions_full_structure-tex.pdf}}%
  \end{picture}%
\endgroup%

%% file: img/simulation_overlap_thresholds_full_structure-tex.pdf_tex
\begingroup%
  \makeatletter%
  \providecommand\color[2][]{%
    \errmessage{(Inkscape) Color is used for the text in Inkscape, but the package 'color.sty' is not loaded}%
    \renewcommand\color[2][]{}%
  }%
  \providecommand\transparent[1]{%
    \errmessage{(Inkscape) Transparency is used (non-zero) for the text in Inkscape, but the package 'transparent.sty' is not loaded}%
    \renewcommand\transparent[1]{}%
  }%
  \providecommand\rotatebox[2]{#2}%
  \newcommand*\fsize{\dimexpr\f@size pt\relax}%
  \newcommand*\lineheight[1]{\fontsize{\fsize}{#1\fsize}\selectfont}%
  \ifx\svgwidth\undefined%
    \setlength{\unitlength}{720bp}%
    \ifx\svgscale\undefined%
      \relax%
    \else%
      \setlength{\unitlength}{\unitlength * \real{\svgscale}}%
    \fi%
  \else%
    \setlength{\unitlength}{\svgwidth}%
  \fi%
  \global\let\svgwidth\undefined%
  \global\let\svgscale\undefined%
  \makeatother%
  \begin{picture}(1,0.7)%
    \lineheight{1}%
    \setlength\tabcolsep{0pt}%
    \put(0,0){\includegraphics[width=\unitlength,page=1]{simulation_overlap_thresholds_full_structure-tex.pdf}}%
  \end{picture}%
\endgroup%

%% file: tables/simulation_ess_table_full.tex
\providecommand{\mediqr}[2]{\begin{tabular}[t]{@{}c@{}}#1\\[-1pt]{\scriptsize[#2]}\end{tabular}}
\begin{table}[t]
\centering
\caption{Median [interquartile range] effective sample size (ESS) over parameters and datasets, full network structure. The reference posterior is sampled exactly for $p \leq 10$ (Exact) and by DMH for $p \geq 15$ (DMH (reference)). DMH as a competing method is run for $p \leq 10$ only.}
\label{tab:ess}
\footnotesize
\setlength{\tabcolsep}{2pt}
\begin{tabular}{@{}lcccccccc@{}}
\toprule
 & \multicolumn{2}{c}{$p = 6$} & \multicolumn{2}{c}{$p = 10$} & \multicolumn{2}{c}{$p = 15$} & \multicolumn{2}{c}{$p = 22$} \\
\cmidrule(lr){2-3}\cmidrule(lr){4-5}\cmidrule(lr){6-7}\cmidrule(lr){8-9}
Method & $n = 500$ & $n = 3{,}000$ & $n = 500$ & $n = 3{,}000$ & $n = 500$ & $n = 3{,}000$ & $n = 500$ & $n = 3{,}000$ \\
\midrule
Exact & \mediqr{3,088}{2991; 3176} & \mediqr{3,285}{3190; 3362} & \mediqr{2,091}{2026; 2148} & \mediqr{2,282}{2216; 2339} & -- & -- & -- & -- \\[2pt]
DMH (reference) & -- & -- & -- & -- & \mediqr{1,440}{1390; 1490} & \mediqr{442}{425; 460} & \mediqr{992}{954; 1031} & \mediqr{220}{209; 232} \\[2pt]
Pseudo & \mediqr{3,271}{3184; 3354} & \mediqr{3,324}{3231; 3399} & \mediqr{2,298}{2234; 2350} & \mediqr{2,327}{2269; 2381} & \mediqr{1,704}{1659; 1745} & \mediqr{1,724}{1680; 1764} & \mediqr{1,247}{1213; 1279} & \mediqr{1,257}{1223; 1288} \\[2pt]
CoRe & \mediqr{3,276}{3187; 3347} & \mediqr{3,318}{3232; 3395} & \mediqr{2,301}{2241; 2353} & \mediqr{2,330}{2271; 2384} & \mediqr{1,705}{1662; 1745} & \mediqr{1,725}{1682; 1765} & \mediqr{1,245}{1212; 1276} & \mediqr{1,258}{1225; 1290} \\[2pt]
AdaCoRe & \mediqr{3,274}{3180; 3351} & \mediqr{3,328}{3237; 3408} & \mediqr{2,299}{2238; 2350} & \mediqr{2,330}{2271; 2382} & \mediqr{1,705}{1662; 1745} & \mediqr{1,724}{1682; 1764} & \mediqr{1,244}{1211; 1276} & \mediqr{1,255}{1221; 1286} \\[2pt]
PH-GHW & \mediqr{3,275}{3185; 3355} & \mediqr{3,316}{3222; 3397} & \mediqr{2,303}{2243; 2356} & \mediqr{2,327}{2266; 2379} & \mediqr{1,704}{1661; 1744} & \mediqr{1,723}{1678; 1764} & \mediqr{1,246}{1213; 1277} & \mediqr{1,256}{1223; 1288} \\[2pt]
PH-RM & \mediqr{3,276}{3183; 3356} & \mediqr{3,316}{3223; 3396} & \mediqr{2,303}{2244; 2355} & \mediqr{2,327}{2266; 2380} & \mediqr{1,705}{1663; 1746} & \mediqr{1,723}{1679; 1764} & \mediqr{1,248}{1215; 1279} & \mediqr{1,256}{1223; 1288} \\[2pt]
PH-MCH & \mediqr{3,275}{3182; 3356} & \mediqr{3,316}{3224; 3397} & \mediqr{2,303}{2244; 2356} & \mediqr{2,327}{2266; 2380} & \mediqr{1,706}{1663; 1746} & \mediqr{1,722}{1679; 1764} & \mediqr{1,248}{1215; 1279} & \mediqr{1,256}{1223; 1288} \\[2pt]
CoRe-RM & \mediqr{3,278}{3187; 3356} & \mediqr{3,320}{3220; 3400} & \mediqr{2,308}{2246; 2360} & \mediqr{2,331}{2271; 2384} & \mediqr{1,708}{1666; 1750} & \mediqr{1,724}{1680; 1765} & \mediqr{1,249}{1216; 1281} & \mediqr{1,256}{1223; 1288} \\[2pt]
CoRe-MCH & \mediqr{3,279}{3187; 3362} & \mediqr{3,327}{3239; 3408} & \mediqr{2,307}{2248; 2360} & \mediqr{2,330}{2270; 2380} & \mediqr{1,710}{1666; 1750} & \mediqr{1,724}{1681; 1765} & \mediqr{1,246}{1214; 1278} & \mediqr{1,257}{1224; 1289} \\[2pt]
DMH & \mediqr{2,949}{2864; 3026} & \mediqr{1,776}{1729; 1823} & \mediqr{1,979}{1917; 2044} & \mediqr{869}{841; 898} & -- & -- & -- & -- \\[2pt]
\bottomrule
\end{tabular}
\end{table}

%% file: tables/simulation_runtime_table_full.tex
\providecommand{\mediqr}[2]{\begin{tabular}[t]{@{}c@{}}#1\\[-1pt]{\scriptsize[#2]}\end{tabular}}
\begin{table}[t]
\centering
\caption{Median [interquartile range] runtime per chain in seconds over datasets, full network structure. Rows are the same as in Table \ref{tab:ess}.}
\label{tab:runtime}
\footnotesize
\setlength{\tabcolsep}{2pt}
\begin{tabular}{@{}lcccccccc@{}}
\toprule
 & \multicolumn{2}{c}{$p = 6$} & \multicolumn{2}{c}{$p = 10$} & \multicolumn{2}{c}{$p = 15$} & \multicolumn{2}{c}{$p = 22$} \\
\cmidrule(lr){2-3}\cmidrule(lr){4-5}\cmidrule(lr){6-7}\cmidrule(lr){8-9}
Method & $n = 500$ & $n = 3{,}000$ & $n = 500$ & $n = 3{,}000$ & $n = 500$ & $n = 3{,}000$ & $n = 500$ & $n = 3{,}000$ \\
\midrule
Exact & \mediqr{1.6}{1.6; 1.7} & \mediqr{1.6}{1.6; 1.6} & \mediqr{2,385}{2326; 2485} & \mediqr{2,383}{2354; 2460} & -- & -- & -- & -- \\[2pt]
DMH (reference) & -- & -- & -- & -- & \mediqr{5,865}{5855; 5874} & \mediqr{5,859}{5850; 5874} & \mediqr{8,941}{8927; 8957} & \mediqr{8,944}{8931; 8960} \\[2pt]
Pseudo & \mediqr{5.7}{5.4; 6.2} & \mediqr{20.2}{19.3; 21.6} & \mediqr{15.9}{15.8; 16.1} & \mediqr{81.1}{78.9; 84.8} & \mediqr{33.1}{33.0; 33.2} & \mediqr{165}{165; 166} & \mediqr{51.0}{50.9; 51.1} & \mediqr{261}{260; 262} \\[2pt]
CoRe & \mediqr{6.0}{5.6; 6.5} & \mediqr{20.5}{19.6; 21.7} & \mediqr{16.1}{15.9; 16.2} & \mediqr{80.7}{78.5; 84.6} & \mediqr{28.3}{28.3; 28.4} & \mediqr{154}{153; 154} & \mediqr{53.2}{53.1; 53.4} & \mediqr{269}{267; 271} \\[2pt]
AdaCoRe & \mediqr{7.7}{7.1; 8.3} & \mediqr{28.7}{27.0; 30.7} & \mediqr{27.6}{27.2; 28.0} & \mediqr{163}{157; 172} & \mediqr{104}{103; 105} & \mediqr{599}{595; 605} & \mediqr{384}{381; 389} & \mediqr{2,513}{2486; 2534} \\[2pt]
PH-GHW & \mediqr{5.7}{5.5; 6.2} & \mediqr{20.3}{19.4; 21.6} & \mediqr{16.0}{15.9; 16.2} & \mediqr{81.5}{79.3; 85.2} & \mediqr{33.5}{33.4; 33.6} & \mediqr{167}{167; 167} & \mediqr{52.6}{52.5; 52.6} & \mediqr{267}{266; 268} \\[2pt]
PH-RM & \mediqr{23.9}{23.6; 24.4} & \mediqr{38.6}{37.6; 40.0} & \mediqr{48.4}{48.3; 48.7} & \mediqr{114}{112; 118} & \mediqr{96.3}{96.1; 96.4} & \mediqr{219}{219; 220} & \mediqr{141}{141; 141} & \mediqr{355}{355; 356} \\[2pt]
PH-MCH & \mediqr{6.1}{5.9; 6.6} & \mediqr{20.7}{19.9; 22.1} & \mediqr{17.1}{16.9; 17.2} & \mediqr{82.5}{80.3; 86.2} & \mediqr{36.8}{36.7; 36.8} & \mediqr{170}{169; 170} & \mediqr{61.0}{61.0; 61.1} & \mediqr{275}{275; 276} \\[2pt]
CoRe-RM & \mediqr{24.0}{23.7; 24.6} & \mediqr{38.7}{37.7; 39.9} & \mediqr{48.5}{48.3; 48.7} & \mediqr{114}{111; 117} & \mediqr{80.1}{80.0; 80.2} & \mediqr{204}{204; 205} & \mediqr{142}{141; 142} & \mediqr{356}{355; 358} \\[2pt]
CoRe-MCH & \mediqr{6.3}{6.0; 6.8} & \mediqr{20.8}{20.0; 22.1} & \mediqr{17.2}{17.0; 17.4} & \mediqr{81.8}{79.2; 85.5} & \mediqr{30.8}{30.7; 30.8} & \mediqr{156}{156; 157} & \mediqr{61.5}{61.4; 61.8} & \mediqr{276}{274; 280} \\[2pt]
DMH & \mediqr{2,159}{2153; 2164} & \mediqr{2,176}{2165; 2184} & \mediqr{3,790}{3782; 3798} & \mediqr{3,792}{3781; 3800} & -- & -- & -- & -- \\[2pt]
\bottomrule
\end{tabular}
\end{table}

%% file: img/simulation_bivariate_overlap_full_structure-tex.pdf_tex
\begingroup%
  \makeatletter%
  \providecommand\color[2][]{%
    \errmessage{(Inkscape) Color is used for the text in Inkscape, but the package 'color.sty' is not loaded}%
    \renewcommand\color[2][]{}%
  }%
  \providecommand\transparent[1]{%
    \errmessage{(Inkscape) Transparency is used (non-zero) for the text in Inkscape, but the package 'transparent.sty' is not loaded}%
    \renewcommand\transparent[1]{}%
  }%
  \providecommand\rotatebox[2]{#2}%
  \newcommand*\fsize{\dimexpr\f@size pt\relax}%
  \newcommand*\lineheight[1]{\fontsize{\fsize}{#1\fsize}\selectfont}%
  \ifx\svgwidth\undefined%
    \setlength{\unitlength}{1008bp}%
    \ifx\svgscale\undefined%
      \relax%
    \else%
      \setlength{\unitlength}{\unitlength * \real{\svgscale}}%
    \fi%
  \else%
    \setlength{\unitlength}{\svgwidth}%
  \fi%
  \global\let\svgwidth\undefined%
  \global\let\svgscale\undefined%
  \makeatother%
  \begin{picture}(1,0.5)%
    \lineheight{1}%
    \setlength\tabcolsep{0pt}%
    \put(0,0){\includegraphics[width=\unitlength,page=1]{simulation_bivariate_overlap_full_structure-tex.pdf}}%
  \end{picture}%
\endgroup%

%% file: tables/simulation_threshold_asymmetry_table_full.tex
\begin{table}[t]
\centering
\caption{Thresholds, full network structure: asymmetry of the reference posterior and displacement of the posterior means. Medians over thresholds and datasets. Except for the SD ratio, each entry is a difference divided by the posterior standard deviation of the threshold under the reference, so that entries are comparable across thresholds and conditions. Mode offset: mode of the pseudo-posterior minus mode of the reference posterior (estimated by Robbins-Monro); asymmetry: posterior mean minus mode of the reference posterior; SD ratio: posterior standard deviation of CoRe relative to the reference; displacement: posterior mean of the method minus that of the reference.}
\label{tab:simulation-threshold-asymmetry}
\small
\begin{tabular}{rr r r r rrr}
\toprule
 & & Mode offset & Asymmetry & SD ratio & \multicolumn{3}{c}{Displacement} \\
\cmidrule(lr){6-8}
$p$ & $n$ & pseudo & reference & CoRe & Pseudo & CoRe & PH-RM \\
\midrule
6 & 500 & 0.01 & $-$0.12 & 0.99 & 0.05 & 0.13 & 0.13 \\
6 & 3,000 & 0.02 & $-$0.05 & 1.00 & 0.03 & 0.06 & 0.05 \\
10 & 500 & 0.06 & $-$0.18 & 0.99 & 0.12 & 0.25 & 0.20 \\
10 & 3,000 & 0.03 & $-$0.08 & 1.00 & 0.05 & 0.11 & 0.08 \\
15 & 500 & 0.07 & $-$0.28 & 0.96 & 0.18 & 0.36 & 0.30 \\
15 & 3,000 & 0.03 & $-$0.11 & 1.00 & 0.07 & 0.15 & 0.12 \\
22 & 500 & 0.07 & $-$0.38 & 0.96 & 0.22 & 0.48 & 0.42 \\
22 & 3,000 & 0.03 & $-$0.15 & 1.00 & 0.09 & 0.19 & 0.16 \\
\bottomrule
\end{tabular}
\end{table}

%% file: img/simulation_scatterplot_bf_reference_vs_pseudo_random_structure-tex.pdf_tex
\begingroup%
  \makeatletter%
  \providecommand\color[2][]{%
    \errmessage{(Inkscape) Color is used for the text in Inkscape, but the package 'color.sty' is not loaded}%
    \renewcommand\color[2][]{}%
  }%
  \providecommand\transparent[1]{%
    \errmessage{(Inkscape) Transparency is used (non-zero) for the text in Inkscape, but the package 'transparent.sty' is not loaded}%
    \renewcommand\transparent[1]{}%
  }%
  \providecommand\rotatebox[2]{#2}%
  \newcommand*\fsize{\dimexpr\f@size pt\relax}%
  \newcommand*\lineheight[1]{\fontsize{\fsize}{#1\fsize}\selectfont}%
  \ifx\svgwidth\undefined%
    \setlength{\unitlength}{720bp}%
    \ifx\svgscale\undefined%
      \relax%
    \else%
      \setlength{\unitlength}{\unitlength * \real{\svgscale}}%
    \fi%
  \else%
    \setlength{\unitlength}{\svgwidth}%
  \fi%
  \global\let\svgwidth\undefined%
  \global\let\svgscale\undefined%
  \makeatother%
  \begin{picture}(1,0.8)%
    \lineheight{1}%
    \setlength\tabcolsep{0pt}%
    \put(0,0){\includegraphics[width=\unitlength,page=1]{simulation_scatterplot_bf_reference_vs_pseudo_random_structure-tex.pdf}}%
  \end{picture}%
\endgroup%

%% file: img/simulation_scatterplot_bf_methods_random_structure-tex.pdf_tex
\begingroup%
  \makeatletter%
  \providecommand\color[2][]{%
    \errmessage{(Inkscape) Color is used for the text in Inkscape, but the package 'color.sty' is not loaded}%
    \renewcommand\color[2][]{}%
  }%
  \providecommand\transparent[1]{%
    \errmessage{(Inkscape) Transparency is used (non-zero) for the text in Inkscape, but the package 'transparent.sty' is not loaded}%
    \renewcommand\transparent[1]{}%
  }%
  \providecommand\rotatebox[2]{#2}%
  \newcommand*\fsize{\dimexpr\f@size pt\relax}%
  \newcommand*\lineheight[1]{\fontsize{\fsize}{#1\fsize}\selectfont}%
  \ifx\svgwidth\undefined%
    \setlength{\unitlength}{720bp}%
    \ifx\svgscale\undefined%
      \relax%
    \else%
      \setlength{\unitlength}{\unitlength * \real{\svgscale}}%
    \fi%
  \else%
    \setlength{\unitlength}{\svgwidth}%
  \fi%
  \global\let\svgwidth\undefined%
  \global\let\svgscale\undefined%
  \makeatother%
  \begin{picture}(1,0.8)%
    \lineheight{1}%
    \setlength\tabcolsep{0pt}%
    \put(0,0){\includegraphics[width=\unitlength,page=1]{simulation_scatterplot_bf_methods_random_structure-tex.pdf}}%
  \end{picture}%
\endgroup%

%% file: tables/application_table_main.tex
\begin{table}[t]
\centering
\caption{Empirical application: efficiency and accuracy relative to the DMH reference. Runtime per chain; ESS: pooled minimum over parameters (four chains); $\hat{R}_{\max}$: largest potential scale reduction factor. Overlap $\eta$ and SD ratio: median [interquartile range] over thresholds ($\mu$) and interactions ($\theta$); bivariate overlap: median over 50 pairs per type.}
\label{tab:application}
\begin{adjustbox}{max width=\textwidth}
\begin{tabular}{l rrrr cc ccc cc}
\toprule
 & \multicolumn{4}{c}{Efficiency} & \multicolumn{2}{c}{Overlap $\eta$} & \multicolumn{3}{c}{Bivariate overlap} & \multicolumn{2}{c}{$\sigma_{\text{method}}/\sigma_{\text{ref}}$} \\
\cmidrule(lr){2-5} \cmidrule(lr){6-7} \cmidrule(lr){8-10} \cmidrule(lr){11-12}
Method & Runtime & ESS & ESS/s & $\hat{R}_{\max}$ & $\mu$ & $\theta$ & $\theta$--$\theta$ & $\mu$--$\theta$ & $\mu$--$\mu$ & $\mu$ & $\theta$ \\
\midrule
Pseudo & 1.2 min & 3,272 & 11.6 & 1.019 & \mediqr{0.82}{0.75, 0.87} & \mediqr{0.69}{0.68, 0.70} & 0.57 & 0.64 & 0.76 & \mediqr{0.90}{0.88, 0.93} & \mediqr{0.68}{0.68, 0.69} \\
PH-RM & 2.3 min & 3,973 & 7.1 & 1.010 & \mediqr{0.77}{0.74, 0.86} & \mediqr{0.96}{0.93, 0.97} & 0.91 & 0.76 & 0.71 & \mediqr{0.99}{0.98, 1.00} & \mediqr{0.99}{0.99, 1.00} \\
CoRe & 1.4 min & 4,111 & 12.5 & 1.003 & \mediqr{0.85}{0.77, 0.89} & \mediqr{0.91}{0.88, 0.94} & 0.85 & 0.81 & 0.78 & \mediqr{1.07}{1.05, 1.12} & \mediqr{1.05}{1.03, 1.08} \\
DMH & 1.8 h & 1,966 & 0.08 & 1.032 & \mediqr{0.97}{0.96, 0.97} & \mediqr{0.97}{0.96, 0.97} & 0.92 & 0.92 & 0.93 & \mediqr{1.00}{0.99, 1.01} & \mediqr{1.00}{0.99, 1.01} \\
DMH (reference) & 12.0 h & 5,376 & 0.03 & 1.002 & -- & -- & -- & -- & -- & -- & -- \\
\bottomrule
\end{tabular}
\end{adjustbox}
\end{table}

%% file: img/application_sd_ratio_main-tex.pdf_tex
\begingroup%
  \makeatletter%
  \providecommand\color[2][]{%
    \errmessage{(Inkscape) Color is used for the text in Inkscape, but the package 'color.sty' is not loaded}%
    \renewcommand\color[2][]{}%
  }%
  \providecommand\transparent[1]{%
    \errmessage{(Inkscape) Transparency is used (non-zero) for the text in Inkscape, but the package 'transparent.sty' is not loaded}%
    \renewcommand\transparent[1]{}%
  }%
  \providecommand\rotatebox[2]{#2}%
  \newcommand*\fsize{\dimexpr\f@size pt\relax}%
  \newcommand*\lineheight[1]{\fontsize{\fsize}{#1\fsize}\selectfont}%
  \ifx\svgwidth\undefined%
    \setlength{\unitlength}{576bp}%
    \ifx\svgscale\undefined%
      \relax%
    \else%
      \setlength{\unitlength}{\unitlength * \real{\svgscale}}%
    \fi%
  \else%
    \setlength{\unitlength}{\svgwidth}%
  \fi%
  \global\let\svgwidth\undefined%
  \global\let\svgscale\undefined%
  \makeatother%
  \begin{picture}(1,0.59895833)%
    \lineheight{1}%
    \setlength\tabcolsep{0pt}%
    \put(0,0){\includegraphics[width=\unitlength,page=1]{application_sd_ratio_main-tex.pdf}}%
  \end{picture}%
\endgroup%

%% file: img/application_sddr_main-tex.pdf_tex
\begingroup%
  \makeatletter%
  \providecommand\color[2][]{%
    \errmessage{(Inkscape) Color is used for the text in Inkscape, but the package 'color.sty' is not loaded}%
    \renewcommand\color[2][]{}%
  }%
  \providecommand\transparent[1]{%
    \errmessage{(Inkscape) Transparency is used (non-zero) for the text in Inkscape, but the package 'transparent.sty' is not loaded}%
    \renewcommand\transparent[1]{}%
  }%
  \providecommand\rotatebox[2]{#2}%
  \newcommand*\fsize{\dimexpr\f@size pt\relax}%
  \newcommand*\lineheight[1]{\fontsize{\fsize}{#1\fsize}\selectfont}%
  \ifx\svgwidth\undefined%
    \setlength{\unitlength}{763bp}%
    \ifx\svgscale\undefined%
      \relax%
    \else%
      \setlength{\unitlength}{\unitlength * \real{\svgscale}}%
    \fi%
  \else%
    \setlength{\unitlength}{\svgwidth}%
  \fi%
  \global\let\svgwidth\undefined%
  \global\let\svgscale\undefined%
  \makeatother%
  \begin{picture}(1,0.40498034)%
    \lineheight{1}%
    \setlength\tabcolsep{0pt}%
    \put(0,0){\includegraphics[width=\unitlength,page=1]{application_sddr_main-tex.pdf}}%
  \end{picture}%
\endgroup%

%% file: img/application_network_main-tex.pdf_tex
\begingroup%
  \makeatletter%
  \providecommand\color[2][]{%
    \errmessage{(Inkscape) Color is used for the text in Inkscape, but the package 'color.sty' is not loaded}%
    \renewcommand\color[2][]{}%
  }%
  \providecommand\transparent[1]{%
    \errmessage{(Inkscape) Transparency is used (non-zero) for the text in Inkscape, but the package 'transparent.sty' is not loaded}%
    \renewcommand\transparent[1]{}%
  }%
  \providecommand\rotatebox[2]{#2}%
  \newcommand*\fsize{\dimexpr\f@size pt\relax}%
  \newcommand*\lineheight[1]{\fontsize{\fsize}{#1\fsize}\selectfont}%
  \ifx\svgwidth\undefined%
    \setlength{\unitlength}{993bp}%
    \ifx\svgscale\undefined%
      \relax%
    \else%
      \setlength{\unitlength}{\unitlength * \real{\svgscale}}%
    \fi%
  \else%
    \setlength{\unitlength}{\svgwidth}%
  \fi%
  \global\let\svgwidth\undefined%
  \global\let\svgscale\undefined%
  \makeatother%
  \begin{picture}(1,0.40584089)%
    \lineheight{1}%
    \setlength\tabcolsep{0pt}%
    \put(0,0){\includegraphics[width=\unitlength,page=1]{application_network_main-tex.pdf}}%
  \end{picture}%
\endgroup%

%% file: appendix.tex
\section{FisherMALA Algorithm and Prior Specification}
\label{appendix:the_fishermala_sampling_algorithm_and_priors}

This appendix specifies the FisherMALA sampler \citep{Titsias2024} used throughout the paper. The method extends the Metropolis-Adjusted Langevin Algorithm by preconditioning the proposal with an online estimate of the Fisher information matrix of the target distribution. The proposal covariance is given by the inverse Fisher information and is updated during sampling.

The sampler updates the full parameter vector ($\bm{\eta}$) jointly and adapts the step size to target an acceptance probability of $\alpha_{\text{target}}=0.574$. It is used for all methods in the main text to ensure comparability of results across correction approaches.

\subsection{FisherMALA Algorithm}

Algorithm~\ref{algorithm:fisher_adaptive_mala} describes a single transition of the FisherMALA Markov chain.

\begin{algorithm}[ht]
    \caption{FisherMALA transition step}
    \label{algorithm:fisher_adaptive_mala}
    \begin{algorithmic}
    \State{\textbf{Input:}} current parameter state $\bm{\eta}$, and log-posterior and gradient $(\log \pi(\bm{\eta}\mid\mathbf{X}), \nabla \log \pi(\bm{\eta}\mid\mathbf{X}))$, target acceptance probability $\alpha_{\text{target}} = 0.574$, learning rate $\rho = 0.015$, damping parameter $\lambda = 10$, $\text{iter}_{\text{MALA}} = 500$, and $d = |\bm{\eta}|$.
    \State{\textbf{Stored from iteration $s-1$:}} step size $\sigma^2$, square root of the inverse Fisher information $\mathbf{R}$, and normalized step size $\sigma^2_\mathbf{R}$.
    \Statex
    \If{$s < \text{iter}_{\text{MALA}}$} \textsc{(Simple MALA phase)}
        \State Propose $\bm{\eta}' = \bm{\eta} + (\sigma^2/2)\nabla \log \pi(\bm{\eta}\mid\mathbf{X}) + \sigma\bm{\zeta}$, \quad $\bm{\zeta} \sim \mathcal{N}(\mathbf{0}, \mathbf{I})$.
        \State Compute $(\log \pi(\bm{\eta}'\mid\mathbf{X}), \nabla \log \pi(\bm{\eta}'\mid\mathbf{X}))$.
        \State Compute $\alpha(\bm{\eta}, \bm{\eta}') = \min\left(1, e^{\log \pi(\bm{\eta}'\mid\mathbf{X}) + \log q(\bm{\eta} \mid \bm{\eta}') - \log \pi(\bm{\eta}\mid\mathbf{X}) - \log q(\bm{\eta}' \mid \bm{\eta})}\right)$.
        \State Adapt step size $\sigma^2 \gets \sigma^2 [1 + \rho(\alpha(\bm{\eta}, \bm{\eta}') - \alpha_{\text{target}})]$.
    \Else \textsc{(Fisher adaptive MALA phase)}
        \If{$s = \text{iter}_{\text{MALA}}$} \textsc{(Initialization of $\mathbf{R}$ and $\sigma^2_\mathbf{R}$)}
            \State Initialize square root of the inverse Fisher information $\mathbf{R}$.
            \State Initialize normalized step size $\sigma^2_\mathbf{R} = \sigma^2 / \frac{1}{d}\mathrm{tr}(\mathbf{R}\mathbf{R}^\intercal)$. 
        \EndIf
        \State Propose $\bm{\eta}' = \bm{\eta} + (\sigma^2_\mathbf{R}/2)\mathbf{R}(\mathbf{R}^\intercal \nabla \log \pi(\bm{\eta}\mid\mathbf{X})) + \sigma_\mathbf{R} \mathbf{R}\bm{\zeta}$, \quad $\bm{\zeta} \sim \mathcal{N}(\mathbf{0}, \mathbf{I})$.
        \State Compute $(\log \pi(\bm{\eta}'\mid\mathbf{X}), \nabla \log \pi(\bm{\eta}'\mid\mathbf{X}))$.
        \State Compute $\alpha(\bm{\eta}, \bm{\eta}') = \min\left(1, e^{\log \pi(\bm{\eta}'\mid\mathbf{X}) + \log q(\bm{\eta} \mid \bm{\eta}') - \log \pi(\bm{\eta}\mid\mathbf{X}) - \log q(\bm{\eta}' \mid \bm{\eta})}\right)$.
        \State Compute adaptation signal $\mathbf{s}^{\delta} = \sqrt{\alpha(\bm{\eta}, \bm{\eta}')}(\nabla \log \pi(\bm{\eta}'\mid\mathbf{X}) - \nabla \log \pi(\bm{\eta}\mid\mathbf{X}))$.
        \State Update square root of the inverse Fisher information $\mathbf{R}$ using $\mathbf{s}^{\delta}$.
        \State Adapt step size $\sigma^2 \gets \sigma^2 [1 + \rho(\alpha(\bm{\eta}, \bm{\eta}') - \alpha_{\text{target}})]$.
        \State Normalize step size $\sigma^2_\mathbf{R} = \sigma^2 / \frac{1}{d}\mathrm{tr}(\mathbf{R}\mathbf{R}^\intercal)$. 
    \EndIf
    \Statex
    \State Accept $\bm{\eta}_{s+1} = \bm{\eta}'$ with probability $\alpha(\bm{\eta}, \bm{\eta}')$; otherwise set $\bm{\eta}_{s+1} = \bm{\eta}$. If accepted, set $(\bm{\eta}_{s+1}, \log \pi(\bm{\eta}_{s+1}), \nabla \log \pi(\bm{\eta}_{s+1})) = (\bm{\eta}', \log \pi(\bm{\eta}'\mid\mathbf{X}), \nabla \log \pi(\bm{\eta}'\mid\mathbf{X}))$.
    \end{algorithmic}
\end{algorithm}

The sampler operates in three phases: an initial simple MALA phase, initialization of the preconditioning matrix, and the adaptive FisherMALA phase. In our implementation, $\sigma^2$ is initialized at $1.0$ for all methods.

\subsection{Likelihood-Specific Log-Posterior and Gradient}

The FisherMALA transition requires evaluation of the log-posterior density $\log{\pi(\bm{\eta}\mid\mathbf{X})}$ and its gradient $\nabla\log{\pi(\bm{\eta}\mid\mathbf{X})}$. These quantities depend on the log-likelihood function $\log{\pi(\mathbf{X}\mid\bm{\eta})}$. Below, we specify their form under each likelihood and describe the Monte Carlo approximation in the DMH variant.

\paragraph{Full likelihood.}
The logarithm of the full likelihood function in Eq. \eqref{eq:sample_model_ordinal_mrf_model} and its gradient are given by
\begin{align*}
     \log{\pi(\mathbf{X}\mid\bm{\eta})} &= \log{f(\mathbf{X};\bm{\eta})}  \\
     \nabla\log{\pi(\mathbf{X}\mid\bm{\eta})} &= \nabla\log{f(\mathbf{X};\bm{\eta})} = \mathbf{s}(\mathbf{X}) - n\sum_{\mathbf{X}'\in\mathcal{X}}{\mathbf{s}(\mathbf{X}')Pr(\mathbf{X} =\mathbf{X}'\mid\bm{\eta})},
\end{align*}
where $\mathbf{s}(\mathbf{X})$ denotes the vector of sufficient statistics, and the second term is the expectation of the sufficient statistics under $Pr(\mathbf{X} =\mathbf{X}'\mid\bm{\eta})$. The probability
\[
Pr(\mathbf{X} =\mathbf{X}'\mid\bm{\eta}) = \exp{\left\lbrace-E(\mathbf{X}';\bm{\eta})\right\rbrace} / \sum_{\mathbf{X}''\in\mathcal{X}}\exp{\left\lbrace-E(\mathbf{X}'';\bm{\eta})\right\rbrace},
\]
is defined over the full state space $\mathcal{X}$.

\paragraph{Pseudo-likelihood.}
The logarithm of the pseudo-likelihood function in Eq. \eqref{eq:sample_pseudo_likelihood_mrf_model} and its gradient are given by
\begin{align*}
     \log{\pi(\mathbf{X}\mid\bm{\eta})} &= \log{\tilde{f}(\mathbf{X};\bm{\eta})} \\
     \nabla\log{\pi(\mathbf{X}\mid\bm{\eta})} &= \nabla\log{\tilde{f}(\mathbf{X};\bm{\eta})} = \mathbf{\tilde{s}}(\mathbf{X}) - \sum_{\nu=1}^{n}\nabla\log{\tilde{Z}(\mathbf{X}_{\nu},\bm{\eta})}
\end{align*}
where $\mathbf{\tilde{s}}(\mathbf{X})$ denotes the vector of sufficient statistics $\mathbf{s}(\mathbf{X})$ with the interaction entries multiplied by two, because each cross product $X_{\nu i}X_{\nu j}$ enters the full conditionals of both variables $i$ and $j$, and the second term is the gradient of the log normalizing constant $\tilde{Z}(\mathbf{X},\bm{\eta})$.

\paragraph{Double Metropolis--Hastings (DMH).}    
In the DMH variant, an inner Gibbs sampler is embedded within the FisherMALA transition to approximate the full likelihood gradient and the ratio of normalizing constants. 
    
The full likelihood gradient is
\[
    \nabla\log{\pi(\mathbf{X}\mid\bm{\eta})} = \mathbf{s}(\mathbf{X}) - n\sum_{\mathbf{X}'\in\mathcal{X}}{\mathbf{s}(\mathbf{X}')Pr(\mathbf{X} =\mathbf{X}'\mid\bm{\eta})}
\]
where the second term is intractable because it requires summation over the full state space. We approximate this expectation using $L$ Monte Carlo samples,
\[
    n\sum_{\mathbf{X'}\in\mathcal{X}}{\mathbf{s}(\mathbf{X}')Pr(\mathbf{X} =\mathbf{X}'\mid\bm{\eta})} \approx n \ \frac{1}{L}\sum_{l=1}^{L}\mathbf{s}(\mathbf{Y}_l)
\]
where each sample $\mathbf{Y}_l$ with $l=1,\ldots,L$ is generated via a Gibbs sampler.

For DMH, the inner Gibbs sampler is run for five iterations per outer step and initialized at an observation $\mathbf{X}_\nu$ drawn at random from the data. Additional implementation details are given in Appendix \ref{appendix:framework_for_dataset_simulation_and_analysis}.

\subsection{Approximation of the Log Normalizing Constant Ratio}

In the acceptance probability
\[
\alpha(\bm{\eta}, \bm{\eta}') = \min\left(1, e^{\log \pi(\bm{\eta}'\mid\mathbf{X}) + \log q(\bm{\eta} \mid \bm{\eta}') - \log \pi(\bm{\eta}\mid\mathbf{X}) - \log q(\bm{\eta}' \mid \bm{\eta})}\right),
\]
the computational complexity arises from the difference between log posterior densities
\[
\log \pi(\bm{\eta}'\mid\mathbf{X}) - \log \pi(\bm{\eta}\mid\mathbf{X}) = \log{\pi(\mathbf{X}\mid\bm{\eta}')} + \log{\pi(\bm{\eta}')} -\log{\pi(\mathbf{X}\mid\bm{\eta})} - \log{\pi(\bm{\eta})}.
\]
The difference between the log-likelihood terms is
\begin{equation*}
    \begin{split}
    \log{\pi(\mathbf{X}\mid\bm{\eta}')}-\log{\pi(\mathbf{X}\mid\bm{\eta})} &= \log{f(\mathbf{X};\bm{\eta}')}-\log{f(\mathbf{X};\bm{\eta})} + n\left(\log{Z(\bm{\eta})}-\log{Z(\bm{\eta}')}\right)  \\ &= (\bm{\eta}'-\bm{\eta})^{\intercal}\mathbf{s}(\mathbf{X}) + n\left(\log{Z(\bm{\eta})}-\log{Z(\bm{\eta}')}\right),    
    \end{split}
\end{equation*}
which requires evaluating the normalizing constant $Z(\bm{\eta})$ at both the proposed parameter $\bm{\eta}'$ and the current parameter $\bm{\eta}$. When the state space $\mathcal{X}$ is large, this evaluation is infeasible. 

We therefore approximate the difference between the log normalizing constants. Following \citet[][p.~128, Eq.~5.9]{Murray2007},
\begin{equation}
   n\left(\log{Z(\bm{\eta})}-\log{Z(\bm{\eta}')}\right)\approx n \ \log{\frac{1}{L}\sum_{l=1}^{L}\exp{\left((\bm{\eta}-\bm{\eta}')^{\intercal}\mathbf{s}(\mathbf{Y}_l)\right)}} 
   \label{eq:murray_log_partition_ratio_approximation}
\end{equation}
where $\mathbf{Y}_{l}$, $l=1,\ldots,L$, are the same Monte Carlo samples used to approximate the full posterior gradient.

\subsection{Prior Distributions and Curvature}
All methods use the same prior distribution for the parameter $\bm{\eta} = (\bm{\mu},\bm{\theta})$, with separate priors for the threshold parameters ($\bm{\mu}$) and the interaction parameters ($\bm{\theta}$). 

Independent beta-prime distributions are placed on the exponentiated threshold parameters $\exp(\mu_{ih})\sim\text{Beta-Prime}(a,b)$. Using a change of variables the induced prior on $\bm{\mu}$ is
\[\pi(\bm{\mu}) = \pi(\exp(\bm{\mu})) \times\left|\text{det}\left(\frac{\partial\exp(\bm{\mu})}{\partial\bm{\mu}}\right)\right|\propto \prod_{i=1}^{p}\prod_{h=1}^{m} \frac{\exp{\left(\mu_{ih}\right)}^{a}}{(1+\exp{\left(\mu_{ih}\right)})^{a+b}}\]
with hyperparameters $a=b=0.5$. 

Independent Cauchy distributions are assigned to the interaction parameters,
\[\pi(\bm{\theta})\propto \prod_{i=1}^{p-1}\prod_{j=i+1}^{p}\left(1+\frac{\theta_{ij}^{2}}{\gamma^{2}}\right)^{-1}\]
with scale parameter $\gamma=2.5$ and location parameter zero. The priors on $\bm{\mu}$ and $\bm{\theta}$ are assumed independent, so that
\[
\log{\pi(\bm{\eta})} = \log{\pi(\bm{\mu})} + \log{\pi(\bm{\theta})}.
\]

The prior curvature matrix is obtained from the Hessian of the log prior
\[
-\mathbf{H}_{\bm{\eta}} 
= - \nabla^2_{\bm{\eta}} \log \pi(\bm{\eta})
=
\begin{pmatrix}
- \nabla^2_{\bm{\mu}} \log \pi(\bm{\mu}) & \mathbf{0} \\
\mathbf{0} & - \nabla^2_{\bm{\theta}} \log \pi(\bm{\theta})
\end{pmatrix}.
\]
The diagonal blocks are
\[
- \nabla^2_{\bm{\mu}} \log \pi(\bm{\mu})
= \mathrm{diag}\!\left( (a+b)\,\frac{e^{\mu_{ih}}}{\left(1 + e^{\mu_{ih}}\right)^2} \right),
\]
and
\[
- \nabla^2_{\bm{\theta}} \log \pi(\bm{\theta})
= \mathrm{diag}\!\left( \frac{2(\gamma^2 - \theta_{ij}^2)}{(\gamma^2 + \theta_{ij}^2)^2} \right).
\]


\section{Post Hoc Calibration: Curvature Estimation and Implementation}
\label{appendix:post_hoc_calibration_methodological_details_and_implementation}

This appendix describes the technical specifications of the post hoc calibration methods, focusing on the estimation of the posterior covariance structure.

All three methods compute the posterior covariance ($\bm{\Gamma\Gamma}^{\intercal}$) by inverting the sum of the calibrated model curvature ($-\mathbf{H}_{\text{calibrated}}$) and the prior curvature ($-\mathbf{H}_{\bm{\eta}}$), that is
\[
    \bm{\Gamma}\bm{\Gamma}^{\intercal} = -(\mathbf{H}_{\text{calibrated}} + \mathbf{H}_{\bm{\eta}})^{-1}.
\]
The methods differ in how the calibrated model curvature ($-\mathbf{H}_{\text{calibrated}}$) is computed and in the choice of centering parameter ($\bm{\eta}^{\star}$) used for rescaling. The draws are standardized around the pseudo-posterior mode ($\hat{\bm{\eta}}_{\text{MAP}}$) and re-centered at $\bm{\eta}^\star$,
\[
    \bm{\eta}_{\text{rescaled}} = \bm{\Gamma} \mathbf{L}^{\intercal } (\bm{\eta} - \hat{\bm{\eta}}_{\text{MAP}}) + \bm{\eta}^{\star}. 
\]
For PH-GHW and PH-MCH $\bm{\eta}^\star=\hat{\bm{\eta}}_{\text{MAP}}$; for PH-RM $\bm{\eta}^\star=\hat{\bm{\eta}}_{\text{RM}}$, the estimate obtained by Robbins-Monro iterations.

Since the prior curvature ($-\mathbf{H}_{\bm{\eta}}$) is defined in Appendix \ref{appendix:the_fishermala_sampling_algorithm_and_priors}, we focus here on the computation of the calibrated model curvature for each method.

\subsection{Post Hoc Calibration via Godambe-Huber-White (PH-GHW)}

The PH-GHW method is based on the Godambe-Huber-White covariance estimator \citep{Godambe1960,Huber1967,White1980}. Posterior draws are centered at the pseudo-posterior mode (MAP), $\bm{\eta}^{\star}=\hat{\bm{\eta}}_{\text{MAP}}$. The calibrated model curvature is defined as
\[
 -\mathbf{H}_{\text{GHW}} = \bm{\Sigma}^{-1}_{\text{GHW}} = \left[\left(-\mathbf{H}^{-1}\right) \times \mathbf{U} \times \left(-\mathbf{H}^{-1}\right)\right]^{-1}
\]
where $\mathbf{H}$ is the Hessian of the pseudo-log-likelihood evaluated at $\bm{\eta}^{\star}$ and 
\[
    \mathbf{U} = \sum_{\nu=1}^{n}\mathbf{u}_{\nu}\mathbf{u}_{\nu}^{\intercal}
\]
is the empirical variance of the score vectors $\mathbf{u}_\nu = \nabla\log\tilde f(\mathbf{X}_\nu;\bm{\eta}^\star)$.
The matrix $\bm{\Sigma}_{\text{GHW}}$ corresponds to the sandwich covariance estimator. This method does not require simulation from the full likelihood.

\subsection{Post Hoc Calibration via Monte Carlo Hessian Approximation (PH-MCH)}

The PH-MCH method rescales the pseudo-posterior draws using a Monte Carlo approximation of the posterior curvature. The draws are centered at the pseudo-posterior mode $\bm{\eta}^{\star}=\hat{\bm{\eta}}_{\text{MAP}}$, and the calibrated model curvature is defined as
\[
-\mathbf{H}_{\text{MCH}} \approx n\left[\mathbb{E}_{\mathbf{X}\mid\bm{\eta}^{\star}}\left[\mathbf{s}(\mathbf{X})\mathbf{s}(\mathbf{X})^{\intercal}\right] - \mathbb{E}_{\mathbf{X}\mid\bm{\eta}^{\star}}\left[\mathbf{s}(\mathbf{X})\right]\mathbb{E}_{\mathbf{X}\mid\bm{\eta}^{\star}}\left[\mathbf{s}(\mathbf{X})\right]^{\intercal}\right].
\] 
The expectations are approximated using $T = 100{,}000$ Monte Carlo samples,
\begin{align*}
\mathbb{E}_{\mathbf{X}\mid\bm{\eta}^{\star}}\left[\mathbf{s}(\mathbf{X})\mathbf{s}(\mathbf{X})^{\intercal}\right] &\approx \frac{1}{T}\sum_{t=1}^{T}\mathbf{s}\left(\mathbf{Y}_{t}\right)\mathbf{s}\left(\mathbf{Y}_{t}\right)^{\intercal}, \\
\mathbb{E}_{\mathbf{X}\mid\bm{\eta}^{\star}}\left[\mathbf{s}(\mathbf{X})\right] &\approx \frac{1}{T}\sum_{t=1}^{T}\mathbf{s}\left(\mathbf{Y}_{t}\right),
\end{align*}
where each $\mathbf{Y}_{t}\mid\bm{\eta}^{\star}\sim p(\cdot\mid\bm{\eta}^{\star})$ is generated by five Gibbs iterations started at a randomly chosen observed response pattern, and $\mathbf{s}\left(\mathbf{Y}_{t}\right)$ denotes the corresponding sufficient statistics.

\subsection{Post Hoc Calibration via Robbins-Monro (PH-RM)}

The PH-RM method follows the Robbins-Monro stochastic approximation approach  \citep{Bouranis2017, RobbinsMonro}. The centering parameter $\hat{\bm{\eta}}_{\text{RM}}$ is obtained via Robbins-Monro iterations, initialized at the pseudo-posterior mode $\hat{\bm{\eta}}_{\text{MAP}}$. 

The calibrated model curvature $-\mathbf{H}_{\text{RM}}$ is the Monte Carlo curvature of PH-MCH, evaluated at $\hat{\bm{\eta}}_{\text{RM}}$ instead of the pseudo-posterior mode.

The posterior covariance is
\[
    \bm{\Gamma}_{\text{RM}}\bm{\Gamma}_{\text{RM}}^{\intercal} = -(\mathbf{H}_{\text{RM}} + \mathbf{H}_{\bm{\eta}})^{-1},
\]
and the rescaled draws are
\[
\bm{\eta}_{\text{rescaled}} = \bm{\Gamma}_{\text{RM}}\mathbf{L}^\intercal(\bm{\eta}-\hat{\bm{\eta}}_{\text{MAP}}) + \hat{\bm{\eta}}_{\text{RM}}.
\]

In the simulation, the Robbins-Monro algorithm starts at the pseudo-posterior mode and updates $\bm{\eta}_{t+1} = \bm{\eta}_t + (t+1)^{-1}\,\mathbf{M}\,\widehat{\nabla}\log\pi(\bm{\eta}_t \mid \mathbf{X})$ for $t = 0, \dots, 49$. Here $\widehat{\nabla}\log\pi(\bm{\eta}_t \mid \mathbf{X})$ is the Monte Carlo gradient of the full log-posterior, computed from $100{,}000$ auxiliary observations, each generated by five Gibbs iterations started at a randomly chosen observed response pattern. $\mathbf{M}$ is the inverse of the negative Monte Carlo Hessian of the full log-posterior at the pseudo-posterior mode. This Newton-type gain \citep{Ruppert1985, GuZhu2001} makes the step size independent of the curvature of the likelihood, which grows with sample size and network size.


\section{Coordinate Rescaling Under a Random-Walk Proposal}
\label{appendix:random_walk_proposal}

\par All methods in the main comparison share the FisherMALA proposal \citep{Titsias2024}, so the comparison may confound the effect of coordinate rescaling with that of the proposal distribution. The rescaling matrix $\mathbf{A}$ defines the target distribution (Section \ref{subsection:the-core-sampler}), whereas the proposal determines only how efficiently the Markov chain explores it. After convergence, the calibration results of the main text hold for any valid proposal. In this appendix, we compare sampling efficiency across two targets, the pseudo-posterior and CoRe, and two proposals, FisherMALA and a random-walk Metropolis (RW) proposal without adaptation.

We consider the full network with $p = 15$ variables and $n = 1{,}000$ observations ($d = 150$ parameters) and reuse $20$ of the $100$ simulated datasets. The FisherMALA sampler is that of the main study (Appendix \ref{appendix:the_fishermala_sampling_algorithm_and_priors}). The RW sampler uses a symmetric Gaussian proposal with a scale fixed before sampling: $\bm{\eta}^\prime = \bm{\eta} + \bm{\epsilon}$ with $\bm{\epsilon} \sim \mathcal{N}\left(\mathbf{0}, \frac{2.38^2}{d}(\mathbf{L}\mathbf{L}^{\intercal})^{-1}\right)$ for the pseudo-posterior, and $\bm{\beta}^\prime = \bm{\beta} + \bm{\epsilon}$ with $\bm{\epsilon} \sim \mathcal{N}\left(\mathbf{0}, \frac{2.38^2}{d}\mathbf{\Gamma}\mathbf{\Gamma}^{\intercal}_{\text{GHW}}\right)$ for CoRe (Section \ref{subsection:the-rescaling-matrix-A}), with $d =|\bm{\eta}|$. Since $\mathbf{A}(\mathbf{L}\mathbf{L}^{\intercal})^{-1}\mathbf{A}^{\intercal} = \mathbf{\Gamma}\mathbf{\Gamma}^{\intercal}_{\text{GHW}}$, each proposal is matched to the Gaussian approximation of its own target. The factor $2.38^2/d$ is the optimal scaling for random-walk Metropolis \citep{Roberts1997, Roberts2001}.

For each dataset and sampler, we run four chains of $25{,}000$ iterations and discard the first $5{,}000$ as burn-in. The chains start from overdispersed values $\bm{\eta}^{\star} + \bm{\xi}$, with $\bm{\xi} \sim \mathcal{N}\left(\mathbf{0}, (\mathbf{L}\mathbf{L}^{\intercal})^{-1}\right)$, shared across the four samplers. We compute the bulk effective sample size (ESS) and the rank-normalized split-$\widehat{R}$ \citep{Vehtari2021} for each parameter from the four chains using the R package \texttt{posterior} \citep{posterior-package}, and the ESS per second from the summed run time of the four chains, including the computation of $\mathbf{A}$ for CoRe. \footnote{All chains ran on the same machine: a MacBook Air with an Apple M3 chip (8 cores: 4 performance and 4 efficiency) and 16 GB of memory, with eight chains in parallel. Runtimes are therefore comparable within this appendix but not with the runtimes reported in the main text.}

\begin{table}[t]
\centering
\caption{Sampling efficiency of the pseudo-posterior (Pseudo) and CoRe targets under the FisherMALA and random-walk (RW) proposals, for a full network with $p = 15$ variables and $n = 1{,}000$ observations ($d = 150$ parameters). Entries are medians across $20$ datasets, with interquartile ranges between brackets. For each dataset, ESS and $\widehat{R}$ are summarized across parameters and are based on four chains of $20{,}000$ draws after burn-in.}
\label{tab:random_walk}
\begin{adjustbox}{max width=\textwidth}
\small
\begin{tabular}{llcccccc}
\toprule
Target & Proposal & Median ESS & Min.\ ESS & ESS/s & Median $\widehat{R}$ & Max.\ $\widehat{R}$ & Acceptance \\
\midrule
Pseudo & FisherMALA & \mediqr{6,609}{6,588; 6,646} & \mediqr{3,214}{2,828; 4,040} & \mediqr{28.7}{28.4; 28.9} & \mediqr{1.001}{1.001; 1.001} & \mediqr{1.002}{1.002; 1.002} & \mediqr{0.57}{0.57; 0.58} \\
Pseudo & RW & \mediqr{179}{179; 181} & \mediqr{55}{49; 73} & \mediqr{2.6}{2.5; 2.7} & \mediqr{1.021}{1.020; 1.021} & \mediqr{1.068}{1.063; 1.075} & \mediqr{0.24}{0.24; 0.24} \\[2pt]
\midrule
CoRe & FisherMALA & \mediqr{6,625}{6,614; 6,658} & \mediqr{5,083}{4,573; 5,646} & \mediqr{27.6}{27.4; 28.0} & \mediqr{1.001}{1.001; 1.001} & \mediqr{1.002}{1.002; 1.002} & \mediqr{0.58}{0.57; 0.58} \\
CoRe & RW & \mediqr{179}{178; 182} & \mediqr{70}{56; 76} & \mediqr{2.4}{2.2; 2.5} & \mediqr{1.021}{1.020; 1.022} & \mediqr{1.068}{1.062; 1.076} & \mediqr{0.24}{0.24; 0.24} \\[2pt]
\bottomrule
\end{tabular}
\end{adjustbox}
\end{table}

Table \ref{tab:random_walk} shows that CoRe and the pseudo-posterior (Pseudo) mix equally well overall under both proposals. The median ESS is $6{,}609$ for Pseudo and $6{,}625$ for CoRe under FisherMALA, and $179$ for both targets under RW. The $\widehat{R}$ summaries coincide. The ESS per second is slightly lower for CoRe ($27.6$ versus $28.7$ under FisherMALA and $2.4$ versus $2.6$ under RW) because of the cost of the transformation at each iteration, not because of a difference in mixing. CoRe also achieves a higher minimum ESS under both proposals ($5{,}083$ versus $3{,}214$ under FisherMALA and $70$ versus $55$ under RW).

In contrast, the choice of the proposal has a large effect on sampling efficiency. For both targets, FisherMALA yields about $37$ times more effective samples than RW and about $11$ times more effective samples per second. All FisherMALA chains reach $\widehat{R} < 1.01$ for every parameter, whereas the RW chains do not meet this threshold (maximum $\widehat{R} \approx 1.07$). Both proposals attain acceptance rates close to their optimal values ($0.234$ for RW and $0.574$ for FisherMALA), so this gap reflects the proposal mechanism rather than poor tuning, in line with optimal-scaling theory \citep{Roberts1997,Roberts1998,Roberts2001}.

Coordinate rescaling thus changes the target, and hence the calibration, but not the efficiency with which it is sampled, under either proposal. By using FisherMALA for all methods in the main comparison, we establish a common baseline that is more efficient than a random-walk proposal, without favoring CoRe.

\section{Simulation Design and Data-Generation Details}
\label{appendix:framework_for_dataset_simulation_and_analysis}

This appendix provides the details on the generation of the datasets of the simulation study (Section \ref{sec:simulation-study}) and justifies the initialization of the Gibbs sampler that generates the auxiliary data of DMH and of the Monte Carlo approximations of the full likelihood (PH-RM, PH-MCH, CoRe-RM and CoRe-MCH).

\begin{algorithm}[ht]
    \caption{Simulation of the datasets of one condition}
    \label{algorithm:simulate_random_networks_with_structure}
    \begin{algorithmic}
     \State{\textbf{Input:}} observed data $\mathbf{X}$ of dimensions $n^{\star} \times p^{\star}$, network size $p < p^{\star}$, sample size $n$, structure type $\mathcal{S}$, number of data-generating models $K_{\text{str}}$, number of random datasets per data-generating model $K_{\text{sample}}$, number of Gibbs iterations $T$.
     \State{\textbf{Output:}} list $\mathcal{D}$ of $K_{\text{str}} \times K_{\text{sample}}$ simulated datasets of dimension $n \times p$.
     \Statex
    \For{$i \gets 1$ to $K_{\text{str}}$}
        \State Sample column indices $\mathbf{c}_i$ of size $p$ without replacement from $\{1,\ldots, p^{\star}\}$.
        \State Generate a network structure $s_i$ of type $\mathcal{S}$ on the $p$ selected variables.
        \State Estimate $\hat{\bm{\eta}}_i \gets \arg\max_{\bm{\eta}}\tilde\pi(\bm{\eta} \mid \mathbf{X}[\,\cdot\,,\mathbf{c}_i])$ subject to $s_i$ (interactions of absent edges fixed at zero).
        \For{$j \gets 1$ to $K_{\text{sample}}$}
            \Repeat
                \State Generate $\mathbf{Y}_{i,j} \sim p(\cdot \mid \hat{\bm{\eta}}_i)$ with a Gibbs sampler run for $T$ iterations from a random starting state.
            \Until{every variable of $\mathbf{Y}_{i,j}$ takes at least two categories}
            \State Append $\mathbf{Y}_{i,j}$ to $\mathcal{D}$.
        \EndFor
    \EndFor
    \end{algorithmic}
\end{algorithm}

Algorithm \ref{algorithm:simulate_random_networks_with_structure} summarizes the data generation for one of the $48$ conditions. The observed data are the RADS-2 responses ($n^{\star} = 917$, $p^{\star} = 25$), and we use $K_{\text{str}} = 10$, $K_{\text{sample}} = 10$ and $T = 5{,}000$. The small-world structures are Watts-Strogatz graphs \citep{Watts1998} whose neighborhood size is drawn uniformly from $\{1, \ldots, \lfloor (p-1)/2 \rfloor\}$. The rewiring probability starts at $0.2$ and is increased in steps of $0.1$ until the graph is more clustered than random graphs with the same numbers of nodes and edges. The random structures are Erd\H{o}s-R\'enyi graphs \citep{Renyi1959} with edge-inclusion probability drawn from $\mathcal{U}(0,1)$, and the full structure contains all $p(p-1)/2$ edges. The random starting state assigns each observation a category drawn uniformly for each variable, and each Gibbs iteration cycles through the $p$ full conditional distributions $f(X_{s} \mid \mathbf{X}_{-s}, \bm{\eta})$ of every observation. A dataset in which a variable takes a single category is regenerated, because its thresholds are not identified. Implementation details of the Gibbs sampler are available in the supplementary material at \zenodorepo.

The same Gibbs sampler generates the auxiliary data of DMH and of the Monte Carlo approximations of the full likelihood. There, each auxiliary observation starts at a randomly chosen observed response pattern and is updated for only $5$ iterations. Although any initialization converges to the same stationary distribution under sufficient mixing, the starting state affects the speed of convergence: starting at observed data places the chain in a high-probability region, which reduces the required burn-in relative to a random start. 

To evaluate this choice, we compared datasets generated from observed data with datasets generated from a random starting state, with $n = 2{,}000$, $p = 10$, a random structure, $K_{\text{str}} = 100$ and $K_{\text{sample}} = 100$. The true parameters $\hat{\bm{\eta}}_i$ are obtained as in Algorithm \ref{algorithm:simulate_random_networks_with_structure}. A random structure without edges was redrawn, since the interaction RRMSE below would then be undefined. For each structure, the observed data $\tilde{\mathbf{X}}_i$ are an $n \times p$ submatrix of $\mathbf{X}[\,\cdot\,,\mathbf{c}_i]$ with rows sampled with replacement. Chains started at $\tilde{\mathbf{X}}_i$ ran for $100$ iterations without burn-in, and chains started at a random state ran for $1{,}100$ iterations, of which the first $1{,}000$ are burn-in. The generated dataset $\mathbf{Y}_{i,j}$ is the final state of the chain. In total, $10{,}000$ datasets were generated per initialization strategy.

\begin{figure}[ht]
\centering
\def\svgwidth{\textwidth}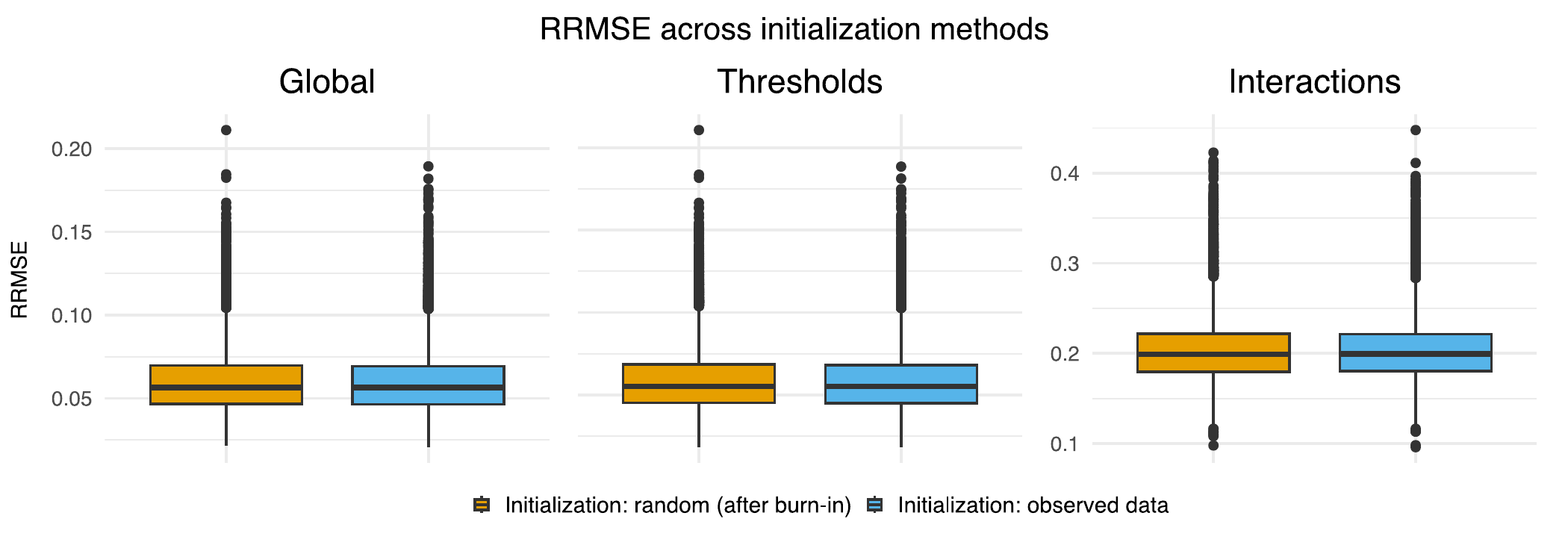
\caption{Relative root mean squared error (RRMSE) of parameter estimates obtained from the generated datasets $\mathbf{Y}_{i,j}$ evaluated against the true parameters $\hat{\bm{\eta}}_i = (\hat{\bm{\mu}}_i,\hat{\bm{\theta}}_i)$. Boxplots display the distribution of global, threshold, and interaction RRMSE across $10{,}000$ datasets under the two initialization strategies.}
\label{fig:rrmse_boxplots_initialization}
\end{figure}

\begin{figure}[ht]
\centering
\begin{center}
\def\svgwidth{\textwidth}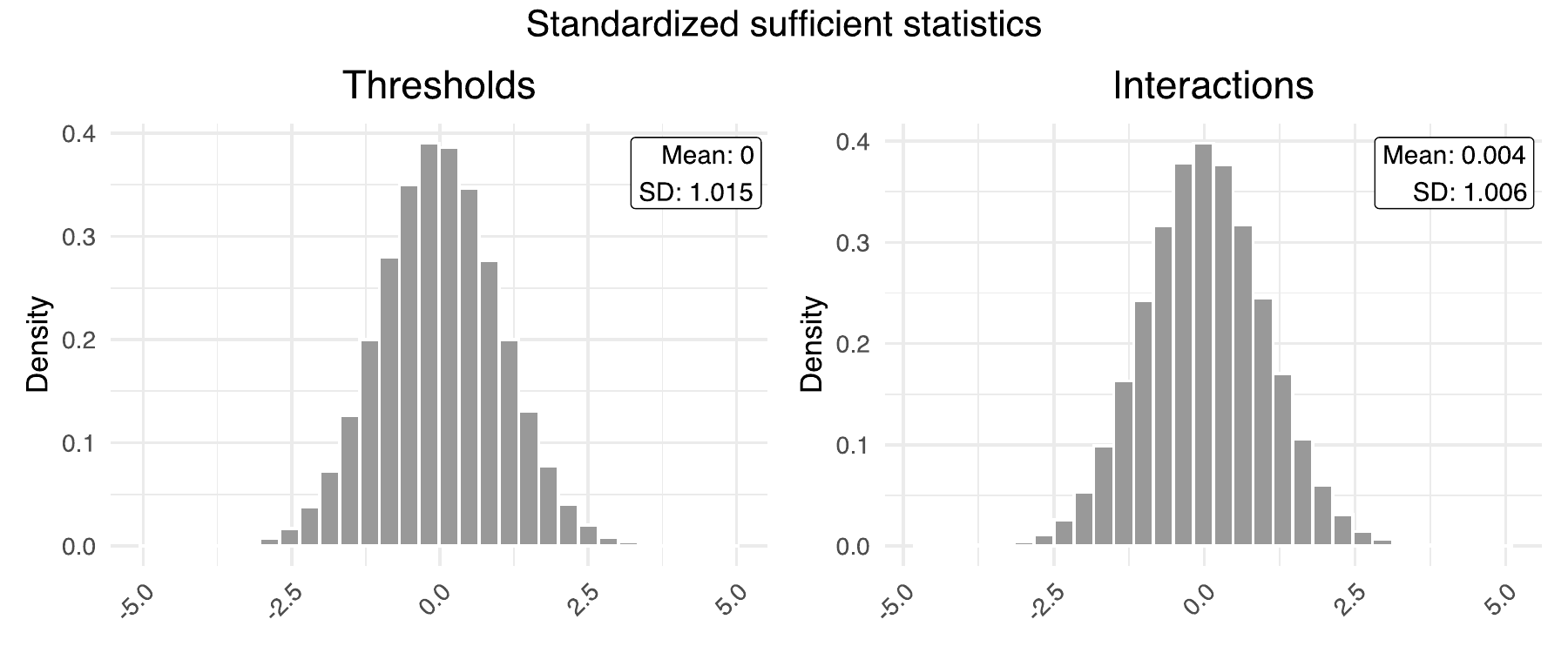
\captionof{figure}{Histograms of the sufficient statistics of the datasets $\mathbf{Y}_{i,j}$ generated under initialization at $\tilde{\mathbf{X}}_i$, standardized with the mean and standard deviation of the datasets generated from a random starting state under the same structure, shown separately for threshold and interaction statistics.}
\label{fig:histograms_standardized_sufficient_statistics}
\end{center}
\end{figure}

Performance was assessed using (i) relative root mean squared error (RRMSE) of the parameter estimates (Figure \ref{fig:rrmse_boxplots_initialization}), (ii) the distribution of sufficient statistics (Figure \ref{fig:histograms_standardized_sufficient_statistics}), and (iii) mixing behavior evaluated via traces of the negative pseudo-log-likelihood (NPL; Figures \ref{fig:npl_iterations} and \ref{fig:npl_iterations_first_10}). The results are presented below.

\begin{figure}[ht]
\begin{center}
\def\svgwidth{\textwidth}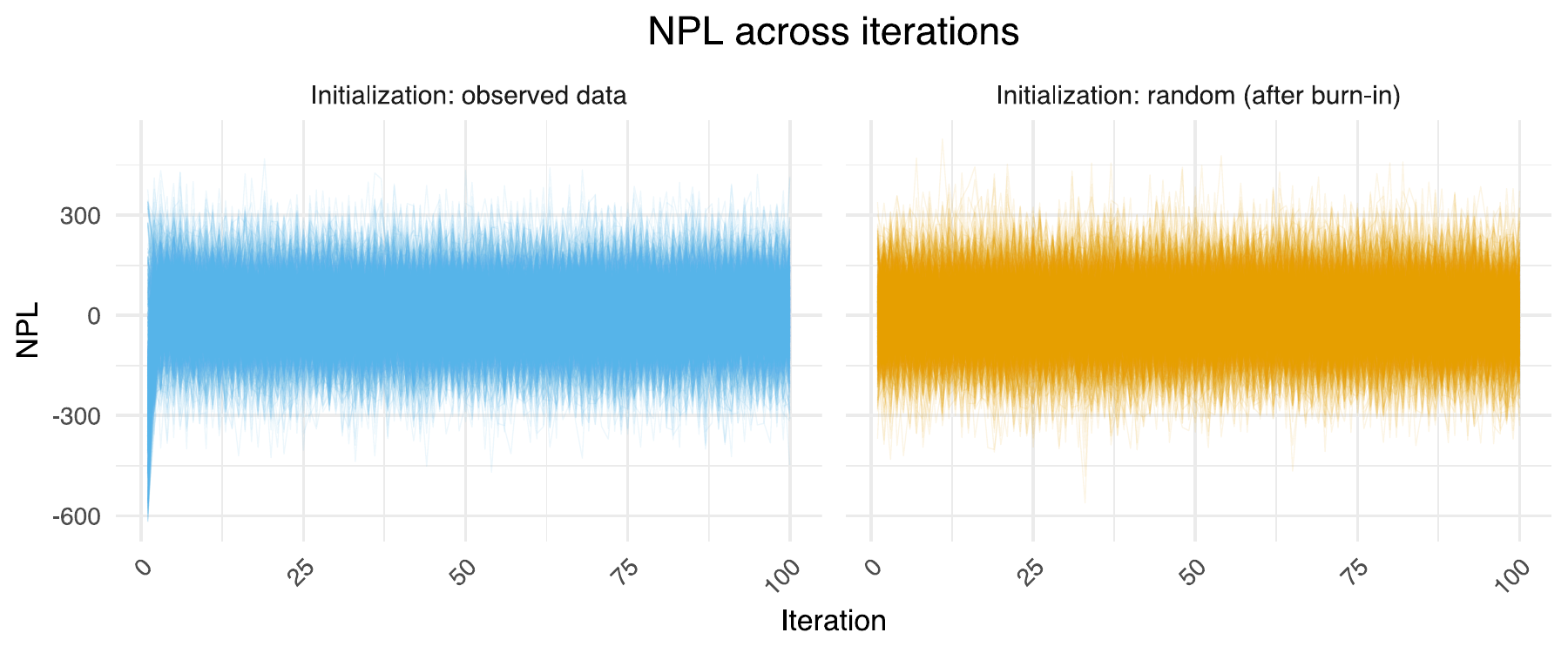
\captionof{figure}{Centered negative pseudo-log-likelihood (NPL) over $100$ Gibbs iterations for $1{,}000$ randomly selected datasets per initialization strategy. Each trajectory is centered by subtracting its mean over the iterations. For initialization at observed data, the first iteration is $\tilde{\mathbf{X}}_i$ itself; for random initialization, the iterations follow a burn-in of $1{,}000$ iterations.}
\label{fig:npl_iterations}
\end{center}
\end{figure}

\begin{figure}[ht]
\begin{center}
\def\svgwidth{\textwidth}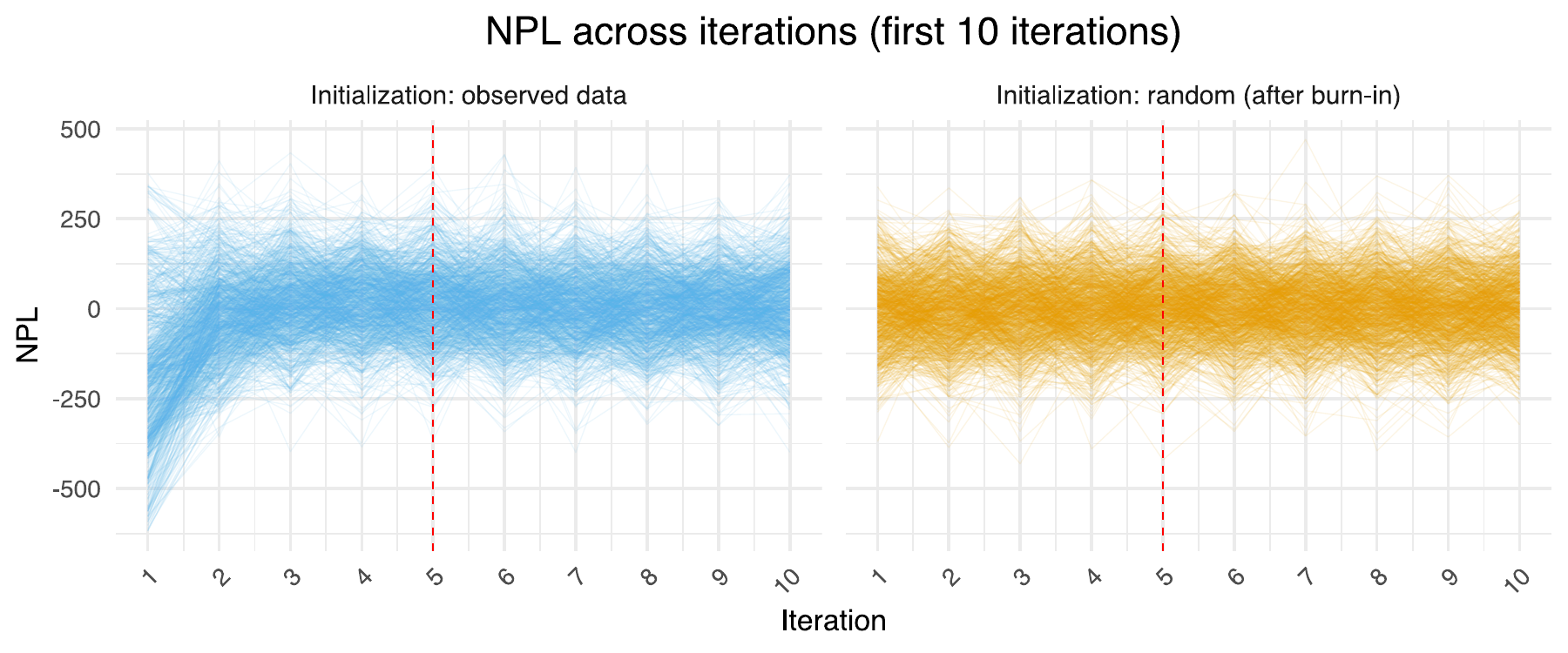
\captionof{figure}{Centered negative pseudo-log-likelihood (NPL) over the first $10$ Gibbs iterations for $1{,}000$ randomly selected datasets per initialization strategy, centered as in Figure \ref{fig:npl_iterations}. The dashed line marks iteration $5$, the number of Gibbs iterations used for the auxiliary draws of DMH and of the Monte Carlo approximations of the full likelihood.}
\label{fig:npl_iterations_first_10}
\end{center}
\end{figure}

Given a generated dataset $\mathbf{Y}_{i,j}$, we computed the RRMSE by comparing the estimated parameters with the true parameter vector $\hat{\bm{\eta}}_i = (\hat{\bm{\mu}}_i,\hat{\bm{\theta}}_i)$. We report a global RRMSE as well as separate RRMSE measures for threshold parameters $\bm{\mu}$ and interaction parameters $\bm{\theta}$. The RRMSEs are defined as
\begin{align*}  
\text{RRMSE}_{\text{global}}(i,j) &= \frac{\lVert \hat{\bm{\eta}}_{i,j} - \hat{\bm{\eta}}_i \rVert}{\lVert \hat{\bm{\eta}}_i \rVert}\text{,} \\
\text{RRMSE}_{\text{thresholds}}(i,j) &= \frac{\lVert \hat{\bm{\mu}}_{i,j} - \hat{\bm{\mu}}_i \rVert}{\lVert \hat{\bm{\mu}}_i \rVert}\text{,} \\
\text{RRMSE}_{\text{interactions}}(i,j) &= \frac{\lVert \hat{\bm{\theta}}_{i,j} - \hat{\bm{\theta}}_i \rVert}{\lVert \hat{\bm{\theta}}_i \rVert}\text{,}
\end{align*}
where $\hat{\bm{\eta}}_{i,j} = (\hat{\bm{\mu}}_{i,j},\hat{\bm{\theta}}_{i,j})$ is the pseudo-posterior mode estimated from $\mathbf{Y}_{i,j}$.
Figure \ref{fig:rrmse_boxplots_initialization} shows the distribution of the RRMSE values across the $10{,}000$ datasets for both initialization strategies. The two strategies give the same median RRMSE, $0.056$ (global), $0.055$ (thresholds) and $0.199$ (interactions), with interquartile ranges that differ by at most $0.001$. Thus, initializing at $\tilde{\mathbf{X}}_i$ preserves estimation accuracy.
    
We next compared the distribution of sufficient statistics obtained under initialization at $\tilde{\mathbf{X}}_i$ with those obtained under random initialization. For each structure $i$, we first computed the mean and standard deviation of the sufficient statistics across the $K_{\text{sample}}$ datasets generated with random starting values. We then standardized the sufficient statistics from the corresponding datasets generated under $\tilde{\mathbf{X}}_i$ using these reference moments. The standardized threshold and interaction statistics are shown in Figure \ref{fig:histograms_standardized_sufficient_statistics}. 

The distributions under $\tilde{\mathbf{X}}_i$-initialization closely overlap with those from random initialization: the standardized statistics have mean $-0.0004$ and standard deviation $1.015$ for the thresholds, and mean $0.004$ and standard deviation $1.006$ for the interactions. This indicates that the Gibbs sampler initialized at $\tilde{\mathbf{X}}_i$ reaches the same stationary distribution as under random initialization.

Finally, we examined the mixing behavior of the Gibbs sampler under both initialization strategies. The NPL is evaluated at the true parameters $\hat{\bm{\eta}}_i$ at every iteration. Figure \ref{fig:npl_iterations} displays the NPL over $100$ Gibbs iterations for $1{,}000$ randomly selected datasets per initialization strategy. For randomly initialized chains, these are the iterations after the burn-in of $1{,}000$ iterations; for chains initialized at $\tilde{\mathbf{X}}_i$, the first iteration is $\tilde{\mathbf{X}}_i$ itself. For visualization purposes, each trajectory was centered by subtracting its mean NPL across iterations.

The NPL stabilizes for both initialization strategies, indicating convergence to a stationary regime. Chains initialized at $\tilde{\mathbf{X}}_i$ start at a lower NPL than at stationarity (median centered value $-183$ at the first iteration), and reach the stationary regime within a few iterations. Figure \ref{fig:npl_iterations_first_10} shows this early behavior over the first $10$ iterations: the standard deviation of the centered NPL across chains falls from $193$ at the first iteration to $109$ at the third and $107$ at the fifth, against $105$--$106$ for the randomly initialized chains after burn-in.

These results justify the initialization of the auxiliary Gibbs draws of DMH, PH-RM, PH-MCH, CoRe-RM and CoRe-MCH at observed response patterns and running only five Gibbs iterations per draw.


\section{The Global Stepsize Parameter \texorpdfstring{$\sigma^2$}{sigma^2}}
\label{appendix:the_global_step_size_parameter}
\begin{center}
\def\svgwidth{\textwidth}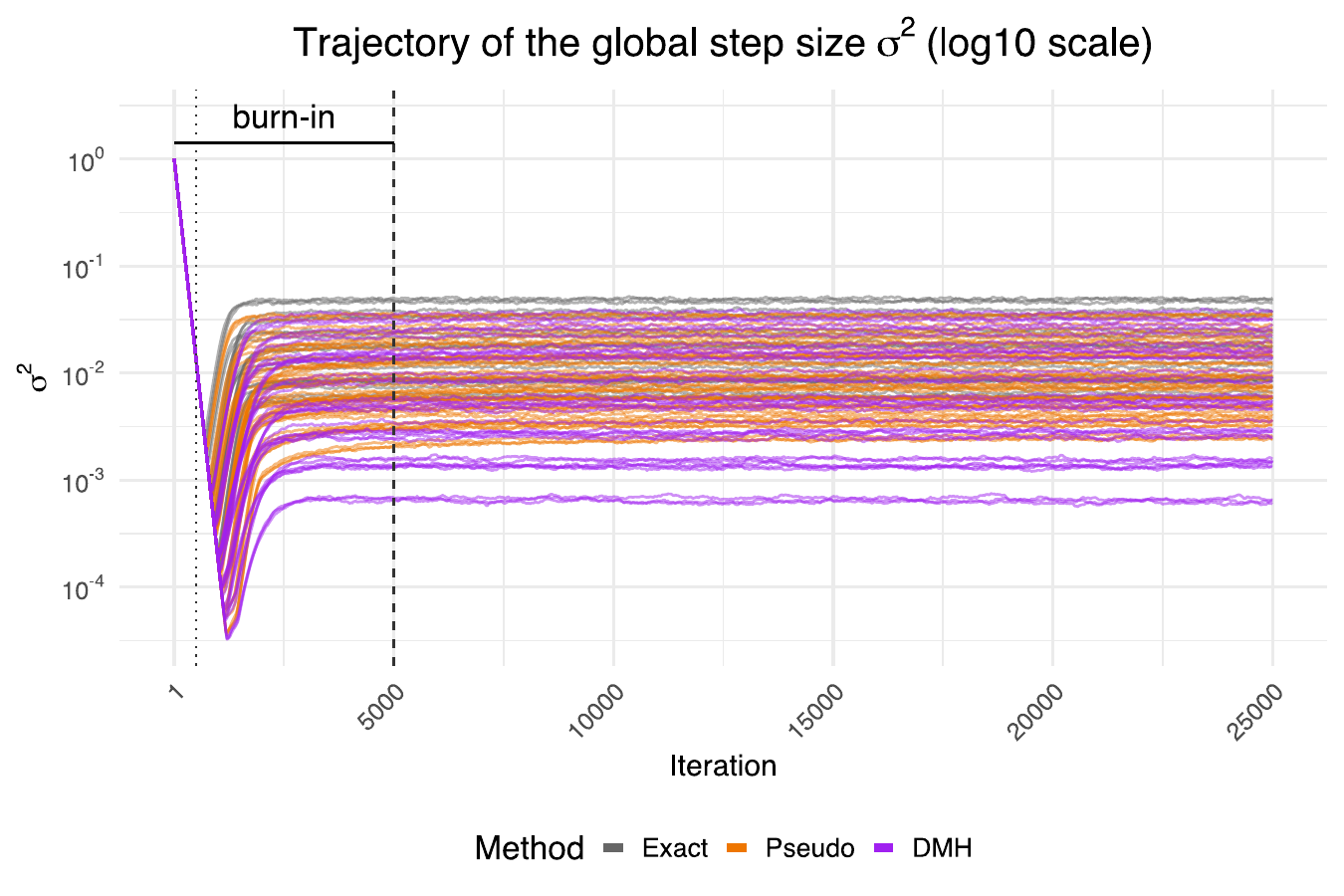
\captionof{figure}{
Trajectory of the global step size $\sigma^2$ of the FisherMALA sampler (Appendix \ref{appendix:the_fishermala_sampling_algorithm_and_priors}) for the pseudo-posterior (orange), the exact posterior (gray) and DMH (purple), on a $\log_{10}$ scale. For all samplers, $\sigma^2$ is initialized at $1.0$ and adapted at every iteration toward the target acceptance probability $\alpha_{\text{target}} = 0.574$. The dotted vertical line marks the end of the initial MALA stage ($500$ iterations) and the dashed line the end of the burn-in ($5{,}000$ iterations). For each simulation condition, five datasets were randomly selected, and at each iteration $\log_{10}\sigma^2$ was averaged across the five datasets, yielding one trajectory per condition and sampler: $48$ for the pseudo-posterior, $24$ for the exact posterior ($p \leq 10$) and $36$ for DMH ($p \geq 10$). After an initial decrease, $\sigma^2$ stabilizes within the first $3{,}500$ iterations for all samplers and conditions, at values between $6 \times 10^{-4}$ and $5 \times 10^{-2}$.}
\end{center}

%% file: img/rrmse_boxplots_initialization-tex.pdf_tex
\begingroup%
  \makeatletter%
  \providecommand\color[2][]{%
    \errmessage{(Inkscape) Color is used for the text in Inkscape, but the package 'color.sty' is not loaded}%
    \renewcommand\color[2][]{}%
  }%
  \providecommand\transparent[1]{%
    \errmessage{(Inkscape) Transparency is used (non-zero) for the text in Inkscape, but the package 'transparent.sty' is not loaded}%
    \renewcommand\transparent[1]{}%
  }%
  \providecommand\rotatebox[2]{#2}%
  \newcommand*\fsize{\dimexpr\f@size pt\relax}%
  \newcommand*\lineheight[1]{\fontsize{\fsize}{#1\fsize}\selectfont}%
  \ifx\svgwidth\undefined%
    \setlength{\unitlength}{1008bp}%
    \ifx\svgscale\undefined%
      \relax%
    \else%
      \setlength{\unitlength}{\unitlength * \real{\svgscale}}%
    \fi%
  \else%
    \setlength{\unitlength}{\svgwidth}%
  \fi%
  \global\let\svgwidth\undefined%
  \global\let\svgscale\undefined%
  \makeatother%
  \begin{picture}(1,0.35714286)%
    \lineheight{1}%
    \setlength\tabcolsep{0pt}%
    \put(0,0){\includegraphics[width=\unitlength,page=1]{rrmse_boxplots_initialization-tex.pdf}}%
  \end{picture}%
\endgroup%

%% file: img/histograms_standardized_sufficient_statistics-tex.pdf_tex
\begingroup%
  \makeatletter%
  \providecommand\color[2][]{%
    \errmessage{(Inkscape) Color is used for the text in Inkscape, but the package 'color.sty' is not loaded}%
    \renewcommand\color[2][]{}%
  }%
  \providecommand\transparent[1]{%
    \errmessage{(Inkscape) Transparency is used (non-zero) for the text in Inkscape, but the package 'transparent.sty' is not loaded}%
    \renewcommand\transparent[1]{}%
  }%
  \providecommand\rotatebox[2]{#2}%
  \newcommand*\fsize{\dimexpr\f@size pt\relax}%
  \newcommand*\lineheight[1]{\fontsize{\fsize}{#1\fsize}\selectfont}%
  \ifx\svgwidth\undefined%
    \setlength{\unitlength}{864bp}%
    \ifx\svgscale\undefined%
      \relax%
    \else%
      \setlength{\unitlength}{\unitlength * \real{\svgscale}}%
    \fi%
  \else%
    \setlength{\unitlength}{\svgwidth}%
  \fi%
  \global\let\svgwidth\undefined%
  \global\let\svgscale\undefined%
  \makeatother%
  \begin{picture}(1,0.41666667)%
    \lineheight{1}%
    \setlength\tabcolsep{0pt}%
    \put(0,0){\includegraphics[width=\unitlength,page=1]{histograms_standardized_sufficient_statistics-tex.pdf}}%
  \end{picture}%
\endgroup%

%% file: img/npl_iterations-tex.pdf_tex
\begingroup%
  \makeatletter%
  \providecommand\color[2][]{%
    \errmessage{(Inkscape) Color is used for the text in Inkscape, but the package 'color.sty' is not loaded}%
    \renewcommand\color[2][]{}%
  }%
  \providecommand\transparent[1]{%
    \errmessage{(Inkscape) Transparency is used (non-zero) for the text in Inkscape, but the package 'transparent.sty' is not loaded}%
    \renewcommand\transparent[1]{}%
  }%
  \providecommand\rotatebox[2]{#2}%
  \newcommand*\fsize{\dimexpr\f@size pt\relax}%
  \newcommand*\lineheight[1]{\fontsize{\fsize}{#1\fsize}\selectfont}%
  \ifx\svgwidth\undefined%
    \setlength{\unitlength}{864bp}%
    \ifx\svgscale\undefined%
      \relax%
    \else%
      \setlength{\unitlength}{\unitlength * \real{\svgscale}}%
    \fi%
  \else%
    \setlength{\unitlength}{\svgwidth}%
  \fi%
  \global\let\svgwidth\undefined%
  \global\let\svgscale\undefined%
  \makeatother%
  \begin{picture}(1,0.41666667)%
    \lineheight{1}%
    \setlength\tabcolsep{0pt}%
    \put(0,0){\includegraphics[width=\unitlength,page=1]{npl_iterations-tex.pdf}}%
  \end{picture}%
\endgroup%

%% file: img/npl_iterations_first_10-tex.pdf_tex
\begingroup%
  \makeatletter%
  \providecommand\color[2][]{%
    \errmessage{(Inkscape) Color is used for the text in Inkscape, but the package 'color.sty' is not loaded}%
    \renewcommand\color[2][]{}%
  }%
  \providecommand\transparent[1]{%
    \errmessage{(Inkscape) Transparency is used (non-zero) for the text in Inkscape, but the package 'transparent.sty' is not loaded}%
    \renewcommand\transparent[1]{}%
  }%
  \providecommand\rotatebox[2]{#2}%
  \newcommand*\fsize{\dimexpr\f@size pt\relax}%
  \newcommand*\lineheight[1]{\fontsize{\fsize}{#1\fsize}\selectfont}%
  \ifx\svgwidth\undefined%
    \setlength{\unitlength}{864bp}%
    \ifx\svgscale\undefined%
      \relax%
    \else%
      \setlength{\unitlength}{\unitlength * \real{\svgscale}}%
    \fi%
  \else%
    \setlength{\unitlength}{\svgwidth}%
  \fi%
  \global\let\svgwidth\undefined%
  \global\let\svgscale\undefined%
  \makeatother%
  \begin{picture}(1,0.41666667)%
    \lineheight{1}%
    \setlength\tabcolsep{0pt}%
    \put(0,0){\includegraphics[width=\unitlength,page=1]{npl_iterations_first_10-tex.pdf}}%
  \end{picture}%
\endgroup%

%% file: img/trend_global_step_size-tex.pdf_tex
\begingroup%
  \makeatletter%
  \providecommand\color[2][]{%
    \errmessage{(Inkscape) Color is used for the text in Inkscape, but the package 'color.sty' is not loaded}%
    \renewcommand\color[2][]{}%
  }%
  \providecommand\transparent[1]{%
    \errmessage{(Inkscape) Transparency is used (non-zero) for the text in Inkscape, but the package 'transparent.sty' is not loaded}%
    \renewcommand\transparent[1]{}%
  }%
  \providecommand\rotatebox[2]{#2}%
  \newcommand*\fsize{\dimexpr\f@size pt\relax}%
  \newcommand*\lineheight[1]{\fontsize{\fsize}{#1\fsize}\selectfont}%
  \ifx\svgwidth\undefined%
    \setlength{\unitlength}{648bp}%
    \ifx\svgscale\undefined%
      \relax%
    \else%
      \setlength{\unitlength}{\unitlength * \real{\svgscale}}%
    \fi%
  \else%
    \setlength{\unitlength}{\svgwidth}%
  \fi%
  \global\let\svgwidth\undefined%
  \global\let\svgscale\undefined%
  \makeatother%
  \begin{picture}(1,0.66666667)%
    \lineheight{1}%
    \setlength\tabcolsep{0pt}%
    \put(0,0){\includegraphics[width=\unitlength,page=1]{trend_global_step_size-tex.pdf}}%
  \end{picture}%
\endgroup%